\documentclass[11pt]{article}

\usepackage[margin=1in]{geometry} 
\usepackage{amsmath} 
\usepackage{amsthm} 
\usepackage{amsfonts, amssymb} 
\usepackage{graphicx} 
\usepackage{hyperref} 
\usepackage{color} 
\usepackage[dvipsnames]{xcolor} 
\usepackage{bm} 
\usepackage{natbib}
\usepackage{bbm}
\usepackage{multirow}
\usepackage{algorithmic} 
\usepackage{algorithm} 
\usepackage{tikz}
\usetikzlibrary{arrows, chains, positioning, quotes, shapes.geometric}
\usepackage{cases}
\usepackage{empheq}
\usepackage{enumitem} 

\usepackage{setspace}
\newtheorem{manualtheoreminner}{Theorem}
\newenvironment{manualtheorem}[1]{%
  \renewcommand\themanualtheoreminner{#1}%
  \manualtheoreminner
}{\endmanualtheoreminner}

\newtheorem{manualpropinner}{Proposition}
\newenvironment{manualprop}[1]{%
  \renewcommand\themanualpropinner{#1}%
  \manualpropinner
}{\endmanualpropinner}

\newtheorem{theorem}{Theorem}[section]

\newtheorem{lemma}{Lemma}[section]

\theoremstyle{definition}
\newtheorem{definition}{Definition}[section]
\newtheorem{assumption}{Assumption}[section]

\theoremstyle{remark}

\newcommand{\F}{\mathcal{F}}
\newcommand{\extendF}{\tilde{\mathcal{F}}}

\newcommand{\bx}{\mathbf{x}}
\newcommand{\bw}{\mathbf{w}}
\newcommand{\by}{\mathbf{y}}
\newcommand{\bX}{\mathbf{X}}
\newcommand{\bW}{\mathbf{W}}

\newcommand{\R}{\mathbb{R}}
\newcommand{\cond}{\,\big\vert\, }
\newcommand{\con}{; }
\newcommand{\T}{ ^{\intercal} }
\newcommand{\sT}{\mid}

\newcommand{\NC}{J}
\newcommand{\tNC}{m}
\newcommand{\NI}{n}
\newcommand{\NIT}{N}

\newcommand{\eji}{ {j(-i)} }
\newcommand{\iv}{Z}
\newcommand{\ivp}{z}
\newcommand{\IV}{\mathbf{Z}}
\newcommand{\IVp}{\mathbf{z}}
\newcommand{\trt}{D}
\newcommand{\trtp}{d}
\newcommand{\TRT}{\mathbf{D}}
\newcommand{\TRTp}{\mathbf{d}}
\newcommand{\pot}[2]{ #1^{(#2)} }
\newcommand{\sumji}{\sum_{ji}}

\newcommand{\randomf}{\widetilde{f}}
\newcommand{\randomC}{\widetilde{C}}
\newcommand{\BrandomC}{\widetilde{C}}

\newcommand{\potY}[2]{ #1^{ \langle #2 \rangle} }

\newcommand{\ITT}{{\rm ITT}}

\newcommand{\B}{\bm{\beta}}
\newcommand{\NT}{{\rm NT}}
\newcommand{\AT}{{\rm AT}}
\newcommand{\CO}{{\rm CO}}
\newcommand{\nNT}{{\rm nNT}}
\newcommand{\ind}{\mathbbm{1}}

\newcommand{\TP}{{\rm TP}} 
\newcommand{\FP}{{\rm FP}} 
\newcommand{\FN}{{\rm FN}} 

\newcommand{\Y}{\mathcal{Y}}

\newcommand{\X}{\mathcal{X}}
\newcommand{\EXP}{{\rm E}}
\newcommand{\VAR}{{\rm Var}}
\newcommand{\mo}{ \mathcal{M} }
\newcommand{\vo}{ \mathcal{V} }

\newcommand{\SolS}{Q}
\newcommand{\randomSolS}{\widetilde{Q}}
\newcommand{\SI}{\mathcal{I}_{c,h}}

\newcommand{\bsS}{\bm{\mathcal{S}}}

\hypersetup{
colorlinks=true,
linkcolor=blue,
filecolor=blue,
urlcolor=blue,
citecolor=blue
}

\graphicspath{ {plot/} }

\definecolor{red1}{RGB}{255,204,204}
\definecolor{blue1}{RGB}{204,204,255}

\numberwithin{table}{section}
\numberwithin{figure}{section}

\usepackage{rotating}
\usepackage{caption}

\usepackage[bottom]{footmisc}

\begin{document}%

\title{Assumption-Lean Analysis of Cluster Randomized Trials in Infectious Diseases for Intent-to-Treat Effects and Network Effects}
\author{Chan Park and Hyunseung Kang\footnote{The research of Hyunseung Kang was supported in part by NSF Grants DMS-1811414. We thank the Associate Editor and two anonymous referees for their valuable comments.}\\Department of Statistics, University of Wisconsin--Madison}
\date{ }
\maketitle

\begin{abstract}
Cluster randomized trials (CRTs) are a popular design to study the effect of interventions in infectious disease settings. However, standard analysis of CRTs primarily relies on strong parametric methods, usually mixed-effect models to account for the clustering structure, and focuses on the overall intent-to-treat (ITT) effect to evaluate effectiveness. The paper presents two assumption-lean methods to analyze two types of effects in CRTs, ITT effects and network effects among well-known compliance groups. For the ITT effects, we study the overall and the heterogeneous ITT effects among the observed covariates where we do not impose parametric models or asymptotic restrictions on cluster size. For the network effects among compliance groups, we propose a new bound-based method that uses pre-treatment covariates, classification algorithms, and a linear program to obtain sharp bounds. A key feature of our method is that the bounds can become narrower as the classification algorithm improves and the method may also be useful for studies of partial identification with instrumental variables. We conclude by reanalyzing a CRT studying the effect of face masks and hand sanitizers on transmission of 2008 interpandemic influenza in Hong Kong.\\
{\it Keywords:}  Bounds, Causal Inference, Noncompliance, Partial Identification, Randomization Inference
\end{abstract}

\section{Introduction}								\label{sec:1}
\subsection{Motivation: Cluster Randomized Trials in Infectious Diseases and the 2008 Interpandemic Influenza in Hong Kong}  \label{sec:1-2}

In infectious diseases, cluster randomized trials (CRTs) are a popular experimental design to study the effect of interventions where an entire cluster of individuals, usually households or villages, are randomized to treatment or control \citep{HongKong2009, China2011, Michigan2012}. CRTs are preferred if (i) clusters occur naturally or (ii) the planned intervention is designed to be implemented at the cluster-level or individual-level intervention is practically infeasible. For example, \citet{HongKong2009} ran a CRT to study the effect of giving free face masks and hand sanitizers on controlling the spread of the influenza A or B virus during the 2008 interpandemic influenza period in Hong Kong. The study randomly assigned treatment at the household level where some households received free face masks and hand sanitizers (i.e. the treated group) and other households did not receive them for free (i.e. the control group). When analyzing data from CRTs in infectious disease settings, investigators primarily use parametric methods, usually a mixed-effect model to adjust for pre-treatment covariates and intra-correlations within clusters, and focus on the overall intent-to-treat (ITT) effect, i.e. the population average effect of the cluster-level intervention on the outcome; see Section \ref{sec:estimand_def} for a formal definition of the overall ITT effect. For example, the Hong Kong study analyzed the overall ITT effect of giving free masks and hand sanitizers on reducing flu cases and adjusted for pre-treatment covariates such as age, sex, and vaccine history using a logistic mixed-effect regression model. Section \ref{sec:EXofParaMethod} of the supplementary materials contains additional examples. While simple, if the parametric models are mis-specified, the results may be misleading. Additionally, studying only the overall ITT effect may hide heterogeneity in the ITT effect in the observed covariates.

Also, individuals may not comply with the cluster-level intervention, potentially inducing meaningful spillover effects. For example, in the Hong Kong study, only 27\% to 47\% of members in the treated clusters actually chose to use the free masks. These mask users may have protected their peers who, for whatever reason, refused to use them. Or, in CRTs of vaccine studies, some may actually not get vaccinated for various reasons (e.g. immunocompromised, severe side effects). But, their vaccinated peers may protect the unvaccinated individuals in the form of herd immunity. In causal inference, this protection is a type of spillover effect \citep{HH2008,TTV2012} and Section \ref{sec:estimand_def} discusses other types of network effects that arise due to noncompliance. 

The main theme of this paper is to propose ``assumption-lean'' methods to analyze these two types of effects, the ITT effects and the network effects induced from noncompliance. That is, we lean towards making fewer assumptions, even if the effects are not point-identifiable, and the maintained assumptions are standard or generally plausible in CRTs; see Section \ref{sec:assup} for the exact assumptions.

\subsection{Our Contributions and Prior Work}
To study the ITT effects in an assumption-lean manner, we propose a modest extension of a nonparametric, regression-esque method based on \citet{Ding2019} that (i) work in CRTs, (ii) are invariant to affine transformations of the outcome, and (iii) have desirable asymptotic properties even when both the cluster size and the number of clusters are growing. To study the network effects induced by noncompliance, \citet{Kang2018} showed that point-identification of these effects is generally infeasible in a CRT without strong assumptions. Instead, we follow an assumption-lean approach where we propose a new method to obtain sharp bounds of these effects. At a high level, our new method combines linear programming (LP) and risk minimization from supervised machine learning (ML) where a trained classifier from risk minimization shrinks the LP bounds for the network effects. Also, compared to existing approaches on bounds under noncompliance (see \citet{Swanson2018} for a review), our bounds (i) use flexible ML classifiers to potentially make the bounds narrower and (ii) irrespective of classifiers' quality, our bounds will always cover the desired effect, with a good classifier leading to shorter bounds. Practically, this means that investigators can potentially get shorter bounds by not only getting good data from a CRT, but also by choosing better classification algorithms from ML. Finally, as a byproduct of our work, we propose extensions of bounds by \citet{GM2008} and \citet{LongHudgens2013} to CRT settings with interference; see Section \ref{sec:5-2-2} for details.

Our contributions fit into an ever-growing work on robust, nonparametric analysis of causal effects under interference, noncompliance, and effect heterogeneity; see \citet{BCD2014}, \citet{HH2016}, \citet{Dorie2019}, and references in these papers. Notably, \citet{Kang2018} showed impossibility results for network effects in CRTs under noncompliance, \citet{Forastiere2016} studied interference in CRTs under a Bayesian framework, and \citet{Forastiere2020arxiv} studied effect heterogeneity under interference, but with perfect compliance. Also, \citet{Kang2016} and \citet{Imai2020} studied interference and noncompliance in non-CRT settings. 

\section{Setup}								\label{sec:2}
\subsection{Review: Notation}
Let $\NC$ be the number of clusters and let each cluster be indexed by $j=1,\ldots,\NC$. Let $\NI_j$ be the number of individuals in cluster $j$ where each individual is indexed by $i=1,\ldots, \NI_j$. Let $\NIT = \sum_{j=1}^\NC \NI_j$ be the total number of individuals in the study. Let $\iv_j \in \{0,1\}$ be the treatment assignment of cluster $j$ where $\iv_j=1$ denotes that cluster $j$ was assigned to treatment and $\iv_{j} = 0$ denotes that cluster $j$ was assigned to control. Let $\trt_{ji} \in \{0,1\}$ be the observed treatment receipt/usage status of individual $i$ in cluster $j$ where $\trt_{ji} = 1$ indicates that individual $i$ used/took the treatment and $\trt_{ji}=0$ otherwise. Unlike $\iv_{j}$, there is an extra subscript $i$ in $\trt_{ji}$ because the decision to take the treatment randomized at the cluster-level occurs at the individual level and this decision is usually confounded.  Let $Y_{ji} \in \R$ be the observed outcome of individual $i$ in cluster $j$. Also, let $\bX_{ji} \in \R^p$ be $p$-dimensional pre-treatment covariates of individual $i$ in cluster $j$. 

In the Hong Kong study, $\iv_j$ represents giving away free masks and hand sanitizers to households, with $\iv_{j} = 1$ if household $j$ got free masks and hand sanitizers and $\iv_{j} = 0$ if household $j$ did not get them. $D_{ji}$ represents whether an individual in a household is using these items, with $D_{ji} = 1$ if the individual used the masks and/or hand sanitizers and $D_{ij} = 0$ if the individual did not use them. $Y_{ji}$ is the influenza status of an individual, with $Y_{ji} = 1$ if the individual did not contract the flu and $Y_{ji} = 0$ if the individual contracted the flu. The list of pre-treatment covariates $\bX_{ji} \in \R^p$ is in Table \ref{tab:HK-HTE}.

Finally, for a vector $\bm{v}$, let $\bm{v}_{(-k)}$ be the vector $\bm{v}$ with the $k$th index removed. For two non-random sequences $a_\NC$ and $b_\NC$, let $a_\NC = O(b_\NC)$ mean $\lim\sup_{\NC \rightarrow \infty} \big| a_\NC / b_\NC \big| < \infty$, $a_\NC = o(b_\NC)$ mean $\lim_{\NC \rightarrow \infty} \big| a_\NC / b_\NC \big| = 0$, and $a_\NC = \Theta(b_\NC)$ mean $a_\NC = O(b_\NC)$ plus $\lim \inf_{\NC \rightarrow \infty} \big| a_\NC / b_\NC \big| > 0$. For notational brevity, let $\sum_{j}$ and $\sum_{ji}$ mean $\sum_{j=1}^\NC$ and $\sum_{j=1}^\NC \sum_{i=1}^{\NI_j}$, respectively.

\subsection{Review: Potential Outcomes in CRTs and Interference}							\label{sec:2-2}

We use the potential outcomes notation of \citet{Neyman1923} and \citet{Rubin1974} to define causal effects. For each $\ivp_j \in \{0,1\}$, let $\trt_{ji}^{(\ivp_j)}$ denote the potential treatment receipt of individual $i$ in cluster $j$ if cluster $j$ were assigned to treatment $\ivp_j$. The vector of potential treatment receipts in cluster $j$ under treatment $\ivp_j$ is denoted as $\TRT_j^{(\ivp_j)} = (\trt_{j1}^{(\ivp_j)} , \ldots , \trt_{j\NI_j}^{(\ivp_j)})$ and the vector of potential treatment receipts in cluster $j$ that excludes individual $i$ is denoted as $\pot{\TRT_\eji}{\ivp_j} = (\pot{\trt_{j1}}{\ivp_j} , \ldots , \pot{\trt_{j,i-1}}{\ivp_j}, \pot{\trt_{j,i+1}}{\ivp_j},\ldots, \pot{\trt_{j \NI_j}}{\ivp_j} )$. Let $\IVp_{(-j)}= (z_1,\ldots z_{j-1}, z_{j+1}, \ldots ,z_\NC) \in \{0,1\}^{\NC-1}$ be the treatment vector of clusters except $j$, $\TRT^{(\IVp_{(-j)})}_{(-j)} = (\TRT_1^{(\ivp_1)}, \ldots, \TRT_{j-1}^{(\ivp_{j-1})}, \TRT_{j+1}^{(\ivp_{j+1})}, \ldots, \TRT_\NC^{(\ivp_\NC)}) \in \{0,1\}^{\NIT - \NI_j}$ be the potential treatment receipts of clusters except $j$.

Let $\pot{Y_{ji}}{\ivp_j, \IVp_{(-j)}, \trtp_{ji}, \TRTp_\eji, \TRTp_{(-j)} }$ be the potential outcome of individual $i$ in cluster $j$ if cluster $j$ and other clusters were assigned to $\ivp_j$ and $\IVp_{(-j)}$, respectively, individual $i$'s treatment receipt were $\trtp_{ji} \in \{0,1\}$, peers of individual $i$ had treatment receipt $\TRTp_{\eji} \in \{0,1\}^{\NI_j - 1}$, and individuals in clusters other than $j$ had treatment receipts $\TRTp_{(-j)} \in  \{0,1\}^{\NIT - \NI_j}$. Let $\potY{Y_{ji}}{ \ivp_j, \IVp_{(-j)} } = \pot{Y_{ji}}{\ivp_j, \IVp_{(-j)}, \pot{\trt_{ji}}{\ivp_j}, \pot{\TRT_{\eji}}{\ivp_j} , \pot{\TRT_{(-j)}}{\ivp_{(-j)} } }$ be the potential outcome of individual $i$ in cluster $j$ if cluster $j$ and other clusters were assigned to $\ivp_j$ and $\IVp_{(-j)}$, respectively, individual $i$'s treatment receipt were $\pot{\trt_{ji}}{\ivp_j}$, peers of individual $i$ had treatment receipt $\pot{\TRT_{\eji}}{\ivp_j}$, and individuals in clusters other than $j$ had treatment receipt $\pot{\TRT_{(-j)}}{\ivp_{(-j)} }$. Finally, let $\mathcal{F}_\NC = \big\{ \pot{Y_{ji}}{\ivp_j, \IVp_{(-j)}, \trtp_{ji}, \TRTp_\eji, \TRTp_{(-j)} }  , \pot{\trt_{ji}}{\ivp_j} , \bX_{ji} \cond \ivp_j \in \{0,1\} , \IVp_{(-j)} \in \{0,1\}^{\NC-1} , \trtp_{ji} \in \{0,1\} , \TRTp_{\eji} \in \{0,1\}^{\NI_j - 1}, \TRTp_{(-j)} \in  \{0,1\}^{\NIT - \NI_j}, i = 1,\ldots, \NI_j , j=1,\ldots, \NC \big\}$ be the collection of all potential outcomes and pre-treatment covariates. We assume a finite population framework where $\mathcal{F}_{\NC}$ is fixed, but unknown and only $\iv_{j}$ is random in the study. 

We make some important remarks about the notation.  First, because CRTs randomize treatment at the cluster-level and everyone in a cluster is assigned treatment or control, we use $\trt_{ji}^{(\ivp_j)}$ instead of $\trt_{ji}^{(\ivp_j, \mathbf{\ivp}_{i(-j)})}$  for the treatment receipt; the latter notation allows the treatment assignment to vary between individuals in the same cluster, which is not a feasible intervention in a CRT. This also means that we still have the four usual principal strata under noncompliance \citep{Angrist1996}; see below for details. Second, the observed data and the potential outcomes are linked through the cluster-level treatment assignment $\iv_{j}$, i.e. $Y_{ji} = \potY{Y_{ji}}{Z_{j},\mathbf{Z}_{(-j)}}$ and $\trt_{ji} = \pot{\trt_{ji}}{Z_{j}}$.

\subsection{Review: Assumptions on Noncompliance and Interference} \label{sec:assup}
Next, we introduce standard working assumptions in noncompliance and interference; see \citet{Angrist1996} and \citet{HH2008} for details.

\begin{enumerate}[itemsep=2pt, topsep=4pt, parsep=0pt]
	\item[\hypertarget{(A1)}{(A1)}] (\textit{Partial Interference}): 
	The potential outcome of individual $i$ in cluster $j$ does not depend on the treatment assignment or treatment receipts from individuals in other cluster $j'$, $j' \neq j$. That is, given $\ivp_j, \trtp_{ji}, \TRTp_\eji$, we have $\pot{Y_{ji}}{\ivp_j, \IVp_{(-j)}, \trtp_{ji}, \TRTp_\eji, \TRTp_{(-j)} } = \pot{Y_{ji}}{\ivp_j, \IVp_{(-j)}', \trtp_{ji}, \TRTp_\eji, \TRTp_{(-j)}' } = \pot{Y_{ji}}{\ivp_j, \trtp_{ji}, \TRTp_\eji} $ for any $\IVp_{(-j)} \neq \IVp_{(-j)}'$ and $\TRTp_{(-j)} \neq \TRTp_{(-j)}'$. 
	
	\item[\hypertarget{(A2)}{(A2)}] (\textit{Cluster Random Assignment}): 
 $\tNC$ clusters are randomly assigned to treatment, i.e. $P \big( \IV_j = \bm{\ivp}\cond \mathcal{F}_\NC , \mathcal{\iv}_\NC \big) ={ \NC \choose \tNC }^{-1}$ where $\mathcal{\iv}_\NC$ is the set of allowable cluster-level treatment assignments with $m$ treated clusters. Also, $m/J$ is bounded away from 0 and $1$. 
 
 	\item[\hypertarget{(A3)}{(A3)}] (\textit{Non-Zero Causal Effects of $\iv$ on $\trt$}): 
	$\sumji \big\{ \pot{\trt_{ji}}{1} - \pot{\trt_{ji}}{0} \big\} \neq 0$.
	
	\item[\hypertarget{(A4)}{(A4)}] (\textit{Network Exclusion Restriction}):
	Given everyone's treatment receipt in cluster $j$, the potential outcome of individual $i$ in cluster $j$ does not depend on the treatment assignment, i.e. given $\trtp_{ji}$, $\TRTp_\eji $, we have	$\pot{Y_{ji}}{\ivp_j=1, \trtp_{ji}, \TRTp_\eji} = \pot{Y_{ji}}{\ivp_j=0, \trtp_{ji}, \TRTp_\eji} = \pot{Y_{ji}}{\trtp_{ji}, \TRTp_\eji}$.
	
	\item[\hypertarget{(A5)}{(A5)}] (\textit{Monotonicity of Treatment Receipt}):
	For every individual $ji$, we have $\pot{\trt_{ji}}{0} \leq \pot{\trt_{ji}}{1}$.
	
	\item[\hypertarget{(A6)}{(A6)}] (\textit{Bounded, Monotonic Outcome}):
	For every individual $ji$ and $\IVp_{(-j)} \in \{0,1\}^{\NC-1}$, we have $0 \leq \potY{Y_{ji}}{0,\IVp_{(-j)}} \leq \potY{Y_{ji}}{1,\IVp_{(-j)}} \leq 1$.
\end{enumerate}
We take a moment to discuss Assumptions  \hyperlink{(A1)}{(A1)}-\hyperlink{(A6)}{(A6)} in the context of CRTs in infectious disease. Assumption \hyperlink{(A1)}{(A1)} holds if the potential outcome of an individual can be affected by his/her peers' treatment usage in the same cluster; in other words, Assumption \hyperlink{(A1)}{(A1)} allows for within-cluster interference. Also, under \hyperlink{(A1)}{(A1)}, we have $\potY{Y_{ji}}{ \ivp_j, \IVp_{(-j)} } = \potY{Y_{ji}}{\ivp_j}$, which equals $\pot{Y_{ji}}{\ivp_j, \trtp_{ji}, \TRTp_\eji}$. For the Hong Kong study, partial interference is a plausible approximation of the interference pattern because (i) the study was limited to households that only included one flu-infected individual (i.e. the index patient), (ii) the secondary attack (i.e. the outcome of interest) was assessed a week after randomization, and (iii) past works state that households are a key unit of influenza transmission due to the closeness between household members; see Section \ref{sec:EXofParaMethod} of the supplementary materials for additional discussions and references. Combined, it is likely that flu transmission primarily occurred through the index patient in the household rather than from others in different households. Nevertheless, if an individual frequently interacted with others outside of their own households during the study period, partial interference may not hold.

Assumptions \hyperlink{(A2)}{(A2)} and  \hyperlink{(A3)}{(A3)} are often satisfied in CRTs. For the Hong Kong study, households are randomly assigned free masks and hand sanitizers (Assumption \hyperlink{(A2)}{(A2)}) and, as mentioned above, some individuals ended up using them (Assumption \hyperlink{(A3)}{(A3)}). Assumption \hyperlink{(A4)}{(A4)} states that conditional on all of the study units' usage of the intervention, the cluster-level intervention does not affect the outcome.  In the Hong Kong study, Assumption \hyperlink{(A4)}{(A4)} implies that the conditional on the entire cluster's usage of face masks and hand sanitizers, each study unit's flu status no longer depends on whether they acquired these items for free (or not) via cluster-level randomization. Assumptions \hyperlink{(A2)}{(A2)}-\hyperlink{(A4)}{(A4)} with a binary outcome are often the ``minimal'' set of assumptions needed for bounds under noncompliance \citep{Swanson2018}. Assumption \hyperlink{(A5)}{(A5)} is often assessed by defining four subgroups of the study population, always-takers (ATs; $\pot{\trt_{ji}}{0}= \pot{\trt_{ji}}{1} = 1$), compliers (COs; $\pot{\trt_{ji}}{0}=0$, $\pot{\trt_{ji}}{1} = 1$), defiers (DFs; $\pot{\trt_{ji}}{0}=1, \pot{\trt_{ji}}{1} = 0$), and never-takers (NTs; $\pot{\trt_{ji}}{0}= \pot{\trt_{ji}}{1} = 0$). In the Hong Kong study, always-takers always use face masks and hand hygiene regardless of whether the items were offered for free or not.  Compliers use these items only if they were offered and defiers do the opposite. Finally, never-takers never use these items even if they are offered.  Under Assumption \hyperlink{(A5)}{(A5)}, there are no defiers. 

Finally, the monotonicity part of Assumption \hyperlink{(A6)}{(A6)} is plausible (or approximately plausible) if the treatment is not harmful to all (or almost all) individuals on the outcome being measured; see \citet{Choi2017} for a similar assumption. The assumption may fail if the treatment harms a subset of study units on the outcome being measured. For example, for a small subset of patients, a new drug could have severe side effects that may lead to a worse outcome than that from the placebo. In the Hong Kong setting, the assumption likely holds as it is unlikely that giving free masks or hand sanitizers is going to reduce protection from the flu compared to not giving them for free. Also, if the outcome is bounded, we can transform the outcome into the 0-1 range to satisfy the boundedness part of Assumption \hyperlink{(A6)}{(A6)}. Note that in the Hong Kong study, the outcome, i.e. the flu status, is already bounded between $0$ and $1$.

Subsequent sections will use different combinations of Assumptions  \hyperlink{(A1)}{(A1)}-\hyperlink{(A6)}{(A6)}. Specifically, when studying the ITT effects, we will only use Assumptions \hyperlink{(A1)}{(A1)} and \hyperlink{(A2)}{(A2)}. When studying network effects among compliance types, we will use all Assumptions  \hyperlink{(A1)}{(A1)}-\hyperlink{(A6)}{(A6)}.

\subsection{Causal Estimands of Interest and Problem Statement} \label{sec:estimand_def}

For the ITT effects, we focus on the overall ITT effect and the heterogeneous ITT effect. The overall ITT effect is defined as the average of individual ITT effects $\tau_{ji} = \potY{Y_{ji}}{1} - \potY{Y_{ji}}{0}$ for everyone in the study; i.e. $\tau_\ITT = \sumji \tau_{ji}/\NIT =  \sumji \big\{  \potY{Y_{ji}}{1} - \potY{Y_{ji}}{0} \big\} /\NIT $. In the Hong Kong study, $\tau_\ITT$ represents the population average effect of giving free masks and hand sanitizers to households on contracting the flu. A positive $\tau_{\ITT}$ would indicate that giving these items for free prevented flu cases. For the heterogeneous ITT effect, we follow \citet{Ding2019} and define it as the best linear approximation of individual ITT effects $\tau_{ji}$ in the observed covariates $\bX_{ji}$, i.e. $\B^* = \arg \min_{\B} \sumji \big( \tau_{ji} - \bX_{ji}\T \B \big)^2$. Broadly speaking, $\bX_{ji}\T \B^*$ is the best linear approximation of the conditional ITT effect in the span of the covariates $\bX_{ji}$ and $\B^*$ measures how much the treatment effect changes as a linear function of $\bX_{ji}$. For example, in the Hong Kong study, suppose $\bX_{ji}$ is equal to age measured in years. Then, a positive $\B^*$ would suggest that giving free masks and hand sanitizers to households becomes more beneficial as the individual gets older while a negative $\B^*$ suggests the opposite. Also, when $\bX_{ji}$ is a collection of dummy variables defining non-overlapping strata, $\bX_{ji}\T \B^*$ is equal to conditional ITT effect among individuals in a stratum. For example, if $\bX_{ji} = (X_{ji}, 1-X_{ji})\T$ and $X_{ji}$ is a binary covariate, $\bX_{ji} \T \B^*$ is equivalent to the conditional ITT effect with $\beta_1^*$ equal to the conditional ITT effect among individuals with $X_{ji}=1$ and $\beta_2^*$ equal to the conditional ITT effect among individuals with $X_{ji}=0$. Finally, as discussed in  \citet{Ding2019}, the decomposition of $\tau_{ji}$ is not a modeling assumption where the treatment effect must be linear and equal to $\bX_{ji}\T \B^*$ plus mean-zero error;  there may be non-mean-zero residual variation in $ \tau_{ji} - \bX_{ji}\T \B^*$ that could be explained by higher-order terms of $\bX_{ji}$s.

For studying the network effects, we focus on the following estimand defined under Assumption \hyperlink{(A4)}{(A4)}. 
\begin{align*}	
	\tau_\NT & = \frac{1}{\NIT_\NT}  \sumji \Big\{ \pot{Y_{ji}}{d_{ji} =0, \pot{\TRT_\eji}{1} } -  \pot{Y_{ji}}{d_{ji} =0, \pot{\TRT_\eji}{0} } \Big\} \NT_{ji}	\ , \\
	\tau_\AT	& = \frac{1}{\NIT_\AT}  \sumji  \Big\{ \pot{Y_{ji}}{d_{ji} =1, \pot{\TRT_\eji}{1} } -  \pot{Y_{ji}}{d_{ji}=1, \pot{\TRT_\eji}{0} } \Big\} \AT_{ji} \ , \\
	\tau_\CO & = \frac{1}{\NIT_\CO} \sumji  \Big\{ \pot{Y_{ji}}{d_{ji} = 1, \pot{\TRT_\eji}{1} } -  \pot{Y_{ji}}{d_{ji} = 0, \pot{\TRT_\eji}{0} } \Big\} \CO_{ji} \ .
\end{align*}
Here, $\NT_{ji} =\ind \{ \pot{D_{ji}}{1} = \pot{D_{ji}}{0} = 0 \}$, $\AT_{ji} =\ind \{ \pot{D_{ji}}{1} =  \pot{D_{ji}}{0} = 1 \}$, and $\CO_{ji} =\ind \{ \pot{D_{ji}}{1} = 1,  \pot{D_{ji}}{0} = 0 \}$ denote whether individual $i$ in cluster $j$ is a NT, AT, or CO, respectively, and $\NIT_\NT, \NIT_\AT$, and $\NIT_\CO$ are the total number of NTs, ATs, and COs in the population, respectively. The definition of $\tau_\NT, \tau_\AT$, and $\tau_{\CO}$ implicitly assumes that there is at least one NT, AT, or CO in the population. But, if one of the compliance types does not exist, we define the effect to be zero, say if $\NIT_\NT = 0$, we let $\tau_{\NT} = 0$. The estimands $\tau_{\NT}$ and $\tau_{\AT}$ are a type of spillover effects among NTs and ATs, respectively \citep{Sobel2006, HH2008, Kilpatrick2020}, but $\tau_\NT$ and $\tau_\AT$ can also interpreted as ITT effects among NTs and ATs, respectively, if Assumption \hyperlink{(A4)}{(A4)} does not hold; see Section \ref{sec:supp-sec2-2} of the supplementary materials. 

In the Hong Kong study, if  $\tau_\NT > 0$, using masks and hand sanitizers among the NTs' peers spilled over and there was a protective effect among the NTs who never wore masks or used hand sanitizers. Similarly, if $\tau_\AT > 0$, using masks and hand sanitizers among the ATs' peers spilled over and there was an additional protective effect among the ATs who always used these items. If $\tau_\CO>0$, the combined net effect of COs and COs' peers using masks and hand sanitizers was positive. 

We conclude with a brief remark between the estimands $\B^*$ and $\tau_t$ for $t \in \{\NT,\AT,\CO\}$. First, the estimand $\B^*$ reflects effect heterogeneity among the observed covariates $\bX_{ji}$ whereas $\tau_t$ reflects effect heterogeneity among compliance types. Relatedly, as we will discuss below, because compliance types are not observed for all study units, $\tau_t$ may not be point-identifiable whereas $\B^*$ can be identified from data. Second, pages 711 and 713 of \citet{Kilpatrick2020} show that if the study unit is blinded to the cluster-level random assignment, say in a CRT trial for vaccines with blinding, it is plausible to assume that COs do not exist and we can then point-identify $\tau_{\NT}$ and $\tau_{\AT}$ from data by examining the observed treatment receipts. Also,  \citet{JS2009} uses principal ignorability to point-identify effects among compliance types. In contrast, our work does not assume principal ignorability or blinding and seeks to obtain bounds of the effects among compliance types.

\section{Analyzing Intent-to-Treat Effects}						\label{sec:3}
As discussed in \citet{Neyman1923} and \citet{Ding2019}, natural estimators of $\tau_{\ITT}$ and $\B^*$ would be the Neyman-type unbiased estimators where we take the difference in the average of the observed outcomes between the treated and control clusters. For example, for $\tau_{\ITT}$, a Neyman-type unbiased estimator would be $\widehat{\tau}_{\ITT} = (\NC/\NIT) \{ \sumji  \iv_j Y_{ji} / \tNC - \sumji (1-\iv_j) Y_{ji} / (\NC-\tNC)  \}$.  However, in clustered settings, these unbiased estimators are sensitive to affine transformations in the outcome variable. As a simple example, if we were to flip a binary outcome, say flu status, from $\{1,0\}$ to $\{0,1\}$, these estimators do not always flip the sign of the estimated effect (i.e. the expected behavior). Ultimately, this is because cluster size $\NI_{j}$ varies between treated and control groups in the observed sample, where the treated clusters, by chance, may be larger than the control clusters; this does not occur in non-clustered settings where every unit is a ``cluster'' of size $1$ and Section \ref{sec:ATvariance} of the supplementary materials contains a more technical reason for this phenomena. 

To resolve this, we forgo unbiasedness at the expense of affine-invariance by using ratio estimators in the sampling literature \citep{Cochran1977, Fuller2009}.
\begin{align*}	
	&	\widehat{\tau}_{\ITT}	= 	\bigg\{ \sum_{j} \iv_j \NI_j \bigg\}^{-1} \bigg\{ \sumji \iv_j Y_{ji} \bigg\} 	- 	\bigg\{ \sum_{j} (1-\iv_j) \NI_j \bigg\}^{-1}	\bigg\{ \sumji (1-\iv_j) Y_{ji} \bigg\}	\ , \\
	&	\widehat{\B}	= 	\bigg\{ \sumji \iv_j \bX_{ji} \bX_{ji}\T \bigg\}^{-1} \bigg\{ 	\sumji \iv_j \bX_{ji} Y_{ji}	\bigg\} 	- 	\bigg\{ 	\sumji (1-\iv_j) \bX_{ji} \bX_{ji}\T 	\bigg\}^{-1} 	\bigg\{ 	\sumji (1-\iv_j) \bX_{ji} Y_{ji}  \bigg\} \ . 
\end{align*}
Section \ref{sec:ATinvariance} of the supplementary materials formally shows that the above estimators are robust to affine transformations. Also, despite the ratio estimators being biased in finite sample, Theorem \ref{thm:302} shows that they are consistent and asymptotically Normal.
\begin{theorem}							\label{thm:302}
	 Suppose Assumptions \hyperlink{(A1)}{(A1)}, \hyperlink{(A2)}{(A2)},  and moment assumptions in Section \ref{sec:TheoryITT} of the supplementary material hold. As $\NC, \NI_j \to \infty$ or $J \to \infty$ while $\NI_j$ is bounded, the limiting distributions of $\widehat{\tau}_\ITT$ and $\widehat{\B}$ are $\sqrt{\NC} \big( \widehat{\tau}_\ITT - \tau_\ITT \big) 	\stackrel{D}{\rightarrow} 	N \big( 0 , \sigma_\ITT^2 \big)$ and $ \sqrt{\NC} \big( \widehat{\B} - \B^* \big) 	\stackrel{D}{\rightarrow} 	N \big( 0 ,\Sigma_{\B} \big) $ for a non-negative constant $\sigma_\ITT^2$ and a positive semi-definite matrix $\Sigma_{\B}$, respectively.
\end{theorem}

Theorem \ref{thm:302} allows us to conduct tests on $\tau_{\ITT}$ and $\B^*$ where the cluster size is bounded (e.g., the Hong Kong study or other household-level CRTs) or where the cluster size is comparable to the number of clusters (e.g., CRTs with a moderate number of villages or clinics), all without making parametric modeling assumptions on the outcome or on the intra-correlation structure. Finally, following \citet{Ding2019}, we use variance estimators $\widehat{\sigma}_\ITT^2$ and $\widehat{\Sigma}_{\B}$ that are conservative for $\sigma_\ITT^2$ and $\Sigma_{\B}$, respectively; see Section \ref{sec:TheoryITT} of the supplementary materials for details.

\section{Analyzing Network Effects Among Compliance Types}							\label{sec:4}
\subsection{Overview}								\label{sec:4.1}
To motivate our analysis of the network effects among compliance types using bounds, consider decomposing the effects into averages of potential outcomes under treatment and control.
\begin{equation}
\tau_\NT = \frac{\potY{S_\NT}{1} - \potY{S_\NT}{0}}{\NIT_\NT} \ , \
\tau_\AT = \frac{\potY{S_\AT}{1} - \potY{S_\AT}{0}}{\NIT_\AT} \ , \
\tau_\CO = \frac{\potY{S_\CO}{1} - \potY{S_\CO}{0}}{\NIT_\CO}
\label{eq-tauNT}
\end{equation}
where $\potY{S_t}{\ivp} = \sumji \potY{Y_{ji}}{\ivp} t_{ji}$, $\NIT_t = \sumji  t_{ji}$, and $t$ is shorthand for $\NT$, $\AT$, and $\CO$, respectively. Theorem 2 in \citet{Kang2018} showed that $\tau_{\NT}$, $\tau_{\AT}$, and $\tau_{\CO}$ cannot be point-identified in a CRT because the compliance status is only partially observed from data. A bit more formally, $\potY{S_\NT}{0}$, $\potY{S_\AT}{1}$, $\potY{S_\CO}{0}$, and $\potY{S_\CO}{1}$ cannot be point-identified and subsequently, $\tau_{\NT}$ is identified up to the bound $[0,\potY{S_\NT}{1}/\NIT_\NT]$ and $\tau_\AT$ is identified up to the bound $[0,1-\potY{S_\AT}{0}/\NIT_\AT]$. Unfortunately, these bounds do not use any covariate information. 
	
Our proposed method aims to tighten this bound by incorporating covariate information inside of classification algorithms from ML and LP. Specifically, in the first step, we build ``compliance type classifiers'' that take in pre-treatment covariates from individual $ji$ and outputs their predicted compliance types. In the second step, we use the classifiers inside an LP to obtain sharp bounds for $\tau_{\NT}$, $\tau_{\AT}$, and $\tau_{\CO}$. An attractive feature of our method is that the classifiers in the first step do not have to be perfect; in the worst case, we can use random classifiers. But, if an investigator builds good classifiers with low mis-classification rates, the resulting bounds will tighten. 

Our approach to bounds differs from other well-known instrumental variables bounds in the literature (e.g. \citet{Balke1997}) in that  (i) prior literature has primarily focused on bounds for the average treatment effect, not local or network effects and (ii) prior literature often ignored pre-treatment covariates and classifiers from ML to sharpen bounds. In particular, as we discuss in Section \ref{Fig:Compliance} of the supplementary materials,  \citet{GM2008} and \citet{LongHudgens2013} proposed bounds for effects defined by principal strata in non-interference, non-CRT settings, but required binary covariates to remove concerns for model mis-specification. In contrast, our bounds work with discrete and continuous covariates. Also, as mentioned earlier, our bounds do not assume that we have a correct classifier for the compliance types or a correct model; a bad classifier will lead to wide bounds, a good classifier will lead to short bounds and any classifier of the form discussed below will always generate a bound that covers the target estimand.

\subsection{Training the NT and AT Classifiers: Constrained Risk Minimization} \label{sec:classifier}

Consider a classifier for the NTs, $C_\NT$, that takes in pre-treatment covariates $\bX_{ji}$ and reports $1$ if $ji$ is predicted to be a NT and $0$ otherwise. The classifier is parametrized by $\eta_\NT=(\theta_\NT,q_\NT)$ and has the form $C_\NT(\bX_{ji} \con \eta_\NT) = \ind \{ f_\NT (\bX_{ji} \con \theta_\NT) \geq q_\NT \} $. Here $f_\NT (\cdot \con \theta_\NT)$ is a $\R^p \to \R$ function that transforms the covariates $\bX_{ji}$ onto the real line and is parametrized by $\theta_\NT$. This type of indicator-based classifier is not new, as it is related to the rectified linear unit common in deep neural networks \citep{DL2016}, the margin condition in support vector machines, or a quantile-threshold classifier in \citet{Kennedy2020}. Similarly, we define a classifier $C_\AT(\bX_{ji} \con \eta_\AT) = \ind  \{ f_\AT (\bX_{ji} \con \theta_\AT) \geq q_\AT \} $ predicting AT status. For notational simplicity, we use $C_t(\bX_{ji} \con \eta_t)$ to denote the classifier for compliance type $t \in \{\NT, \AT\}$, $\NIT_{t}$ to denote the total number of study units for compliance type $t$, and $t_{ji}$ to denote the compliance type of individual $ji$, i.e. $t_{ji} \in \{\NT_{ji}, \AT_{ji}\}$.

To train the classifier, let $L : \R \otimes \R \to [0,\infty)$ be a loss function, say squared error loss or logistic/cross-entropy loss with a penalty, and consider a variant of constrained risk minimization.
\begin{equation} \label{eq:margin_1}
\theta_t^* =\arg\min_{\theta} \sumji L \big( t_{ji}, f_t (\bX_{ji} \con \theta) \big)\ , \
q_t^* \in \bigg\{ q \, \bigg| \,  \NIT_t = \sumji \ind \big\{ f_t (\bX_{ji} \con \theta_t^* ) \geq q \big\} \bigg\}
\ .
\end{equation}
The first part of equation \eqref{eq:margin_1} is a risk minimizer and as such, we can use a large library of classifiers based on risk minimization. Some well-known examples include:

\begin{enumerate}[itemsep=2pt, topsep=4pt, parsep=0pt]

	\item[\hypertarget{(Linear)}{(Linear)}] (\textit{Multiple linear regression: linear $f_t$, squared error loss}):
	A linear learner $f_t(\bX_{ji} \con \theta_t) = \bX_{ji}\T \theta_t$ and a square-error loss function $L (t_{ji}, f_t(\bX_{ji} \con \theta_t) ) = \big\{ t_{ji} - f_t(\bX_{ji} \con \theta_t) \big\}^2$. 
	
	\item[\hypertarget{(Logistic)}{(Logistic)}] (\textit{Penalized logistic regression: logistic $f_t$, $\ell^2$-regularized logistic loss}):
	 A logistic learner $f_t(\bX_{ji} \con \theta_t) =\{ 1+ \exp ( - \bX_{ji}\T \theta_t ) \}^{-1}$ and a logistic loss where $ L (t_{ji}, f_t(\bX_{ji} \con \theta_t) )  =  - t_{ji} \log \big\{ f_t (\bX_{ji} \con \theta_t)\big\} 	- \big( 1-t_{ji} \big) \log \big\{ 1 - f_t(\bX_{ji} \con \theta_t )\big\} + \lambda\|\theta_t\|_2^2/2$ with $\lambda>0$ as a regularization parameter. 
	\end{enumerate}
The second part of \eqref{eq:margin_1} calibrates $f_t$ obtained from risk minimization so that $C_t(\bX_{ji} \con \eta_t)$  correctly estimates of the total number of NTs or ATs. Combined, the two parts of \eqref{eq:margin_1} are designed to find the best $\eta_t$ given the investigator's choice of the loss function $L$ and learner $f_t$; as mentioned earlier, $C_t(\bX_{ji} \con \eta_t)$ does not have to be a perfect classifier of NTs or ATs for our procedure to work. 

Now, \eqref{eq:margin_1}, as written, is impossible to use with data because (i) the term $q_t^*$ in \eqref{eq:margin_1} may not always exist in finite samples, (ii) the indicator function $\ind$ to solve for $q_t^*$ is not smooth, posing a computational challenge to find (ideally a unique) $q_t^*$, and (iii) $t_{ji}$s are not observed for every $ji$. To resolve issue (i), we leverage data augmentation techniques in ML where we perturb the original learner $f_t$ by adding independent and identically distributed random noise and solve $q_t$ based on the new, randomized learner.
\begin{equation*}
q_t^* \in \bigg\{ q \, \bigg| \, \NIT_t = \sumji \ind \big\{ \randomf_t (\bX_{ji} \con \theta_t^*) \geq q \big\} \bigg\} \ , \ \randomf_t(\bX_{ji} \con \theta_t^*) = f_t(\bX_{ji} \con \theta_t^*) + e_{tji} \ .
\end{equation*}
Here,  $\randomf_t$ denotes the new randomized learner with the random noise generated from $e_{tji} \sim {\rm Unif}(-r,r)$ $(r>0)$. The choice to use a uniform distribution as the randomizer is out of convenience and other distributions are possible. The choice of $r$ governing the uniform distribution depends on the original $f_t$ and under some conditions, $r$ should be of order $\Theta(\NC^{-1})$; see Section \ref{sec:choice-r} of the supplementary materials for details.  Using the randomized learner $\randomf_t$ always guarantees a solution for $q_t^*$ in finite sample and we can re-define the classifier as $\randomC_t (\bX_{ji} \con \eta_t) = \ind \big\{ \randomf_t(\bX_{ji} \con \theta_t^* ) \geq q_t \big\}$ based on  $\randomf_t$. 

Next, to resolve (ii), we use a trick from optimization where the indicator function is replaced with a continuous surrogate indicator function $\mathcal{I}$, say a scaled hyperbolic tangent or a logistic function. For our setting, we use the following surrogate indicator function  $\SI$ parameterized by $c, h>0$; this surrogate function is not only continuous but also continuously differentiable and strictly increasing.
\begin{align*}
& \SI(v) =  \left\{	\begin{array}{lllll}
	1 - c \cdot \exp \big\{ - \frac{1-2c}{2ch}(v-h) \big\} & (h \leq v) \\[-0.1cm]
	\frac{1-2c}{2h}(v+h)	+c & (-h \leq v < h ) \\[-0.1cm]
	c \cdot \exp \big\{\frac{1-2c}{2ch}(v+h) \big\} & (v < -h)
	\end{array} 	\right. 	\ .
\end{align*}
The exact choice of $c$ and $h$ depends on the randomized learner $\randomf_t$ and broadly speaking, $c$ and $h$ should roughly be of order $c = \Theta\big( (\log \NC)^{-1} \big)$ and $h = \Theta(\NC^{-1})$; see Section \ref{sec:choice-ch} of the supplementary materials. After addressing issues (i) and (ii), the original optimization problem \eqref{eq:margin_1} becomes 
\begin{align}									\label{eq:margin_1_welldefined}
	\theta_t^* =\arg\min_{\theta} \sumji L \big( t_{ji}, f_t (\bX_{ji} \con \theta) \big) \ , \
	\NIT_t = \sumji \SI \big( \randomf_t(\bX_{ji} \con \theta_t^* ) - q_t^* \big) \ ,
\end{align}
and the classifier based on $\eta_t^*=(\theta_t^*,q_t^*)$ has the form $\randomC_t(\bX_{ji} \con \eta_t^*)=\ind \big\{ \randomf_t(\bX_{ji} \con \theta_t^* ) \geq q_t^* \big\}$.

Finally, to resolve issue (iii), under Assumption \hyperlink{(A5)}{(A5)}, the NT status is known for treated clusters as $\NT_{ji} = 1-D_{ji}$ and the AT status is known for control clusters as $\AT_{ji} = D_{ji}$. Also, under Assumption \hyperlink{(A2)}{(A2)}, the characteristics of each compliance type are similar between the control and treatment arms. Therefore, the estimation of the NT classifier using only the treated clusters can be used to predict NT status in the control clusters; a similar argument holds for the AT classifier. Similarly, because of  Assumption \hyperlink{(A2)}{(A2)}, we have roughly equal numbers of NT or AT individuals in the treated and control clusters.

Combining (i)-(iii), our estimated classifier is based on the following.
{\small
\begin{align}
&
\widehat{\theta}_\NT =\arg\min_{\theta} \sum_{ji: \iv_j=1} L (1- D_{ji} , f_\NT (\bX_{ji} \con \theta) ), 
&&
\hspace*{-0.4cm}
\sum_{ji: \iv_j=1} (1-D_{ji}) =
\hspace*{-0.1cm}
\sum_{ji: \iv_j=1} \SI (\randomf_\NT (\bX_{ji} \con \widehat{\theta}_\NT) - \widehat{q}_\NT) 
\ , 
\nonumber
\\
&
\widehat{\theta}_\AT =\arg\min_{\theta} \sum_{ji: \iv_j=0} L ( D_{ji} , f_\AT (\bX_{ji} \con \theta) ), 
&& 
\hspace*{-0.4cm}
\sum_{ji: \iv_j=0} D_{ji} = 
\hspace*{-0.1cm}
\sum_{ji: \iv_j=0} \SI (\randomf_\AT (\bX_{ji} \con \widehat{\theta}_\AT) - \widehat{q}_\AT) 
\ .
\label{eq:margin_2}
\end{align}}
We denote the estimated classifier as $\randomC_t (\bX_{ji} \con  \widehat{\eta}_t) = \ind  \big\{ \randomf_t (\bX_{ji} \con \widehat{\theta}_t) \geq \widehat{q}_t \big\}$ where $\widehat{\eta}_t = (\widehat{\theta}_t,\widehat{q}_t)$.

\subsection{Training the CO Classifier} \label{sec:classifierCO}
To train the classifier for the COs, denoted as $\randomC_{\rm CO}$, we simply use the NT and AT classifiers from above. Specifically, we define the learner for COs, denoted as $f_{\CO}(\bX_{ji} \con \theta_{\CO}^*)$, to be a weighted combination of learners from the NTs, i.e.  $f_{\NT}$, and the ATs, i.e. $f_{\AT}$. For example, for the linear and penalized logistic examples from Section \ref{sec:classifier},  $f_\CO$ is defined as

\begin{itemize}[itemsep=2pt, topsep=4pt, parsep=0pt]

	\item[(Linear)] (\textit{Multiple linear regression}):
	$f_\CO (\bX_{ji} \con \theta_\CO^*) = - w_\NT^* \bX_{ji}\T \theta_\NT^* - w_\AT^* \bX_{ji} \T \theta_\AT^*$,

	\item[(Logistic)] (\textit{Penalized logistic regression}):
	$f_\CO (\bX_{ji} \con \theta_\CO^*) = \big\{ 1+ \exp ( w_\NT^* \bX_{ji}\T \theta_\NT^* + w_\AT^* \bX_{ji}\T \theta_\AT^* ) \big\}^{-1}$,
\end{itemize}
where $w_{\NT}^* = \NIT_\NT / \NIT$, $w_{\AT}^* = \NIT_\AT/\NIT$, and $\theta_\CO^* = (w_\NT^*, w_\AT^*, \theta_\NT^*, \theta_\AT^*)$. Given the CO's learner, we define the threshold parameter $q_\CO^*$ similarly as before, i.e. a value of $q_{\CO}^*$ that satisfies $\NIT_\CO = \sumji  \SI \big( \randomf_\CO (\bX_{ji} \con \theta_\CO^* ) - q_\CO^* \big)$. To estimate the unknown parameters $\theta_{\rm CO}^*$ and $q_{\CO}^*$, we use the plug-in estimates from Section \ref{sec:classifier}, i.e. $\widehat{\theta}_\CO = \big( \widehat{w}_\NT, \widehat{w}_\AT, \widehat{\theta}_\NT , \widehat{ \theta}_\AT \big)$ where $\widehat{\NIT}_\NT = \NIT \sum_{ji:\iv_j = 1} (1-D_{ji}) / \sum_{j: \iv_j=1} \NI_j $, $\widehat{\NIT}_\AT = \NIT \sum_{ji: \iv_j = 0} D_{ji}/ \sum_{j: \iv_j=0} \NI_j  $ and $\NIT_{\CO}$ is replaced by the estimated number of COs, i.e. $\widehat{\NIT}_{\CO} = N - \widehat{\NIT}_{\NT} - \widehat{\NIT}_{\AT}$. In the end, the estimated CO classifier is denoted as $\randomC_\CO(\bX_{ji} \con \widehat{\eta}_\CO) = \ind  \big\{ \randomf_\CO (\bX_{ji} \con \widehat{\theta}_\CO) \geq \widehat{q}_\CO \big\}$ where $\widehat{\eta}_\CO = (\widehat{\theta}_\CO,\widehat{q}_\CO)= (\widehat{w}_\NT, \widehat{w}_\AT, \widehat{\theta}_\NT, \widehat{\theta}_\AT, \widehat{q}_\CO)$.

\subsection{Bounds with Linear Program and Classifiers} \label{sec:lp}
To construct a sharp bound using the classifiers, we first consider the ``population-level''/``true'' classifiers $\randomC_t (\bX_{ji} \con \eta_t^*)$; again, for notational convenience, we use the shorthand $t$ to denote NT, AT, or CO. For each compliance type $t$, let $R_t = \sumji t_{ji}  \big\{ 1 - \randomC_t (\bX_{ji} \con \eta_t^*) \big\}$ be the number of mis-classified cases for the classifier $\randomC_t$. Also, for each effect among compliance type $t$ (i.e. $\tau_t$), consider a proxy estimate of $\potY{S_t}{\ivp}$ in equation \eqref{eq-tauNT} by replacing the true compliance type with the predicted compliance type from the classifier; we denote this as $\potY{S_{\BrandomC,t}}{\ivp}$.
\begin{align*}
\potY{S_{\BrandomC,t}}{\ivp} &= \sumji  \potY{Y_{ji}}{\ivp}  \randomC_t (\bX_{ji} \con \eta_t^*) = \underbrace{ \sumji \potY{Y_{ji}}{\ivp} t_{ji} \randomC_t (\bX_{ji} \con \eta_t^*)  }_{ \equiv \potY{\TP_t}{\ivp} } + \underbrace{ \sumji \potY{Y_{ji}}{\ivp} (1 - t_{ji}) \randomC_t (\bX_{ji} \con \eta_t^*) }_{ \equiv \potY{\FP_t}{\ivp} } \ .   \\[-1.7cm]
\end{align*}
The term $\potY{\TP_t}{\ivp}$ making up $\potY{S_{\BrandomC,t}}{\ivp}$ is the sum of potential outcomes among compliance type $t$ that were correctly classified by the classifier, or the true-positives of the classifier. The term $\potY{\FP_t}{\ivp}$ is the sum of potential outcomes that were incorrectly classified, specifically the false-positives of the classifier. Similarly, we can decompose $\potY{S_t}{\ivp}$ as $\potY{S_t}{\ivp} = \potY{\TP_t}{\ivp} + \potY{\FN_t}{\ivp} $ where $\potY{\FN_t}{\ivp} =  \sumji \potY{Y_{ji}}{\ivp} t_{ji}  \big\{ 1 - \randomC_t (\bX_{ji} \con \eta_t^*) \big\}$ is the sum of potential outcomes among the false-negatives of the classifier. Combined, the decompositions allow us to re-express the network effect as $\tau_t =  \{( \potY{\TP_t}{1} + \potY{\FN_t}{1}) - (\potY{\TP_t}{0} + \potY{\FN_t}{0})\} / \NIT_t $. Also, the decompositions reveal a set of linear relationships between the terms $\potY{\TP_t}{\ivp}$, $\potY{\FP_t}{\ivp}$, and $\potY{\FN_t}{\ivp}$, and the terms $\NIT_t$, $\potY{S_\NT}{1}$, $\potY{S_\AT}{0}$, $\potY{S_{\BrandomC,t}}{\ivp}$, and $R_t$; as we show below, the latter terms can be estimated from data.

Our proposed LP uses these decompositions and Assumptions \hyperlink{(A1)}{(A1)}-\hyperlink{(A6)}{(A6)} to find the upper and lower bounds for $\tau_{t}$ $(t \in \{\NT, \AT, \CO\})$: 
\begin{subequations}
\begin{align}
\begin{split}
\text{Min/Max } \tau_t = \big\{ ( \potY{\TP_t}{1} + \potY{\FN_t}{1} ) - (\potY{\TP_t}{0} + \potY{\FN_t}{0}) \big\}/\NIT_t \text{ over } \potY{\TP_t}{\ivp} , \potY{\FP_t}{\ivp} , \potY{\FN_t}{\ivp}
\label{eq-LP-Population}
\end{split}
	\end{align}
	\vspace*{-1cm}
\begin{empheq}[left=\text{ subject to } \empheqlbrace\, ]{align}
 		&
 		\potY{\TP_\NT}{\ivp} + \potY{\FN_\NT}{\ivp}
 		+
 		\potY{\TP_\AT}{\ivp} + \potY{\FN_\AT}{\ivp}
 		+
 		\potY{\TP_\CO}{\ivp} + \potY{\FN_\CO}{\ivp}  = \potY{S}{\ivp}
 		 \label{eq-LP-restrict1}
 		\\[-0.2cm]
 		&
		\potY{\TP_\NT}{1} + \potY{\FN_\NT}{1} = \potY{S_\NT}{1} \ , \ 
		\potY{\TP_\AT}{0} + \potY{\FN_\AT}{0} = \potY{S_\AT}{0}   \label{eq-LP-restrict2}
		\\[-0.2cm]
		&
		\potY{\TP_t}{\ivp} + \potY{\FP_t}{\ivp} = \potY{S_{\BrandomC,t}}{\ivp}   \label{eq-LP-restrict3}
		\\[-0.2cm]
		&
		\potY{\TP_t}{0} \leq \potY{\TP_t}{1} \ , \ 
		\potY{\FP_t}{0} \leq \potY{\FP_t}{1} \ , \ 
		\potY{\FN_t}{0} \leq \potY{\FN_t}{1}    \label{eq-LP-restrict4}
		\\[-0.2cm]
		&
		\potY{\TP_t}{1} \leq \NIT_t - R_t \ , \
		\potY{\FP_t}{1} \leq R_t \ , \
		\potY{\FN_t}{1} \leq R_t  \label{eq-LP-restrict5}
		\\[-0.2cm]
		&
		0 \leq \potY{\TP_t}{\ivp} , \potY{\FP_t}{\ivp}, \potY{\FN_t}{\ivp}  \ , \ t \in \{\NT,\AT,\CO\}, \ivp \in \{0,1\}  \label{eq-LP-restrict6}
	\end{empheq}
	\end{subequations}
Minimizing and maximizing the LP give us the lower and upper bounds for each network effect $\tau_t$. Constraint \eqref{eq-LP-restrict1} is from the compliance types being mutually exclusive. Constraints \eqref{eq-LP-restrict2} and  \eqref{eq-LP-restrict3} are based the decompositions of $\potY{S_\NT}{1}$, $\potY{S_\AT}{0}$, and $\potY{S_{\BrandomC,t}}{\ivp}$. Constraint \eqref{eq-LP-restrict4} is from the monotonicity of the outcome in Assumption \hyperlink{(A6)}{(A6)}. Finally, constraints \eqref{eq-LP-restrict5} and  \eqref{eq-LP-restrict6} are from the boundedness of the outcome and the definition of the number of mis-classified cases $R_t$. 

Theorem \ref{thm:402} characterizes the solution to this LP, including the bound's sharpness. 
\begin{theorem}										\label{thm:402}
	Suppose Assumptions \hyperlink{(A1)}{(A1)}-\hyperlink{(A6)}{(A6)} hold.  For each compliance type $t \in \{ \NT, \AT, \CO\}$ and the classifier $\BrandomC_t$, let ${\rm LB}_{\BrandomC,t}$ and ${\rm UB}_{\BrandomC,t}$ be the minimizing and maximizing solutions to the LP of $\tau_t$, respectively. Then, $\big[{\rm LB}_{\BrandomC,t} , {\rm UB}_{\BrandomC,t} \big]$ are the sharp bounds for $\tau_t$, i.e. they are the narrowest possible bounds given $\potY{S_{\BrandomC,t}}{\ivp}$ and $R_t$.
\end{theorem}
\noindent 
If the classifiers has a 100\% mis-classification rate and everyone in compliance type $t$ is mis-classified, i.e. $R_t = \NIT_t$, our estimated bound for $\tau_t$ would be equivalent to the bounds that are achievable without classifiers in \citet{Kang2018}. However, as an investigator chooses better classifiers with low $R_t$, we can tighten the bound on $\tau_t$.  In the ideal case when $R_t$s are zero, every classifier perfectly classifies the compliance type for all individuals and our bounds reduce to $\tau_t$s. 

We also make two important remarks about interpreting Theorem \ref{thm:402}. First, Theorem \ref{thm:402}, like most theorems in the bound literature, does not quantify how far the true effect for a particular study is away from the lower and upper ends of the bounds. It may be possible that the lower bound  may be farther away from the true effect compared to the upper bound. Second, Theorem \ref{thm:402} does not say that the bound is the shortest possible bound given the covariates $\bX_{ji}$; it only states that the bound is the shortest possible given the classifier $\BrandomC_{t}$. It may be possible to obtain shorter bounds by using a better classifier with a lower mis-classification rate; see Section \ref{sec:HKdata} for additional discussions.

Now, the LP as written above cannot be used because it requires the true classifiers and population-level terms $\NIT_t$, $\potY{S_\NT}{1}$, $\potY{S_\AT}{0}$, $\potY{S_{\BrandomC,t}}{\ivp}$, and $R_t$. To resolve this, we can replace $\NIT_t$ with ratio estimators $\widehat{\NIT}_t$,  replace the population-level classifiers with the estimated classifiers $\randomC_t ( \bX_{ji} \con \widehat{\eta}_t)$, and use the following plug-in estimators for the rest:
\begin{align*}
	& \potY{ \widehat{S} }{\ivp}  = \NIT \cdot \frac{\sumji \ind( \iv_j = \ivp) Y_{ji} }{\sum_{j} \ind( \iv_j = \ivp)  \NI_j} 	\ , \ \ivp \in \{0,1\} \ , \ \\
	& \potY{ \widehat{S}_\NT }{1} = \widehat{\NIT}_\NT \cdot \frac{\sumji \ind( \iv_j = 1) Y_{ji} \NT_{ji} }{\sumji \ind( \iv_j = 1) \NT_{ji} } \ , \
	\potY{ \widehat{S}_\AT }{0} = \widehat{\NIT}_\AT \cdot \frac{\sumji \ind( \iv_j = 0) Y_{ji} \AT_{ji} }{\sumji \ind( \iv_j = 0) \AT_{ji} }	\ ,	\\
	&	\potY{ \widehat{S}_{\BrandomC,t}}{\ivp} = \widehat{\NIT}_t \cdot \frac{\sumji \ind(\iv_j=\ivp) Y_{ji} \randomC_t (\bX_{ji} \con \widehat{\eta}_t) }{ \sumji \ind (\iv_j = \ivp) \randomC_t (\bX_{ji} \con \widehat{\eta}_t) }  \ , \ t \in \{\NT,\AT,\CO\} \ , \ \ivp \in \{0,1\} \ , \ \\
	& 	\widehat{R}_\NT = \widehat{\NIT}_\NT \cdot \frac{\sumji \ind( \iv_j = 1)  (1-\NT_{ji}) \randomC_\NT (\bX_{ji} \con \widehat{\eta}_\NT) }{\sumji \ind( \iv_j = 1) \randomC_\NT (\bX_{ji} \con \widehat{\eta}_\NT) }	\ , \  \\
	& \widehat{R}_\AT = \widehat{\NIT}_\AT \cdot \frac{\sumji \ind( \iv_j = 0) (1- \AT_{ji})  \randomC_\AT (\bX_{ji} \con \widehat{\eta}_\AT) }{\sumji \ind( \iv_j = 0)  \randomC_\AT (\bX_{ji} \con \widehat{\eta}_\AT) } 	\ , \ \\
	& 	\widehat{R}_\CO	=	\widehat{\NIT}_\CO \cdot \Bigg\{ \frac{\sumji \ind( \iv_j = 1) 	\NT_{ji} \randomC_\CO (\bX_{ji} \con \widehat{\eta}_\CO) }{\sumji \ind( \iv_j = 1) \randomC_\CO(\bX_{ji} \con \widehat{\eta}_\CO) } 	+ \frac{\sumji \ind( \iv_j = 0) 	\AT_{ji} \randomC_\CO (\bX_{ji} \con \widehat{\eta}_\CO) }{\sumji \ind( \iv_j = 0) \randomC_\CO(\bX_{ji} \con \widehat{\eta}_\CO) } \Bigg\} \ .
\end{align*}
Let $\widehat{{\rm LB}}_{\BrandomC,t}$ and $\widehat{{\rm UB}}_{\BrandomC,t}$ denote the solutions to the LP from plugging these estimators. Section \ref{Fig:Compliance} of the supplementary materials contains additional discussions about the LP, specifically on implementation and other numerical considerations.

\subsection{Theoretical Properties}								\label{sec:asymptotic}
Before we state the asymptotic properties of the estimated sharp bounds  $\big[ {\rm LB}_{\BrandomC,t} , {\rm UB}_{\BrandomC,t} \big]$, we summarize some interesting theoretical challenges as well as insights about dealing with a randomized classifier $\randomf_t$ and a surrogate indicator function in a finite sample, randomization inference framework; to the best of our knowledge, using these two popular concepts in ML and optimization under a randomization inference framework is new. Broadly speaking, consistency of the estimated parameter $\widehat{\eta}_t$ based on a randomized $\randomf_t$ and a smoothed indicator function $\SI$ does not necessarily imply consistency of the estimated classifier due, in part, to the non-smoothness of the indicator function. A related concern is that the set of $q_t^*$s that satisfy \eqref{eq:margin_1_welldefined} may be different than the set of $\widehat{q}_t$s that satisfy \eqref{eq:margin_2}. Not surprisingly, most of these assumptions are variations of familiar assumptions in supervised ML, such as the compactness of the space of $\eta_t$, the true $\eta_t^*$ being in the interior of the parameter space, and smoothness of $f_t$ and $L$ over the parameter space;  Section \ref{Fig:Compliance} of the supplementary materials contains the exact assumptions on $f_t$ and $L$. Overall, except for some important pathological cases which the supplementary materials detail, these assumptions on $f_t$ and $L$ will hold for classifiers like \hyperlink{(Linear)}{(Linear)}  and \hyperlink{(Logistic)}{(Logistic)} in most real data. 

Let $\widetilde{\mathcal{F}}_\NC = \mathcal{F}_\NC \cup \{ e_{tji} \cond t\in\{\NT,\AT, \CO\},  j=1,\ldots,\NC, i = 1,\ldots,\NI_j \}$ be the extended set of $\mathcal{F}_\NC$ that includes the randomization term $e_{tji}$ from $\randomf_t$. We treat $e_{tji}$ as fixed after being randomly generated once and condition on $\widetilde{\mathcal{F}}_\NC$; this allows the randomness in the study to still be from the treatment assignment $\iv_j$ only and is in alignment with finite-sample/randomization inference framework. Theorem \ref{thm:403} shows that the estimated sharp bounds using the estimated classifiers based on the randomized classifier and the surrogate indicator function are consistent to the population-level sharp bounds. 
\begin{theorem}				\label{thm:403}
	Suppose Assumptions \hyperlink{(A1)}{(A1)}-\hyperlink{(A6)}{(A6)} and Assumption \ref{assp:4-1} in Section \ref{sec:generalfL} of the supplementary materials concerning $f_t$ and $L$ hold. Then, the estimated sharp bounds are consistent, i.e. for any $ \epsilon > 0$,  $\lim_{\NC \rightarrow \infty} P \big\{ \big| \widehat{{\rm UB}}_{\BrandomC,t} - {\rm UB}_{\BrandomC,t} \big| > \epsilon \, \big| \, \widetilde{\mathcal{F}}_\NC , \mathcal{\iv}_\NC \big\} = 0$ and $\lim_{\NC \rightarrow \infty} P \big\{ \big| \widehat{{\rm LB}}_{\BrandomC,t} - {\rm LB}_{\BrandomC,t} \big| > \epsilon \, \big| \, \widetilde{\mathcal{F}}_\NC , \mathcal{\iv}_\NC \big\} = 0$.
\end{theorem}
Finally, to construct confidence sets on the bounds, we use the resampling approaches of \citet{Bootstrap1993} and \citet{Romano2010}, modified for cluster-level resampling; see Section \ref{sec:ConfidenceSet} of the supplementary materials for details. Also, Section \ref{sec:ConfidenceSet} of the supplementary materials shows that the estimated bounds satisfy the affine-invariance property of the ITT estimators.

\section{Simulation}								\label{sec:5-2}

\subsection{Model}							\label{sec:5-2-1}

We conduct a simulation study to examine the performance of our method. To make our simulation as realistic as possible to real data, we mimic the Hong Kong study design in that we use the same (i) subset of pre-treatment covariates (i.e. sex, age, vaccination history), (ii) cluster structure (i.e. number of clusters, size of clusters), and (iii) randomization probabilities from the study; the exact model is stated below. 
\begin{align}															\label{eq-simulation}
	&
	\bX_{ji} = \big[ 1, X_{ji, {\rm male}}, X_{ji, {\rm age}}, X_{ji, {\rm age}}^2, X_{ji,{\rm vaccine}} \big]\T \ ,
	\\
	& 
	\big[ P (ji = \NT)	, P (ji = \AT)	, P (ji = \CO)	 \big]
	\nonumber
	\\
	&
		\propto
	\begin{cases}
	\Big[ 1 , 
	\exp \big\{
	- \frac{( X_{ji, {\rm age}} - 40 ) ( X_{ji, {\rm age}} - 65 ) }{150}
		\big\} ,
		\exp \big\{
	- \frac{( X_{ji, {\rm age}} - 20 ) ( X_{ji, {\rm age}} - 50 ) }{150}
		 \big\}
	\Big]
	&
	\text{ if }
	X_{ji,{\rm male}}=1
	\\
	\Big[ 1 , 
	\exp \big\{
	-  \frac{( X_{ji, {\rm age}} - 40 )( X_{ji, {\rm age}} - 65 )}{20}
		\big\} ,
		\exp \big\{
	-  \frac{( X_{ji, {\rm age}} - 20 )( X_{ji, {\rm age}} - 50 )}{20}
		 \big\}
	\Big]
	&
	\text{ if }
	X_{ji,{\rm male}}=0
	\end{cases}
	 \ ,
	\nonumber
	\\
	 &
	 \potY{ Y_{ji} }{0} \sim  {\rm Ber} \big\{ p_Y(\pot{\trt_{ji}}{0}, \bX_{ji})  \big\}
	 , \
	 \potY{ Y_{ji} }{1} \sim 
	 	\max \Big[
	 	\potY{ Y_{ji} }{0} , {\rm Ber} \big\{ p_Y(\pot{\trt_{ji}}{1}, \bX_{ji})  \big\} 
	 	\Big]
	 	 \ ,
	 \nonumber
	  \\
 	 &
 	 p_Y(\pot{\trt_{ji}}{\ivp}, \bX_{ji}) 
 	 =
 	\Bigg\{
 	\begin{array}{ll}
 	 	{\rm expit} \big\{ -3 +  2 \pot{ \overline{\trt}_\eji }{\ivp} \big\} & \quad \text{if } \pot{\trt_{ji}}{\ivp} = 0 
 	 	\\
 	 	{\rm expit} \big( -3 +  2 + 4 X_{ji,{\rm vaccine}} \big) & \quad \text{if } \pot{\trt_{ji}}{\ivp} = 1 
 	 \end{array}
 	 \ .
 	 \nonumber
\end{align}
Here, $\pot{\overline{\trt}_{\eji}}{\ivp} = \sum_{\ell: \ell\neq i} \pot{\trt_{j\ell}}{\ivp} / ( \NI_j - 1)$ is the average number of $ji$'s peers who are using face masks and hand sanitizers when $ji$'s household is randomized to $\ivp$. In words, the model states that females are likely to be ATs if $X_{ji,{\rm age}} \in [45,65]$, COs if $X_{ji,{\rm age}} \in [20,45]$, or NTs otherwise. In contrast, males are less likely to be ATs or COs in the same age intervals; see Section \ref{sec:addsim} of the supplementary material for the graphical illustration. The outcome model is designed to have larger potential outcomes among the ATs. Also, the effect among the COs is greater than that among the NTs and ATs. The effect also varies depending on the individual's vaccination history.  Overall, the true overall ITT effect is $\tau_\ITT = 0.236$ and the true network effects among the compliance types are $\tau_{\NT} = 0.155$, $\tau_{\AT} = 0.148$, and $\tau_{\CO}=0.347$, respectively. Once the potential outcomes are generated, we follow the original design of the study where we randomly assign 72 clusters to treatment and the rest 79 clusters to control. We repeat the treatment randomization 1,000 times.

\subsection{Results}							\label{sec:5-2-2}

Table \ref{Tab:HongKong1} shows the estimation results of the overall ITT effect $\tau_{\ITT}$ and the heterogeneous ITT effect $\B^*$. Overall, we see that our estimators have negligible biases for $\tau_{\ITT}$ and $\B^*$.

\begin{table}[!htp]
		\renewcommand{\arraystretch}{1.2} \centering
		\setlength{\tabcolsep}{2pt}
		\fontsize{9}{11}\selectfont
\begin{tabular}{|c|c|c|c|c|c|c|}
\hline
Target estimand & \multirow{2}{*}{\begin{tabular}[c]{@{}c@{}}Overall\\ ITT $(\tau_\ITT)$\end{tabular}} & \multicolumn{5}{c|}{Heterogeneous ITT effect coefficient $(\B^*)$}                                                    \\ \cline{1-1} \cline{3-7} 
                      Variable &                                                                                                     & Constant & Gender & Age & Age$^{2}$ & Vaccination \\ \hline
      True value &     $\phantom{-}2.36 \times 10^{-1}$ & $\phantom{-}5.15 \times 10^{-2}$ & $5.47 \times 10^{-2}$ & $\phantom{-}1.16 \times 10^{-2}$ & $-1.69 \times 10^{-4}$ & $\phantom{-}1.92 \times 10^{-1}$ \\ \hline
        Estimate &     $\phantom{-}2.35 \times 10^{-1}$ & $\phantom{-}4.92 \times 10^{-2}$ & $5.79 \times 10^{-2}$ & $\phantom{-}1.16 \times 10^{-2}$ & $-1.68 \times 10^{-4}$ & $\phantom{-}1.91 \times 10^{-1}$ \\ \hline
            Bias &  $-2.91 \times 10^{-4}$ & $-2.29 \times 10^{-3}$ & $3.23 \times 10^{-3}$ & $-0.51 \times 10^{-5}$ & $\phantom{-}0.60 \times 10^{-6}$ & $-1.53 \times 10^{-3}$ \\ \hline
  Standard error &      $\phantom{-}2.74 \times 10^{-2}$ & $\phantom{-}7.15 \times 10^{-2}$ & $5.72 \times 10^{-2}$ & $\phantom{-}4.39 \times 10^{-3}$ & $\phantom{-}6.72 \times 10^{-5}$ & $\phantom{-}7.52 \times 10^{-2}$ \\ \hline
 Average p-value &                                                                                 $\phantom{-}5.66 \times 10^{-8}$ & $0.475$ & $0.404$ & $0.089$ & $0.107$ & $0.111$ \\ \hline
        Coverage &                                                                                                $0.983$ & $0.976$ & $0.970$ & $0.974$ & $0.974$ & $0.977$ \\ \hline
\end{tabular}
\caption{\footnotesize Estimation Results for ITT Effects. Each column represents a target quantity of interest. Bias is the mean difference between the estimate and the true value. Standard error is the standard deviation of the estimates. Average p-value is the average of p-values from Wald-type tests. Coverage is the proportion of CIs that include the true value.}
\label{Tab:HongKong1}
\vspace*{-0.3cm}
\end{table}

For the bounds, we use the linear classifier in \hyperlink{(Linear)}{(Linear)} and the penalized logistic classifier in \hyperlink{(Logistic)}{(Logistic)}. The noise for the randomized learner $\randomf_t$ is generated from a uniform distribution ${\rm Unif}(-10^{-10}, 10^{-10})$. Also, for comparison, we compute the bounds based on \citet{GM2008} and \citet{LongHudgens2013}, who proposed bounds on effects defined by principal strata using a collection of binary covariates $X_{ji,{\rm male}}$ and $X_{ji,{\rm vaccine}}$. We remark that we have extended the original methods to account for interference and CRTs. Notably, we show in Section \ref{sec:LH} of the supplementary materials that our extension of them, which we refer to as ``extended bounds,'' (i) cover $\tau_t$, (ii) can be consistently estimated from data arising from CRTs, and (iii) the estimators of these bounds are affine-invariant. However, a notable limitation of the extended bounds is that they can only incorporate binary or discrete covariates with mutually exclusive levels and it may not be sharp. 

Table \ref{Tab:HongKong2} summarizes the population-level bounds, i.e. bounds if we had population-level classifiers and other population-level quantities. Among our bounds, the bound based on the penalized logistic classifier is the narrowest, in part, because the classifier is similar to the true model for compliance types. Also, our bounds are shorter than the extended bounds by 10\% to 71.4\%, in part because our bounds can use both  continuous and discrete covariates. Having said that, because all the bounds are theoretically guaranteed to cover the target parameter, taking the intersection among them can lead to another, shorter bound for the target parameter. 

\begin{table}[!htp]
		\renewcommand{\arraystretch}{1.2} \centering
		\fontsize{9}{11}\selectfont
\begin{tabular}{|c|c|c|c|c|}
\hline
\multirow{3}{*}{Estimand} & \multicolumn{2}{c|}{\multirow{2}{*}{Classifier-based Bound}}      & \multirow{3}{*}{\begin{tabular}[c]{@{}c@{}}Extended Bound of\\ \citet{GM2008} and \\ \citet{LongHudgens2013}\end{tabular}} & \multirow{3}{*}{Intersection Bound} \\
                          & \multicolumn{2}{c|}{}                            &                                                                    &                                     \\ \cline{2-3}
                          & Linear            & Penalized Logistic &                                                                    &                                     \\ \hline
 $\tau_\NT=0.155$ & $[0.110 , 0.216]$ & $[0.110 , 0.216]$ & $[0.102 , 0.220]$ & $[0.110 , 0.216]$ \\ \hline
 $\tau_\AT=0.148$ & $[0.000 , 0.463]$ & $[0.000 , 0.324]$ & $[0.000 , 0.602]$ & $[0.000 , 0.324]$ \\ \hline
 $\tau_\CO=0.347$ & $[0.160 , 0.450]$ & $[0.218 , 0.450]$ & $[0.118 , 0.931]$ & $[0.218 , 0.450]$ \\ \hline
\end{tabular}
\caption{\footnotesize Sharp Bound Using Different Classifiers and Methods. Each row corresponds to the compliance group effects. Each column shows the classifiers used in our bounds, the extended bound of \citet{GM2008} and \citet{LongHudgens2013}, and the intersection bound created by taking the intersection of our bounds and the extended bound.}
\label{Tab:HongKong2}
\vspace*{-0.3cm}
\end{table}

Table \ref{Tab:HongKong3} reports biases, standard errors, and coverages of the 95\% confidence sets with $B = 1,000$ resamples. We see that the biases are negligible, numerically verifying Theorem \ref{thm:403} and the theoretical properties of the extended bounds. Similarly, the confidence sets achieve nominal coverage, numerically validating the coverage properties. But, our confidence sets are conservative; see Section \ref{sec:addsim} of the supplementary materials for additional details.

\begin{table}[!htp]
		\renewcommand{\arraystretch}{1.2} \centering
		\small
		\setlength{\tabcolsep}{5pt}
		\fontsize{9}{11}\selectfont
\begin{tabular}{|c|c|c|c|c|c|c|c|c|c|}
\hline
\multirow{4}{*}{Estimand}   & \multirow{4}{*}{Statistic} & \multicolumn{4}{c|}{\multirow{2}{*}{Classifier-based Bound}}                            & \multicolumn{2}{c|}{\multirow{3}{*}{\begin{tabular}[c]{@{}c@{}}Extended Bound of\\ \citet{GM2008} and \\ \citet{LongHudgens2013}\end{tabular}}} & \multicolumn{2}{c|}{\multirow{3}{*}{Intersection Bound}} \\
                            &                             & \multicolumn{4}{c|}{}                                                       & \multicolumn{2}{c|}{}                                                                   & \multicolumn{2}{c|}{}                                     \\ \cline{3-6}
                     & & \multicolumn{2}{c|}{Linear} & \multicolumn{2}{c|}{Penalized Logistic} &  \multicolumn{2}{c|}{}   & \multicolumn{2}{c|}{} \\ \cline{3-10} 
                           &  & LB & UB & LB & UB  & LB & UB & LB & UB \\ \hline
                                                                             \multirow{3}{*}{$\tau_\NT$} & Bias $(\times 10^3)$ & $0.8$ & $-8.7$ & $2.4$ & $-10.6$ & $13.0$ & $-1.0$ & $11.1$ & $-11.3$ \\ \cline{2-10}
                                                                                                                      & SE $(\times 10^2)$ & $4.0$ & $\phantom{-}3.6$ & $4.0$ & $\phantom{-}3.6$ & $\phantom{0}3.2$ & $\phantom{-}2.9$ & $\phantom{0}3.4$ & $\phantom{-0}3.6$ \\ \cline{2-10}
                                                                       & Coverage & \multicolumn{2}{c|}{$0.968$} & \multicolumn{2}{c|}{$0.961$} & \multicolumn{2}{c|}{$0.962$} & \multicolumn{2}{c|}{$0.957$} \\ \hline
                                                                               \multirow{3}{*}{$\tau_\AT$} & Bias $(\times 10^3)$ & $0.0$ & $\phantom{0}7.9$ & $19.1$ & $-4.2$ & $0.1$ & $-11.7$ & $19.1$ & $-5.1$ \\ \cline{2-10}
                                                                                                                   & SE $(\times 10^2)$ & $0.0$ & $10.3$ & $\phantom{0}3.7$ & $\phantom{,}10.5$ & $0.1$ & $\phantom{-0}7.7$ & $\phantom{0}3.7$ & $\phantom{,}10.3$ \\ \cline{2-10}
                                                                       & Coverage & \multicolumn{2}{c|}{$0.993$} & \multicolumn{2}{c|}{$0.999$} & \multicolumn{2}{c|}{$0.981$} & \multicolumn{2}{c|}{$0.999$} \\ \hline
                                                                             \multirow{3}{*}{$\tau_\CO$} & Bias $(\times 10^3)$ & $5.6$ & $-1.7$ & $11.5$ & $-10.4$ & $6.5$ & $-4.1$ & $12.2$ & $-10.4$ \\ \cline{2-10}
                                                                                                                      & SE $(\times 10^2)$ & $4.7$ & $\phantom{0}4.7$ & $\phantom{0}4.8$ & $\phantom{-0}4.2$ & $4.0$ & $\phantom{-}5.3$ & $\phantom{0}4.6$ & $\phantom{-0}4.2$ \\ \cline{2-10}
                                                                       & Coverage & \multicolumn{2}{c|}{$0.974$} & \multicolumn{2}{c|}{$0.972$} & \multicolumn{2}{c|}{$0.983$} & \multicolumn{2}{c|}{$0.972$} \\ \hline
\end{tabular}		
\caption{\footnotesize Summary of Estimated Bounds. Each row corresponds to the compliance group effect and relevant statistics.  Each column shows the classifiers used in our bounds, the extended bound of \citet{GM2008} and \citet{LongHudgens2013}, and the intersection bound created by taking the intersection of our bounds and the extended bound.}
\label{Tab:HongKong3}
\end{table}

\section{Application}												\label{sec:HKdata}
We re-analyze the original Hong Kong study using our proposed methods. Our replication analysis focuses on the authors' analysis of the secondary attack rate, which is the infection rate excluding the index individual among households that are assigned to either the control intervention or the face masks plus hand hygiene intervention (i.e. the treatment condition). Ultimately, this led to a total of 96 households, of which 45 households were assigned treatment and the other 51 households were assigned control. 
	
The first row of Table \ref{tab:HK-HTE} shows the results of the overall ITT effect. We see that the cluster-level interventions of giving free masks and hand sanitizers had a statistically significant positive effect (at $\alpha = 0.05$) in decreasing flu cases in Hong Kong, reducing flu cases by 7.77\%p on average (95\% CI: $[ 0.0035,0.1520]$). We remark that this result agrees with the authors' original analysis in their Table 3. Remarkably, our approach, despite not assuming a parametric model for the cluster correlation, has the power to detect a significant effect and thus, strengthens the existing causal conclusion from the authors. 

For heterogeneous ITT effects, we use individual-level, cluster-level, and index individual's/peer's characteristics as covariates. In total, with a constant intercept term, $\bX_{ji}$ consists of 13 components (intercept, four individual-level characteristics, four index individual characteristics, three dummy variables about $\NI_j$, and house size);  see Table \ref{tab:HK-HTE} and Section \ref{sec:addsim} of the supplementary materials for details. From  Table \ref{tab:HK-HTE}, the null hypothesis $H_{0}:\beta_{11}^*=0$ concerning the term $\ind(\NI_j \geq 5)$  is rejected at level $\alpha=0.05$ with a positive $\widehat{\beta}_{11}$, implying that the cluster-level intervention of giving free masks and hand sanitizers was more effective at decreasing infection rates among households with many individuals. Also, the null hypothesis $H_{0}: \beta_{12}^*=0$ concerning house size is rejected at level $\alpha=0.1$ with a negative $\widehat{\beta}_{12}$, implying that the intervention was more effective among those who lived in small houses. Combined, the two results suggest that giving free masks and hand sanitizers was more effective among individuals living in dense households. In addition, the p-value of rejecting the overall non-intercept null hypothesis $H_{0}: \B_{(-0)}^* = 0$ is $3.75 \times 10^{-6}$, implying that the ITT effect is heterogeneous among the covariates. We remark that the original analysis by \citet{HongKong2009} did not analyze heterogeneous ITT effects and missed out on these important insights about the ITT effect. Also, like before, our analysis of heterogeneous ITT effects did not rely on parametric modeling assumptions. 

\begin{table}[!htp]
		\renewcommand{\arraystretch}{1.2} \centering
		\fontsize{9}{11}\selectfont
		\setlength{\tabcolsep}{2pt}
\begin{tabular}{|c|c|c|c|c|c|c|}
\hline
\multirow{2}{*}{Variable}   & \multirow{2}{*}{Estimate} & \multirow{2}{*}{SE}   & \multicolumn{2}{c|}{95\% CI}                   & \multirow{2}{*}{$\chi_1^2$-statistic} & \multirow{2}{*}{p-value} \\ \cline{4-5}
 &                           &                       & LB            & UB           &                                       &                          \\ \hline
Overall ITT $(\tau_\ITT)$ & $\phantom{-}7.77 \times 10^{-2}$ & $3.79 \times 10^{-2}$  & $\phantom{-}3.46 \times 10^{-3}$  & $1.52 \times 10^{-1}$ & $4.207$ & $0.040$ \\ \hline
Intercept ($\beta_0^*$) & $\phantom{-}3.41 \times 10^{-1}$ & $2.00 \times 10^{-1}$  & $-5.19 \times 10^{-2}$  & $7.34 \times 10^{-1}$ & $2.894$ & $0.089$ \\ \hline
Gender ($\beta_1^*$) & $\phantom{-}2.96 \times 10^{-2}$ & $6.03 \times 10^{-2}$  & $-8.86 \times 10^{-2}$  & $1.48 \times 10^{-1}$ & $0.241$ & $0.623$ \\ \hline
Age ($\beta_2^*$) & $-4.48 \times 10^{-3}$ & $6.45 \times 10^{-3}$  & $-1.71 \times 10^{-2}$  & $8.16 \times 10^{-3}$ & $0.483$ & $0.487$ \\ \hline
Age$^2$ ($\beta_3^*$) & $\phantom{-}3.22 \times 10^{-5}$ & $7.31 \times 10^{-5}$  & $-1.11 \times 10^{-4}$  & $1.76 \times 10^{-4}$ & $0.194$ & $0.660$ \\ \hline
Vaccination ($\beta_4^*$) & $-5.43 \times 10^{-2}$ & $6.76 \times 10^{-2}$  & $-1.87 \times 10^{-1}$  & $7.82 \times 10^{-2}$ & $0.645$ & $0.422$ \\ \hline
Index ind. gender ($\beta_5^*$) & $-8.85 \times 10^{-2}$ & $5.98 \times 10^{-2}$  & $-2.06 \times 10^{-1}$  & $2.87 \times 10^{-2}$ & $2.188$ & $0.139$ \\ \hline
Index ind. age ($\beta_6^*$) & $-6.38 \times 10^{-3}$ & $8.76 \times 10^{-3}$  & $-2.35 \times 10^{-2}$  & $1.08 \times 10^{-2}$ & $0.531$ & $0.466$ \\ \hline
Index ind. Age$^2$ ($\beta_7^*$) & $\phantom{-}1.66 \times 10^{-4}$ & $1.52 \times 10^{-4}$  & $-1.31 \times 10^{-4}$  & $4.63 \times 10^{-4}$ & $1.199$ & $0.273$ \\ \hline
Index ind. vaccination ($\beta_8^*$) & $\phantom{-}7.39 \times 10^{-2}$ & $5.81 \times 10^{-2}$  & $-3.99 \times 10^{-2}$  & $1.88 \times 10^{-1}$ & $1.619$ & $0.203$ \\ \hline
$\ind(\NI_j=3)$ ($\beta_9^*$) & $\phantom{-}3.29 \times 10^{-2}$ & $8.99 \times 10^{-2}$  & $-1.43 \times 10^{-1}$  & $2.09 \times 10^{-1}$ & $0.134$ & $0.714$ \\ \hline
$\ind(\NI_j=4)$ ($\beta_{10}^*$) & $-3.34 \times 10^{-2}$ & $7.56 \times 10^{-2}$  & $-1.82 \times 10^{-1}$  & $1.15 \times 10^{-1}$ & $0.195$ & $0.659$ \\ \hline
$\ind(\NI_j\geq 5)$ ($\beta_{11}^*$) & $\phantom{-}2.76 \times 10^{-1}$ & $1.09 \times 10^{-1}$  & $\phantom{-}6.27 \times 10^{-2}$  & $4.90 \times 10^{-1}$ & $6.429$ & $0.011$ \\ \hline
House size ($\beta_{12}^*$) & $-1.46 \times 10^{-4}$ & $7.57 \times 10^{-5}$  & $-2.94 \times 10^{-4}$  & $2.24 \times 10^{-6}$ & $3.726$ & $0.054$ \\ \hline
\end{tabular}
\caption{\footnotesize ITT Effects from the Hong Kong Study. Each row represents an estimand and each column shows relevant statistics.}
\label{tab:HK-HTE}
\vspace*{-0.3cm}
\end{table}

Finally, we estimate the bounds among the compliance types. Similar to the simulation study, we use the linear and penalized logistic classifiers for our bounds. We also compare our bounds to the extended bounds that only use binary covariates, specifically the eight indicator variables generated by $\{ X_{ji,{\rm male}}$, $\ind(\NI_j = 3)$, $\ind(\NI_j=4)$, $\ind(\NI_j \geq 5) \}$; each stratum created by these four covariates contains at least one control and treated clusters. 

Table \ref{tab:HK-NTSE} summarizes the results from the bounds. Because the estimated number of ATs in the population is three we only present the results about $\tau_\NT$ and $\tau_\CO$; see Section \ref{sec:addsim} of the supplementary materials for the full results. The linear classifier and the penalized logistic classifier produced similar bounds for all the effects. Also, our classifier-based bounds are narrower than the extended bounds using only binary covariates. Specifically, the classifier-based bounds of $\tau_\NT$ and $\tau_\CO$ are 13.5\%  and 21.5\% narrower than the extended bounds of $\tau_\NT$ and $\tau_\CO$, respectively. But, the lower bound estimate of $\tau_\NT$ under the extended bound is positive whereas our classifier-based bound touches $0$; see the next paragraph for additional discussion. The intersection bound for the NTs suggests a small, but statistically insignificant (at $\alpha = 0.05$ level) spillover effect among NTs, ranging from 5.4\%p to 17.3\%p reduction in infection rates among NTs from their NT peers wearing masks and using hand sanitizers. Similarly, the intersection bound for the COs suggests a small, but statistically insignificant (at $\alpha = 0.05$ level) total effect, ranging from 0\%p to 14.6\%p.

\begin{table}[!htp]
		\renewcommand{\arraystretch}{1.2} \centering
		\fontsize{9}{11}\selectfont
\begin{tabular}{|c|c|c|c|c|c|}
\hline
\multirow{3}{*}{Estimand}   & \multirow{3}{*}{Statistic} & \multicolumn{2}{c|}{\multirow{2}{*}{Classifier-based Bound}} & \multirow{3}{*}{\begin{tabular}[c]{@{}c@{}}Extended Bound of\\ \citet{GM2008} and \\ \citet{LongHudgens2013}\end{tabular}} & \multirow{3}{*}{Intersection Bound} \\
                            &                             & \multicolumn{2}{c|}{}                            &                                                                      &                                     \\ \cline{3-4}
                            &                             &              Linear            &    Penalized Logistic                 &                                                                      &                                     \\ \hline
\multirow{2}{*}{$\tau_\NT$}                  & Bound                      & $ [0.000 , 0.173]$      & $[0.000 , 0.173]$      & $[0.054 , 0.254]$                                                    & $[0.054 , 0.173]$                   \\ \cline{2-6}
                            & 95\% CI                     & $ [0.000 , 0.374]$      & $[0.000 , 0.375]$      & $[0.000 , 0.395]$                                                    & $[0.000 , 0.374]$                   \\ \hline
\multirow{2}{*}{$\tau_\CO$} & Bound                      & $ [0.000 , 0.146]$      & $[0.000 , 0.146]$      & $[0.000 , 0.186]$                                                    & $[0.000 , 0.146]$                   \\ \cline{2-6} 
                            & 95\% CI                     & $ [0.000 , 0.299]$      & $[0.000 , 0.288]$      & $[0.000 , 0.297]$                                                    & $[0.000 , 0.288]$                   \\ \hline
\end{tabular}		
\caption{\footnotesize Network Effects Among Compliance Types in the Hong Kong Study. Each row corresponds to the compliance group effects and relevant statistics.  Each column shows the classifiers used in our bounds, the extended bound of \citet{GM2008} and \citet{LongHudgens2013}, and the intersection bound created by taking the intersection of our bounds and the extended bound.}
\label{tab:HK-NTSE}
\vspace*{-0.3cm}
\end{table}

We take a moment to describe why the extended bound for the NTs is longer than the classifier-based bounds for the NTs, but the extended bound produced a lower bound that is greater than the classifier-based lower bounds. As discussed in Section \ref{sec:lp}, Theorem \ref{thm:402} does not say how far the lower and upper bounds are away from the true effect. As such, the extended bound, despite its longer length, could produce a lower bound that is closer to the true effect than the classifier-based bounds. Thankfully, because we showed that all the bounds, including our extension of the existing bounds, must cover the true effect, we can take the intersection of the bounds to obtain a more informative bound of the treatment effect.

\section{Conclusion}

This paper presents assumption-lean methods to analyze two types of causal effects in CRTs from infectious diseases, the ITT effects and the network effects among different compliance types. For the ITT effects, we make a modest extension of \citet{Ding2019} to CRT settings where we propose estimators that are (i) robust to affine transformations, (ii) do not require parametric modeling assumptions, and (iii) allow the cluster size to potentially grow to infinity. For the network effects, we present a new method to obtain sharp bounds by using LP and supervised ML where using a good classifier for a compliance type can tighten bounds. This new approach gives investigators more ways to tune and shorten bounds using  ML methods instead of only relying on having good data and may have broader applications in other areas of instrumental variables. 

We end by offering some advice on using our methods in practice. First, we believe our approach to analyzing the ITT effects is promising, especially since our method was able to detect significant heterogeneous effects without making explicit modeling assumptions on the cluster structure, say Normally distributed random effects. Second, while a bound-based analysis is often considered conservative, as shown in the empirical example, investigators can fine-tune classifiers to narrow bounds. Third, as the empirical examples showed, while the classifier-based bounds are shorter than the extended bounds based only on binary covariates, intersecting both types of bounds may provide more information about the treatment effect than using one of them alone.

\newpage

\section*{Supplementary Material}

	Section \ref{sec:A} presents additional results related to the main paper. 
	Section \ref{sec:Lemma} presents lemmas that facilitates the proofs in Section \ref{sec:B} and \ref{sec:C}.
	Section \ref{sec:B} proves theorems stated in the paper and the supplementary material. 
	Section \ref{sec:C} proves lemmas introduced in the supplementary material.

\appendix

\section{Additional Results}									\label{sec:A}

\subsection{Details of Section \ref{sec:1} in the Main Paper}					\label{sec:EXofParaMethod}
In Table \ref{tab:refs}, we provide a selective overview of empirical work concerning CRTs for infectious diseases where treatment spillover effects were discussed and noncompliance was reported; for additional studies, see the references cited in these works. Overall, every study relied on some form of parametric modeling, some with random effect models and some using cluster bootstrap standard errors. Additionally, every study had more study units than the number of clusters, with some clusters having 400 study units per cluster \citep{WORMS} while others 1.35 study unit per cluster \citep{Duflo2012}.

\subsection{Details of Section \ref{sec:2} in the Main Paper}					\label{sec:supp-sec2}

\subsubsection{Discussion about Assumptions \protect\hyperlink{(A1)}{(A1)} in the context of the Hong Kong study} \label{sec:supp-sec2-1}

For the Hong Kong study, Assumption \hyperlink{(A1)}{(A1)} (i.e. partial interference) is a plausible, first-order approximation of the interference pattern for the following reasons. First, the study was limited to households that already included the flu-infected individual (i.e. the index patient) and the secondary attack rate (i.e. the outcome among the index patient's peers) was assessed a week after randomization and within 36 hours after the discovery of the index patient. Second, past works studying influenza transmissions state that households are a key unit of influenza transmission due to the closeness between household members \citep{Flu1,Flu4,Flu2,Flu3}. Nevertheless, if a study unit frequently interacted with other study units in different households, especially between the time of randomization and the outcome was recorded, partial interference may be a poor approximation of the interference pattern in the Hong Kong study.

\begin{sideways}%
\begin{minipage}{0.95\textheight}
 \resizebox{.95\textheight}{!}{
\renewcommand{\arraystretch}{1.5}
		\setlength{\tabcolsep}{4pt}
		\fontsize{11}{9} \selectfont
\begin{tabular}{|c|c|c|c|c|c|c|c|c|}
\hline
Reference & Intervention & Primary Outcome & Cluster (size) & Individual (size) & Analytic Method
\\ \hline
\citet{WORMS} & Having deworming drug & Helminth infection & School (75) & Student ($\sim$30,000) &  \begin{tabular}[c]{@{}c@{}}Cluster-robust\\[-0.2cm] probit regression\end{tabular}
\\ \hline
\citet{HongKong2009} & \begin{tabular}[c]{@{}c@{}}Using masks\\[-0.2cm] and hand sanitizer\end{tabular}  & Influenza infection & Household (259) & \begin{tabular}[c]{@{}c@{}}Household member\\[-0.2cm] (794)\end{tabular} & \begin{tabular}[c]{@{}c@{}}Logistic model with\\[-0.2cm] cluster-robust bootstrap\\[-0.2cm] Generalized estimating equations\end{tabular} 
\\ \hline
\citet{France2010} & Using masks  & Influenza-like illness & Household (105) & \begin{tabular}[c]{@{}c@{}}Household member\\[-0.2cm] (306)\end{tabular} & \begin{tabular}[c]{@{}c@{}}Logistic model with\\[-0.2cm] cluster-robust bootstrap\end{tabular}
\\ \hline
\citet{US2011} &  \begin{tabular}[c]{@{}c@{}}Giving hand sanitizer\\[-0.2cm] with education\end{tabular}  & Influenza infection & School (10) & Student (3,360) & \begin{tabular}[c]{@{}c@{}}Poisson/negative binomial\\[-0.2cm] mixed effect model\end{tabular} 
\\ \hline
\citet{China2011} &  \begin{tabular}[c]{@{}c@{}}Wearing masks\\[-0.2cm] or respirators\\[-0.2cm] during the work shift\end{tabular}  & \begin{tabular}[c]{@{}c@{}}Influenza-like illness\\[-0.2cm] Influenza infection\end{tabular} & Hospital (15) & \begin{tabular}[c]{@{}c@{}}Health care workers\\[-0.2cm] (1,441)\end{tabular} & \begin{tabular}[c]{@{}c@{}}Logistic mixed\\[-0.2cm] effect model\end{tabular}
\\ \hline
\citet{Michigan2012} &  \begin{tabular}[c]{@{}c@{}}Using masks\\[-0.2cm] and hand sanitizer\end{tabular}  & \begin{tabular}[c]{@{}c@{}}Influenza-like illness\\[-0.2cm] Influenza infection\end{tabular} &  \begin{tabular}[c]{@{}c@{}}University residence\\[-0.2cm] hall (37)\end{tabular} & \begin{tabular}[c]{@{}c@{}}University student\\[-0.2cm] (1,178)\end{tabular} &  \begin{tabular}[c]{@{}c@{}}Discrete-time\\[-0.2cm] survival analysis\end{tabular}
\\ \hline
\citet{Duflo2012} &  \begin{tabular}[c]{@{}c@{}}Providing information \\[-0.2cm] about and assisting \\[-0.2cm] the connection to\\[-0.2cm] the water network\end{tabular}  & \begin{tabular}[c]{@{}c@{}}Water usage\\[-0.2cm] Child health\\[-0.2cm] Overall well-being\end{tabular}  &  \begin{tabular}[c]{@{}c@{}}Clustered by\\[-0.2cm] location (626)\end{tabular} & Household (845) &  \begin{tabular}[c]{@{}c@{}}Cluster-robust\\[-0.2cm] linear regression\end{tabular}
\\ \hline
\citet{Duflo2015} & \begin{tabular}[c]{@{}c@{}}Providing education subsidy\\[-0.2cm] and/or HIV education program\end{tabular} & \begin{tabular}[c]{@{}c@{}}School attendance\\[-0.2cm] Early marriage/pregnancy\\[-0.2cm] HIV/Herpes infection\end{tabular} & School (328) & Student (19,289) &  \begin{tabular}[c]{@{}c@{}}Cluster-robust\\[-0.2cm] linear regression\end{tabular}
\\ \hline
\citet{China2016} &  Using medical masks & \begin{tabular}[c]{@{}c@{}}Clinical respiratory illness\\[-0.2cm] Influenza-like illness\\[-0.2cm] Respiratory infection\end{tabular} &  Hospital (6) & \begin{tabular}[c]{@{}c@{}}Household\\[-0.2cm] (245)\end{tabular} &  \begin{tabular}[c]{@{}c@{}}Survival analysis\end{tabular}
\\ \hline
\end{tabular}}
\renewcommand\thetable{A.1}
\captionof{table}{Summary of the references in Section \ref{sec:1} in the main paper. Both noncompliance and spillover effects are present in all references.}
\label{tab:refs}
\end{minipage}
\end{sideways}

\subsubsection{Relationship between the ITT Effect and the Average Individual Causal Effects}
\label{sec:supp-sec2-2}

We introduce the relationship between the ITT effect among NT/ATs and the average indirect causal effects  defined in \citet{HH2008}. For NT, the individual treatment effect is written as
	\begin{align}							\label{eq-spillover}
		\Big\{ \potY{Y_{ji}}{1} - \potY{Y_{ji}}{0} \Big\} \NT_{ji}
		 =
		\Big\{ \pot{Y_{ji}}{d_{ji} = 0 , \pot{\trt_\eji}{1}}
		-
		\pot{Y_{ji}}{d_{ji} = 0 , \pot{\trt_\eji}{0}}\Big\} \NT_{ji} \ .
	\end{align}
	The equality is from the definition of $\potY{Y_{ji}}{\ivp}$ and the exclusion restriction assumption (i.e. Assumption \hyperlink{(A4)}{(A4)}. In  \citet{HH2008}, the individual average indirect causal effect comparing treatment policy $\psi$ to $\psi'$ is re-written in our notation as follows.
	\begin{align}							\label{eq-HH}
		&
		\overline{Y}_{ji} (0 \con \psi) 
		-
		\overline{Y}_{ji} (0 \con \psi') 
		=
		\hspace*{-0.1cm}
		\sum_{ \TRTp_\eji  }
		\hspace*{-0.1cm}
		\pot{ Y_{ji} }{ d_{ji} = 0 , \TRTp_\eji } 	
		\bigg\{
			\begin{array}{l}
				P_{\psi} \big( \TRT_\eji = \TRTp_\eji \cond \TRT_{ji} = 0 \big)
				\\[-0.2cm]
				\quad \quad
				-
				P_{\psi'} \big( \TRT_\eji = \TRTp_\eji \cond \TRT_{ji} = 0 \big)
			\end{array} \bigg\} \ .
	\end{align}
	We take $\psi$ as the policy where all individuals are encouraged to take the treatment, and $\psi'$ as the policy where all individuals are assigned to control. Since the treatment receipt is determined for each individual under our setup, $P_{\psi} \big( \TRT_\eji = \pot{\TRT_\eji}{1} \cond \TRT_{ji} = 0 \big)=1$ for NTs, and this leads $\overline{Y}_{ji} (0 \con \psi)  = \pot{ Y_{ji} }{ d_{ji} = 0 , \pot{\TRT_\eji}{1} }$. Similarly, we obtain $\overline{Y}_{ji} (0 \con \psi')  = \pot{ Y_{ji} }{ d_{ji} = 0 , \pot{\TRT_\eji}{0} }$ for NTs. Therefore, \eqref{eq-spillover} is a special case of \eqref{eq-HH} when comparing the all-encouragement policy against a no-encouragement policy. By averaging the individual average indirect causal effect among NTs, we find that the NT group average indirect causal effect under $\psi_1$ and $\psi_0$ is equivalent to $\tau_\NT$ as follows.
	\begin{align*}
		&
		\frac{1}{\NIT_\NT}
		\sumji
		\big\{ \overline{Y}_{ji} (0 \con \psi_1)  - \overline{Y}_{ji} (0 \con \psi_0)  \big\} \NT_{ji}
		=
		\frac{1}{\NIT_\NT}
		\sumji
		\Big\{ \pot{ Y_{ji} }{ d_{ji} = 0 , \pot{\TRT_\eji}{1} } - \pot{ Y_{ji} }{ d_{ji} = 0 , \pot{\TRT_\eji}{0} } \Big\} \NT_{ji}
		=
		\tau_\NT \ .
	\end{align*}
	Therefore, $\tau_\NT$ can be understood as the NT group average indirect causal effect comparing $\psi_1$ to $\psi_0$. By a similar manner, $\tau_\AT$ can be interpreted as the AT group average indirect causal effect comparing $\psi_1$ to $\psi_0$.

\subsection{Details of Section \ref{sec:3} in the Main Paper}					\label{sec:supp-sec1}

\subsubsection{Notation}

We introduce two operators on tensors, the sample mean operator $\mathcal{M}$ and the sample covariance operator $\mathcal{V}$. Let $\bm{\mathcal{S}} = \{ \mathcal{S}_1 , \ldots, \mathcal{S}_\NC \}$ and $\bm{\mathcal{T}} = \{ \mathcal{T}_1 , \ldots, \mathcal{T}_\NC \}$ be collections of $k_1 \times k_2$ matrices where the $\ell$th column of $\mathcal{S}_j$ and $\mathcal{T}_j$ are denoted as $\bm{s}_{j\ell} \in \R^{k_1}$ and $\bm{t}_{j\ell} \in \R^{k_1}$, respectively, for $\ell = 1,\ldots, k_2$. Then, the sample mean operator  $\mo: \{ \R^{k_1 \times k_2} \} ^{\otimes \NC} \rightarrow \R^{k_1 \times k_2}$ on $\bm{\mathcal{S}}$ and the sample covariance operator $\vo: \{ \R^{k_1 \times k_2} \} ^{\otimes \NC} \otimes \{ \R^{k_1 \times k_2} \} ^{\otimes \NC} \rightarrow \R^{k_1k_2 \times k_1k_2}$ on $\bm{\mathcal{S}}$ and $\bm{\mathcal{T}}$ are defined as
\begin{align*}
	\mo \big( \bm{\mathcal{S}} \big)
	= \frac{1}{\NC} \sum_j \mathcal{S}_j, \
	\vo ( \bm{\mathcal{S}}, \bm{\mathcal{T}} ) 
	=
	\frac{1}{\NC-1} \sum_j \big[ \textsf{vec}(\mathcal{S}_j) - \textsf{vec}\big\{ \mo( \bm{\mathcal{S}} ) \big\} \big] \big[ \textsf{vec}(\mathcal{T}_j) - \textsf{vec}\big\{ \mo( \bm{\mathcal{T}} ) \big\} \big] \T
	\ .
\end{align*}
Here $\textsf{vec}(\bm{A}) = \big( \bm{a}_{1}\T , \ldots, \bm{a}_{k_2} \T \big)\T$ is the vectorization of a $k_1\times k_2$ matrix $\bm{A} = [ \bm{a}_1,\ldots, \bm{a}_{k_2} ]$ where $\bm{a}_{\ell} \in \R^{k_1}$ is the $\ell$th column vector of $\bm{A}$. Also, we denote the covariance matrix of $\bm{\mathcal{S}}$ as $\vo(\bm{\mathcal{S}}) = \vo (\bm{\mathcal{S}}, \bm{\mathcal{S}} )$. When $k_2=1$, $\mo \big( \bm{\mathcal{S}} \big)$ and $\vo \big( \bm{\mathcal{S}} \big)$ reduce to the familiar sample mean and the covariance matrix of $J$ vectors of $\mathcal{S}_1,\ldots,\mathcal{S}_\NC$ with $\mathcal{S}_j \in \R^{k_1}$.  Let $\potY{\Y_j}{\ivp} = \big({\NC}/{\NIT} \big) \sum_{i=1}^{\NI_j} \potY{Y_{ji}}{\ivp}$ be the scaled average of potential outcomes in cluster $j$ under treatment arm $\ivp \in \{0,1\}$ and let $\potY{\bm{\Y}}{\ivp} = \big\{ \potY{\Y_1}{\ivp} , \ldots , \potY{\Y_\NC}{\ivp} \big\} $ be the collection of $\potY{\Y_j}{\ivp}$s across $\NC$ clusters.  Let $\potY{ \Y_{X,j}}{\ivp} = (\NC / \NIT) \sum_{i=1}^{\NI_j} \bX_{ji} \potY{Y_{ji}}{\ivp}$, $\X_j = (\NC/\NIT) \sum_{i=1}^{\NI_j} \bX_{ji} \bX_{ji}\T$,  $\potY{\bm{\Y}_X}{\ivp} = \big\{ \potY{\Y_{X,1}}{\ivp} , \ldots , \potY{\Y_{X, \NC}}{\ivp} \big\}$, and  $\bm{\X} = \big\{ \X_1,\ldots,\X_\NC \big\}$. Then, we obtain
$\tau_\ITT = \sumji \tau_{ji}/\NIT = \mo \big\{ \potY{ \bm{\Y} }{1} \big\} - \mo \big\{ \potY{ \bm{\Y} }{0} \big\}$ and $\B^*$ becomes $\B^* = \big\{ \mo \big( \bm{\X} \big)  \big\}^{-1}
	 \big[ \mo \big\{ \potY{\bm{\Y}_X }{1} \big\} - \mo \big\{ \potY{\bm{\Y}_X }{0} \big\} \big] $ if $ \NIT^{-1} \sumji \bX_{ji} \bX_{ji}\T $ is invertible.

We consider $\widehat{\mo}_\ivp (\bm{\mathcal{S}}) $ and $\widehat{\vo}_\ivp (\bm{\mathcal{S}}) $, which are estimators of the mean and covariance operators under treatment arm $\ivp \in \{0,1\}$, i.e.
\begin{align*}
	&\widehat{\mo}_\ivp (\bm{\mathcal{S}}) 
	= \frac{  \sum_j\ind (\iv_j=\ivp)  \mathcal{S}_j }{\sum_j \ind (\iv_j=\ivp)}
	\  ,  \
	&
	\widehat{\vo}_\ivp (\bm{\mathcal{S}}) 
	=
	\frac{\sum_j \ind( \iv_j = \ivp)\big[ \textsf{vec}(\mathcal{S}_j) - \textsf{vec}\big\{ \widehat{\mo}_\ivp(\bm{\mathcal{S}}) \big\} \big]^{\otimes 2}}{ \sum_j \ind(\iv_j = \ivp) - 1}	
	\ .
\end{align*}
Let $\bm{\Y} =\big \{ \potY{\Y_1}{\iv_1},\ldots, \potY{\Y_\NC}{\iv_\NC} \big\}$ and $\bm{\Y}_X = \big\{ \potY{ \Y_{X,1}}{\iv_1},\ldots,\potY{ \Y_{X,\NC}}{\iv_\NC} \big\}$ be the observed values of $\potY{\Y_j}{\ivp_j}$ and $\potY{ \Y_{X,j}}{\ivp_j}$, respectively. Also, let $\bm{\mathcal{N}} = \big\{ (\NC/\NIT) \NI_1, \ldots, (\NC/\NIT) \NI_\NC \big\}$ be the collection of scaled cluster sizes. Then, the ratio estimators of the overall and heterogenous ITT effects are represented as $	\widehat{\tau}_{\ITT} = \big\{ \widehat{\mo}_1 \big( \bm{\mathcal{N}} \big) \big\}^{-1}
		\widehat{\mo}_1 \big( \bm{\Y} \big)
		-
		\big\{ \widehat{\mo}_0 \big( \bm{\mathcal{N}} \big) \big\}^{-1}
		\widehat{\mo}_0 \big( \bm{\Y} \big)$ and $
	\widehat{\B}
	=
	\big\{ \widehat{\mo}_1 \big( \bm{\X} \big) \big\}^{-1}
	\widehat{\mo}_1 \big( \bm{\Y}_X \big)
	-
	\big\{ \widehat{\mo}_0 \big( \bm{\X} \big) \big\}^{-1}
	\widehat{\mo}_0 \big( \bm{\Y}_X \big)$.

\subsubsection{Dependence of Difference-in-means Estimator on Affine Transformations}							\label{sec:ATvariance}					

In this subsection, we show the Neyman-type unbiased estimators $\widehat{\tau}_{\ITT} = \widehat{\mo}_1 \big( \bm{\Y} \big) - \widehat{\mo}_0 \big( \bm{\Y} \big)$ and $\widehat{\B} = \big\{ \mo \big( \bm{\X} \big)  \big\}^{-1}
	\big\{	\widehat{\mo}_1 ( \bm{\Y}_X )
	-	\widehat{\mo}_0 ( \bm{\Y}_X )	\big\} $ are sensitive to transformation in the outcome variable. Suppose the outcome variable is transformed as $Y_{ji}' = c + d Y_{ji}$. The corresponding estimators associated with $Y_{ji}'$ are $ \widehat{\tau}_\ITT '
	=
	\widehat{\mo}_1 ( c\bm{1}_\NC + d \bm{\Y} )
	-
	\widehat{\mo}_0 ( c\bm{1}_\NC + d \bm{\Y} )$ and $ \widehat{\B} '
	=
	\big\{ \mo \big( \bm{\X} \big)  \big\}^{-1}
	\big\{
	\widehat{\mo}_1 ( c\bm{ \mathcal{C} } + d \bm{\Y}_X )
	-
	\widehat{\mo}_0 ( c\bm{ \mathcal{C} } + d \bm{\Y}_X )
	\big\}$ where $\bm{1}_\NC = \{ 1, \ldots , 1 \} $ and $\bm{\mathcal{C}} = \{ \mathcal{C}_1,\ldots, \mathcal{C}_\NC \} $ with $\mathcal{C}_j = (\NC/\NIT) \sum_{i=1}^{\NI_j}\bX_{ji}$. We observe that ${\widehat{\mo}_\ivp (c\bm{1}_{\NC} + d \bm{\Y} )
	= c  \NC \NIT_\ivp / \{ \NIT \sum_j \ind(\iv_j=\ivp) \}	+	d\widehat{\mo}_\ivp (\bm{\Y} )}$ $(\ivp=0,1)$ where ${\NIT_z = \sum_j \ind(\iv_j =z) \NI_j}$.
	
Thus, the dependence of $\widehat{\tau}_\ITT'$ on $c$ does not vanish because $\widehat{\tau}_\ITT' = (c\NC/\NIT) \big\{ {\NIT_1}/{\tNC} - {\NIT_0}/{(\NC - \tNC)} \big\}
	+
	 d	 \widehat{\tau}_\ITT$.
In particular, we are left with the contrast $\NIT_1/\tNC - \NIT_0/(\NC - \tNC)$. Now, if every cluster has the same number of people, say $\NI$, $\NIT_0 = (\NC-\tNC) \NI$, $\NIT_1= \tNC \NI$ and this contrast is equal to zero. But, if the cluster size is heterogeneous where some clusters are larger than other clusters and as such, the treated clusters may be larger (or smaller) than the control clusters, this contrast is no longer zero. In other words, the Neyman-type does not account for the imbalances in cluster size between the treated and control clusters. Another way to interpret the non-vanishing contrast is that while the target estimand is averaging across all units in the study, the Neyman-type unbiased estimators focuses on the contrasts at the cluster-level by using only cluster-level treatment assignment and as such. Similarly,  $\widehat{\mo}_1 (c\bm{\mathcal{C}} + d \bm{\Y}_X )$ and $\widehat{\mo}_0 (c\bm{\mathcal{C}} + d \bm{\Y}_X )$ are given as ${ \widehat{\mo}_\ivp (c\bm{\mathcal{C}} + d \bm{\Y}_X )
	= c \NC \sum_{ji} \ind( \iv_j = \ivp)  \bX_{ji} / \{ \NIT \sum_j \ind( \iv_j = \ivp) \} + 	d\widehat{\mo}_\ivp (\bm{\Y}_X )}$.  Similar to the overall ITT effect, the dependence of $\widehat{\B}'$ on $c$ does not vanish. 
In summary, the Neyman-type unbiased estimators are sensitive to transformation in the outcome variable.

\subsubsection{Independence of Ratio Estimator on Affine Transformations}							\label{sec:ATinvariance}

In this subsection, we show the ratio estimators	$\widehat{\tau}_{\ITT}
	= \big\{ \widehat{\mo}_1 \big( \bm{\mathcal{N}} \big) \big\}^{-1}
		\widehat{\mo}_1 \big( \bm{\Y} \big)
		-
		\big\{ \widehat{\mo}_0 \big( \bm{\mathcal{N}} \big) \big\}^{-1}
		\widehat{\mo}_0 \big( \bm{\Y} \big)$ and $	\widehat{\B}
	=
	\big\{ \widehat{\mo}_1 \big( \bm{\X} \big) \big\}^{-1}
	\widehat{\mo}_1 \big( \bm{\Y}_X \big)
	-
	\big\{ \widehat{\mo}_0 \big( \bm{\X} \big) \big\}^{-1}
	\widehat{\mo}_0 \big( \bm{\Y}_X \big)$ are insensitive to transformation in the outcome variable. Suppose the outcome variable is transformed as $Y_{ji}' = c + d Y_{ji}$. The corresponding estimators associated with $Y_{ji}'$ are $\widehat{\tau}_{\ITT} ' = \widehat{\tau}_{1,\ITT} ' - \widehat{\tau}_{0,\ITT} '$  and $\widehat{\B} '
	=
	\widehat{\B}_1' - \widehat{\B}_0'$ where $\widehat{\tau}_{\ivp,\ITT} ' 
	= \{ \sum_j \ind(\iv_j=\ivp) \NI_j \}^{-1} \{ \sum_{ji} \ind(\iv_j=\ivp) (c + dY_{ji}) \}$ and $\widehat{\B}_\ivp'
	=
	\big\{
		\sum_{ji} \ind(\iv_j=\ivp) \bX_{ji} \bX_{ji}\T
	\big\}^{-1}
	\big\{
		\sum_{ji}  \ind(\iv_j=\ivp)  \bX_{ji} (c + dY_{ji})
	\big\}$.
For the ITT estimator, it is straightforward to find $\widehat{\tau}_\ITT' = d \widehat{\tau}_\ITT$. At a high level, this is because the ratio estimator now takes into consideration the number of treated units in each cluster through its denominator. This has the effect of cancelling out the dependence on the shifting term $c$ when taking the difference in the outcome between treated and control units. Similarly, to show that $\widehat{\B}$ is invariant to affine transformations, we use $\big\{
		\sum_{ji} \ind( \iv_j = \ivp ) \bX_{ji} \bX_{ji}\T
	\big\}^{-1} \big\{ \sum_{ji} \ind( \iv_j = \ivp ) \bX_{ji} \big\} = \bm{u}_1 = (1,0,\ldots, 0)\T \in \R^p$ for $\ivp=0,1$  so long as the first element of $\bX_{ji}$ is 1 (i.e. $\bX_{ji}$ includes the intercept). As a result, we find $\widehat{\B}' = d \widehat{\B}$.


\subsubsection{Asymptotic Properties of the Overall and Heterogeneous ITT Effect Estimators}			\label{sec:TheoryITT}				

In this subsection, we present the details of Theorem \ref{thm:302} in the main paper. First, we introduce the finite moment assumption. 
\begin{assumption}[Finite Moment Assumption] 						\label{assp:3-1}
Let $\potY{\bm{ \mathcal{R} }}{z} = \big\{ \potY{\mathcal{R}_{1}}{z},\ldots, \potY{\mathcal{R}_{\NC}}{z} \big\} $ and $\potY{\bm{ \mathcal{R} }_X}{z}  = \big\{ \potY{\mathcal{R}_{X,1}}{z},\ldots, \potY{\mathcal{R}_{X,\NC}}{z} \big\} $ with $\potY{ \mathcal{R}_{j}}{z} = \potY{\Y_j}{\ivp} - \mathcal{N}_j \mo \big\{ \potY{ \bm{\Y} }{z} \big\}$, and $\potY{ \mathcal{R}_{X,j} }{\ivp} = \potY{ \Y_{X,j}}{\ivp} - \X_j \big\{ \mo \big( \bm{\X} \big) \big\}^{-1}  \mo \big\{ \potY{ \bm{\Y}_X }{z} \big\} $ for $\ivp \in \{0,1\}$. The following condition holds.
	\begin{itemize}
		\item[(i)] (\textit{Asymptotic rate of $\tNC$ and $\NC$}): As $J \to \infty$, we have $\tNC/\NC \rightarrow p_{\rm trt} \in (0,1)$.
		
		\item[(ii)] (\textit{First Moments and full rank}): For each $\ivp = 0,1$, $\mo \big\{ \potY{ \bm{\mathcal{R}} }{z} \big\}$ and $\mo \big\{ \potY{ \bm{ \mathcal{R}}_{X}}{z}  \big\}$ have finite limits as $\NC \rightarrow \infty$. Also,  $\mo (\bm{\X}  )$ is full rank and $\lim_{\NC \rightarrow \infty} \mo ( \bm{\X}  )$ converges to a full-rank matrix.

		\item[(iii)] (\textit{Second Moments}): For each $\ivp = 0,1$, $\vo \big\{\potY{ \bm{\mathcal{R}}}{z} \big\}$, $\vo \big\{ \potY{ \bm{ \mathcal{R}}_{X}}{z} \big\}$, $\vo \big\{ \potY{ \bm{\mathcal{R}}}{1},\potY{ \bm{\mathcal{R}}}{0}\big\}$, $\vo \big\{ \potY{ \bm{ \mathcal{R}}_{X}}{1} , \potY{ \bm{ \mathcal{R}}_{X}}{0} \big\}$, and $\vo ( \bm{\X} )$ have nonzero finite limits as $\NC \rightarrow \infty$. 
		
		\item[(iv)] (\textit{Tightness}) As $\NC \rightarrow \infty$, $\big\| \potY{ \mathcal{R}_{j}}{z}  - \mo\big\{ \potY{ \bm{\mathcal{R}}}{z}  \big\} \big\|_2^2/ \NC$, $\big\|\potY{ \mathcal{R}_{X,j} }{\ivp}- \mo\big\{ \potY{ \bm{ \mathcal{R}}_{X}}{z}  \big\} \big\|_2^2/ \NC$, and $\big\| \X_{j} - \mo ( \bm{\X} ) \big\|_2^2/ \NC$ converge to zero for all $j=1,\ldots,\NC$ and $\ivp=0,1$.
	\end{itemize}

\end{assumption}
\noindent We remark that $\NC$ and $\NI_j$ can grow so long as (i)-(iv) hold. Also, if $\NI_j$ is bounded, the result will still hold so long as (i)-(v) are satisfied. As a simple example, bounded $\NI_j$ implies conditions (iii) and (iv) if $\| \bX_{ji} \|$ is bounded. Using new notations, we restate Theorem \ref{thm:302} in the main paper as Theorem \ref{Supp-thm:302} below; the proof is in Section \ref{sec:proof-thm1}. 
\begin{theorem}							\label{Supp-thm:302}
	Suppose Assumptions \hyperlink{(A1)}{(A1)} and \hyperlink{(A2)}{(A2)} in the main paper hold. Then, under Assumption \ref{assp:3-1}, as $\NC \to \infty$, the limiting distributions of $\widehat{\tau}_\ITT$ and $\widehat{\B}$ are $\sqrt{\NC} \big( \widehat{\tau}_\ITT - \tau	_\ITT \big)
		\stackrel{D}{\rightarrow}
		N \big( 0 , \sigma_\ITT^2 \big)$ and $\sqrt{\NC} \big( \widehat{\B} - \B^* \big)
		\stackrel{D}{\rightarrow}
		N \big( 0 , \Sigma_{\B} \big)$ where the variance terms are given by
\begin{align*}
	\sigma_\ITT^2
	 &=
 \lim_{\NC \rightarrow \infty} \NC \big[
	 \mathcal{V} \big\{ \potY{\bm{ \mathcal{R} }}{1} \big\} / \tNC
	+
	\mathcal{V} \big\{ \potY{\bm{\mathcal{R}}}{0} \big\} / (\NC - \tNC)
	-  \mathcal{V} \big\{ \potY{\bm{\mathcal{R}}}{1} - \potY{\bm{\mathcal{R}}}{0} \big\} / \NC
		 \big] \ ,
\\
	 \Sigma_{\B}
	 &=
	\lim_{\NC \rightarrow \infty} \NC 
	\big\{ \mo \big( \bm{\X} \big)  \big\}^{-1} 
	\big[
	\mathcal{V} \big\{ \potY{\bm{ \mathcal{R}}_X}{1} \big\} / \tNC
	+
	\mathcal{V} \big\{ \potY{\bm{\mathcal{R}}_X}{0} \big\} / (\NC - \tNC)
	- \mathcal{V} \big\{ \potY{\bm{\mathcal{R}}_X}{1} - \potY{\bm{\mathcal{R}}_X}{0} \big\} / \NC \big] 
	\big\{ \mo \big( \bm{\X} \big)  \big\}^{-1}  \ . 
\end{align*}
\end{theorem}
\noindent 

Let $\mathcal{R}_{j} = \potY{\Y_j}{\iv_j} - \mathcal{N}_j \widehat{\mo}_{\iv_j} (\bm{\Y})$ and $\mathcal{R}_{X,j} = \potY{ \Y_{X,j}}{\iv_j} - \X_j \big\{ \widehat{\mo}_{\iv_j} (\bm{\X})  \big\}^{-1}  \widehat{\mo}_{\iv_j} ( \bm{\Y}_X  ) $ be the observed values of $\potY{ \mathcal{R}_{j}}{\ivp}$ and $\potY{ \mathcal{R}_{X,j} }{\ivp}$, respectively. Let $\bm{\mathcal{R}} = \big\{ \mathcal{R}_{1},\ldots,\mathcal{R}_{\NC} \big\}$ and $\bm{ \mathcal{R}}_X = \big\{ \mathcal{R}_{X,1} ,\ldots,\mathcal{R}_{X,\NC} \big\}$. To estimate the asymptotic variance, we use the following variance estimators. 
{\fontsize{10}{12} \selectfont
\begin{align*}
&
\widehat{\sigma}_\ITT^2 
=  \NC \Bigg\{ 
\frac{\widehat{\mathcal{V}}_1 (\bm{\mathcal{R}})}{\tNC}+ 
\frac{\widehat{\mathcal{V}}_0 (\bm{\mathcal{R}})}{\NC - \tNC}
\Bigg\} \ ,
\
\widehat{\Sigma}_{\B}
=
\NC
\Bigg[
  \frac{\big\{\widehat{\mo}_1 (\bm{\X})  \big\}^{-1}\widehat{\mathcal{V}}_1(  \bm{ \mathcal{R}}_X ) \big\{ \widehat{\mo}_1 (\bm{\X})  \big\}^{-1}}{ \tNC} + 
 \frac{ \big\{ \widehat{\mo}_0 (\bm{\X})  \big\}^{-1} \widehat{\mathcal{V}}_0(  \bm{ \mathcal{R}}_X )  \big\{ \widehat{\mo}_0 (\bm{\X})  \big\}^{-1} }{\NC- \tNC}
 \Bigg] \ .
\end{align*}}
See \citet{Fogarty2018-2} for similar estimators. Under Assumption \ref{assp:3-1} and Assumptions \hyperlink{(A1)}{(A1)} and \hyperlink{(A2)}{(A2)} in the main paper, $\widehat{\sigma}_\ITT^2$ and $\widehat{\Sigma}_{\B}$ have the following asymptotic representation.
\begin{align*}
	&
	\widehat{\sigma}_\ITT^2 
	=
	 \NC \bigg[
	 \frac{\mathcal{V} \big\{ \potY{\bm{ \mathcal{R} }}{1} \big\}}{\tNC}
	+
	\frac{\mathcal{V} \big\{ \potY{\bm{\mathcal{R}}}{0} \big\}}{\NC -\tNC}
	\bigg]  + o_P(1)
		 \ , \
		 \widehat{\Sigma}_{\B}
	= \NC 
	\big\{ \mo \big( \bm{\X} \big)  \big\}^{-1} 
	\bigg[	
	\frac{\mathcal{V} \big\{ \potY{\bm{ \mathcal{R}}_X}{1} \big\}}{\tNC}
	+
	\frac{\mathcal{V} \big\{ \potY{\bm{\mathcal{R}}_X}{0} \big\} }{\NC - \tNC} \bigg]
	\big\{ \mo \big( \bm{\X} \big)  \big\}^{-1} 
	+ o_P(1) \ .
\end{align*}
Therefore, $\widehat{\sigma}_\ITT^2  $ and $ \widehat{\Sigma}_{\B}$ are conservative in the sense that $\lim_{\NC \rightarrow \infty} ( \widehat{\sigma}_\ITT^2 -  \sigma_\ITT^2 )$ and $\lim_{\NC \rightarrow \infty} ( \widehat{\Sigma}_{\B} - \Sigma_{\B} ) $ are non-negative and positive semi-definite, respectively.

\subsection{Details of Section \ref{sec:classifier} in the Main Paper}								\label{sec:detail42}

\subsubsection{Choice of the Random Noise Parameter $r$}							 \label{sec:choice-r}

In this subsection, we introduce a practical guide for choosing the random noise parameter $r$. We only use the NT classifier, but the same approach can be applied to AT/CO classifiers.

Let  $\SolS$ be the collection of $q$ satisfying the second part of equation \eqref{eq:margin_1} in the main paper:
\begin{align}									\label{eq-401}
	\SolS = \Big\{ q \, \Big| \, 
	\NIT_\NT 
	= 	\sumji \ind \big\{ f_\NT (\bX_{ji} \con \theta_\NT^*) \geq q \big\} \Big\} \ .
\end{align}
As discussed in the main paper,  $q_\NT^*$ is chosen so that the number of NTs in the population is the same as the number of individuals classified as NT; i.e. the classifier is strength-calibrated \citep{Kennedy2020}.

$\SolS$ should satisfy two desirable conditions to theoretically guarantee consistent estimation of the sharp bound of $\tau_\NT$. The first condition is non-empty $\SolS$, which implies that the quantile-threshold classifier based on the ``optimal learner'' $f_\NT(\cdot \con \theta_\NT^*)$ (with respect to the loss function $L$) can be strength-calibrated. That is, the optimal leaner can be used to classify individuals into NT and non-NT while the numbers of true NTs and the number of individuals classified as NT are the same. If $\SolS$ is non-empty, it has a form of $( f_{(N_\text{nNT})}, f_{(N_\text{nNT}+1)} ]$ where $f_{(k)}$ is $k$-th order statistic of $f_\NT(\bX_{ji}\con\theta_\NT^*)$ among $\NIT$ optimal learners  and $\NIT_\text{nNT} =\NIT -\NIT_\NT$ is the number of non-NTs. The second condition is that the length of $\SolS$ is $\Theta(\NC^{-1})$ provided that $\SolS$ is not empty. This implies that all possible calibration values converge to a single value. As a result, all quantile-threshold classifiers based on the optimal learner satisfying \eqref{eq-401} are asymptotically equivalent. The convergence rate $\Theta(\NC^{-1})$ is the same as two adjacent order statistics of a continuous distribution whose density is upper-bounded. 

Motivated by the above conditions, we define \textit{the optimal solution set} $\SolS$ as follows.
\begin{definition}							\label{def:401}
	A solution set $\SolS \subset \R$ is optimal if $\SolS$ is not empty with a form $\SolS=(q_1, q_2]$ $(q_1,q_2 \in \R)$ and the length of $\SolS$ is $\Theta(\NC^{-1})$. 
\end{definition}

As discussed in the main paper, the solution set $\SolS$ may be non-optimal. Among possible violations, we consider two cases which mostly happens in practice. 
\begin{enumerate}
	\item[(a)] (\textit{Violation 1: Empty $\SolS$}): The set $\SolS$ may empty if the unique values of the optimal proxies are much fewer than $\NC$. This happens it all of the covariates are discrete variables. For example, suppose that the observed dataset has a binary covariate $\bX_{ji} \in \{0,1\}$ and that the linear learner $f_\NT(\bX_{ji}\con \theta_\NT) = \bX_{ji}\T \theta_\NT$ is used. Then, the optimal learner $f_\NT (\bX_{ji}\con\theta_\NT^*)$ can only take two values corresponding to $\bX_{ji}=0$ or $1$, say $f_\NT(1\con \theta_\NT^*)=0.7$ and $f_\NT(0\con \theta_\NT^*)=0.3$. Thus, the possible number of NTs based on $	\sumji \ind \big\{ f_\NT (\bX_{ji}\con \theta_\NT^*) \geq q_\NT \big\} $ are only three: $N$ if $q\leq 0.3$, $\sumji \ind ( \bX_{ji} =1 )$ if $0.3 < q \leq 0.7$, $0$ otherwise. 
Unless the number of individuals whose covariate is 1 equals to $\NIT_\NT$, $\SolS$ is empty. 

	\item[(b)] (\textit{Violation 2: Asymptotically non-shrinking $\SolS$}): Asymptotically non-shrinking $\SolS$ may occur if NT membership is systematically determined by discrete covariates. Continuing the illustration in (a), suppose that a binary covariate $\bX_{ji}$ is the immunocompromise status of individual $ji$ which perfectly indicates NT membership of $ji$; i.e, $\ind \big( \bX_{ji} = 1 ) = \NT_{ji}$. The corresponding $\SolS$ is non-empty with a form $(0.3, 0.7]$. If the proportion of NTs and non-NTs do not vanish as $\NC$ goes to infinity, the length of $\SolS$ also does not shrink.

\end{enumerate}

To obtain an optimal the solution set under the above cases, we use the randomized learner $\randomf_\NT (\bX_{ji} \con \theta_\NT^*) = f_\NT (\bX_{ji} \con \theta_\NT^*) + e_{\NT ji}$ where $e_{\NT ji} \sim {\rm Unif}(-r,r)$. The corresponding solution set $\randomSolS$ involving the noise-added optimal proxies is defined as 
	\begin{align}									\label{eq-403-supp}
	\randomSolS 
	:= \Big\{ q \, \Big| \, 
	\NIT_\NT 
	= 	\sumji \ind \big\{ \randomf_\NT (\bX_{ji} \con \theta_\NT^*) \geq q \big\} \Big\} \ .
\end{align}
In the following paragraphs, we show that $\randomSolS$ is an optimal set under carefully chosen $r$ when either one of the violations occurs. Briefly speaking, due to the distribution of $e_{\NT ji}$, the noise-added optimal proxies are distinct and, as a result, $\randomSolS$ is non-empty. Furthermore, the length of $\randomSolS$ is $\Theta(\NC^{-1})$ when the empirical probability density function of $\bX_{ji}$s is uniformly upper-bounded for all $\NC$. This is because the distance between two adjacent order statistics of a uniform distribution is proportional to the reciprocal of the sample size almost surely. 

\begin{itemize}
	\item[(a)] (Remedy for Empty $\SolS$): When $\SolS$ is empty, there exists a value $q_M$ so that $q_M = f_{(N_\text{nNT})}= f_{(N_\text{nNT}+1)}$. Let $F_1 = \big\{ f_\NT (\bX_{ji} \con \theta_\NT^*) \cond f_\NT (\bX_{ji} \con \theta_\NT^*) > q_M \big\} $ (optimal learners that are larger than the upper candidate threshold), $
		F_2 = \big\{ f_\NT (\bX_{ji} \con \theta_\NT^*) \cond f_\NT (\bX_{ji} \con \theta_\NT^*) < q_M \big\}$ (optimal learners that are smaller than the lower candidate threshold), and $F_3
		= \big\{ f_\NT (\bX_{ji} \con \theta_\NT^*) \cond f_\NT (\bX_{ji} \con \theta_\NT^*) = q_M \big\}$ (optimal learners that have the same value as the candidate thresholds). Note that an individual whose proxy belongs to $F_1$ and $F_2$ is classified as NT and non-NT, respectively.
	
	Let $q_U = \min F_1$ and $q_L = \max F_2$. In words, $q_U$ is the smallest value of the proxies that are larger than $q_M$ and $q_L$ is the largest value of the proxies that are smaller than $q_M$. Let $r$ be  $r=  \min\{ q_U -q_M,q_M- q_L \} / 4$. Under this construction, $r$ is smaller than the distance between $F_1 \cup F_2$ and $F_3$. The corresponding randomized learner $\randomf_\NT(\cdot \con \theta_\NT^*)$ is constructed from $\randomf_\NT (\bX_{ji} \con \theta_\NT^*) = f_\NT (\bX_{ji} \con \theta_\NT^*) + e_{\NT ji}$ where $e_{\NT ji} \sim {\rm Unif}(-r,r)$. Because every randomized learners are different from the others, $\randomSolS$ in \eqref{eq-403-supp}  is not empty.  For any $q_\NT^* \in \randomSolS$, we  find that $\randomf_\NT (\bX_{ji} \con \theta_\NT^*) > q_\NT^*$ for any $f_\NT (\bX_{ji} \con \theta_\NT^*) \in F_1$ and $f_\NT (\bX_{ji} \con \theta_\NT^*) < q_\NT^*$ for any $f_\NT (\bX_{ji} \con \theta_\NT^*) \in F_2$. This implies that the random noises $e_{\NT ji}$ do not affect the classification for individuals whose optimal learners belong to $F_1 \cup F_2$. That is, the random noises only randomize the individuals whose optimal learner belong to $F_3$ and the modified solution set $\randomSolS$ is the interval between two adjacent order statistic of $\randomf_\NT (\bX_{ji} \con \theta_\NT^*)$ whose original proxy $f_\NT (\bX_{ji} \con \theta_\NT^*)$ belongs to $F_3$; i.e. $f_\NT (\bX_{ji} \con \theta_\NT^*) = q_M$. Thus, the randomized learners $\randomf_\NT (\bX_{ji} \con \theta_\NT^*)$ corresponding to $F_3$ follow a uniform distribution $\text{Unif}(q_M - r, q_M + r)$. This implies that the length of $\randomSolS$ has the same asymptotic order of the difference of two adjacent order statistics of a uniform distribution. Thus, $|\randomSolS| = \Theta(\NC^{-1})$ almost surely.

\item[(b)] (Remedy for Non-shrinking $\SolS$): When $\SolS$ does not shrink, it implies that $q_L = f_{(N_\text{nNT})}$ and $q_U = f_{(N_\text{nNT}+1)}$ are different. Let $r= (q_U - q_L)/4$ and we generate $f_\ell^{\text{aux}}$ ($\ell = 1,\ldots,\NC$) from a uniform distribution $\text{Unif}(q_L+r, q_U - r)$ for the data augmentation. Next we choose $\NT_\ell^{\text{aux}} \in \{0,1\}$ so that $ \sum_{\ell=1}^\NC \ind ( f_\ell^{\text{aux}} \geq q_L+r ) 
	<
	\sum_{\ell=1}^\NC \NT_\ell^{\text{aux}}
	<
	\sum_{\ell=1}^\NC \ind ( f_\ell^{\text{aux}} \geq q_U-r )$. We consider the following augmented equation with the randomized learner $\randomf_\NT (\bX_{ji} \con \theta_\NT^*)= f_\NT (\bX_{ji} \con \theta_\NT^*) + e_{\NT ji}$ where $e_{\NT ji} \sim \text{Unif}(-r,r)$.
\begin{align}									\label{eq-tech1}
	& \NIT_\NT + \sum_{\ell=1}^\NC \NT_\ell^{\text{aux}}
	=
	\sumji \ind \big\{ \randomf_\NT (\bX_{ji} \con \theta_\NT^*)\geq q \big\}+ \sum_{\ell=1}^\NC \ind ( f_\ell^{\text{aux}} \geq q )  \ .
\end{align}
Note that $\sumji \ind \big\{ \randomf_\NT (\bX_{ji} \con \theta_\NT^*) \geq q_L + r \big\} = \sumji \ind \big\{ \randomf_\NT (\bX_{ji} \con \theta_\NT^*) \geq q_U - r \big\} = \NIT_\NT$ from the construction of $r$. Thus, the left hand side of  \eqref{eq-tech1} is larger than the right hand side of  \eqref{eq-tech1} from  $q=q_L + r$. Similarly, the left hand side of \eqref{eq-tech1} is smaller than the right hand side of \eqref{eq-tech1} from $q=q_U-r$. Since $f_\ell^{\text{aux}}$s are distinctive and belong to the interval $[q_L+r, q_U-r]$, the solution to the above equation exists. The collection of the solution has a form of $\randomSolS = (f_{(k)}^\text{aux}, f_{(k+1)}^\text{aux}] \subset [q_L, q_U]$ for some $k$. Since $f_{(k+1)}^\text{aux} - f_{(k)}^\text{aux}$ is the difference of two adjacent order statistics of a uniform distribution, $|\randomSolS| = \Theta (\NC^{-1})$ almost surely.


\end{itemize}

Despite the above remedies, we need to know NT membership of each individual to obtain the valid parameter $r$, which is impossible. As a practical guide for choosing $r$, we can observe the behavior of the estimated solution set $ \widehat{\SolS} = \big\{ q \, \big| \, 
	 \sum_{ji} \iv_j \big( 1- \trt_{ji} \big)
	= 	\sum_{ji} \iv_j \ind \big\{ f_\NT (\bX_{ji} \con \widehat{\theta}_\NT) \geq q \big\}
	\big\}$, where $\widehat{\theta}_\NT$ is an estimator of $\theta_\NT^*$ obtained from equation \eqref{eq:margin_2} in the main paper. Based on $\widehat{\SolS}$, we check whether either violation 1 or 2 occurs. It is straightforward to check whether $\widehat{\SolS}$ is empty. However, it is impossible to check whether the length of $\widehat{\SolS}$ is asymptotically non-vanishing based only on the fixed number of $\NC$ clusters. As a practical guide, we consider that violation 2 happens if $\widehat{\SolS}$ is non-empty and all covariates are discrete. When either of violations occurs, we may follow the above remedies	to obtain $\widehat{r}$, the sample counterpart of $r$, based on $\widehat{\SolS}$. We generate the noise $e_{ji}$ from a uniform distribution ${\rm Unif}(-\widehat{r}, \widehat{r})$ and obtain the estimated noise-added proxies $\randomf_\NT (\bX_{ji} \con \widehat{\theta}_\NT) = f_\NT (\bX_{ji} \con \widehat{\theta}_\NT) + e_{\NT ji}$. 
	
	\subsubsection{Choice of the Surrogate Indicator Function Parameter $c$ and $h$}							 \label{sec:choice-ch}
	
We discuss the choice of $c$ and $h$. Replacing the indicator function in \eqref{eq-403-supp} with the surrogate indicator function, we obtain the following equation.
\begin{align}					\label{eq-404}
	\NIT_\NT 
	= 	\sumji \SI \big( \randomf_\NT (\bX_{ji} \con \theta_\NT^*) - q  \big) \ .
\end{align}
Let $q_\NT^*$ be the solution to equation \eqref{eq-404}, which is uniquely determined because $\SI$ is a continuous and strictly decreasing function in $q$.  The equation \eqref{eq-404} may or may not be a reasonable surrogate for equation \eqref{eq-403-supp} depending on $c$ and $h$. A good surrogate equation should render a solution $q_\NT^*$ that belongs to $\randomSolS$ so that the corresponding quantile-threshold classifier $\ind \big( \randomf_\NT (\bX_{ji} \con \theta_\NT^*) \geq q_\NT^* \big)$ is strength-calibrated; i.e. $	\NIT_\NT 
	= 	\sumji \ind \big\{ \randomf_\NT (\bX_{ji} \con \theta_\NT^*)  \geq q_\NT^* \big\}$. Thus, we choose $c$ and $h$ to guarantee $q_\NT^* \in \randomSolS$. Lemma \ref{lem:401} formally shows the result and its proof is in Section \ref{sec:C-3}.

\begin{lemma}									\label{lem:401}
	Suppose that $\randomSolS$ is optimal and that $c$ and $h$ belong to the following set
	\begin{align*}
		\mathcal{J}(\mathcal{D}, N_\NT, N_\nNT) = 
		\bigg\{ 
		(c,h)
	\, \bigg| \,
		0 < h \leq  \frac{\mathcal{D} }{2} 
		\ , \
		0 < c < \min \bigg[ 0.5 , \frac{\mathcal{D} - h }{2h \max \big\{ \log(\NIT_\NT) , \log(\NIT_\nNT) \big\} } \bigg]
		\bigg\} 
	\end{align*}
	where $\mathcal{D} = \randomf_{(\NIT_\nNT+1)} - \randomf_{(\NIT_\nNT)}$. Then, the unique solution to \eqref{eq-404} $q_\NT^*$ belongs to $\randomSolS$ and, as a result, $q_\NT^*$ makes the classifier strength-calibrated.
\end{lemma}

	Among choices satisfying the condition of Lemma \ref{lem:401}, we choose $c =[ \max \{ \log \NIT_\NT , \log \NIT_\nNT\} ]^{-1}$ and $h = \mathcal{D}/4$. We study the rates of $c$ and $h$ an asymptotic regime where $\NC$ goes to infinity with finite cluster size. When $\randomSolS$ is optimal (which is true under most of the cases in practice), the rate of $h$ is $\Theta(\NC^{-1})$. Furthermore, $c= \Theta(( \log \NC)^{-1})$ if the proportions of NT and non-NT do not vanish as $\NC$ increases; i.e. 	$\NIT_\NT = \Theta(\NC)$ and $\NIT_\nNT = \Theta(\NC)$.  
	
	In practice, we do not know $\NIT_\NT$ and $\NIT_\nNT$ and, as a result, we cannot choose the parameters $c$ and $h$ from Lemma \ref{lem:401}. Nonetheless, we present a practical guide for choosing $c$ and $h$ from available data by replacing the unobservable quantities $\{\mathcal{D}, \NIT_\NT,\NIT_\nNT\}$ in Lemma \ref{lem:401} with reasonable estimates. First, we use the ratio estimate $\widehat{\NIT}_\NT$. Accordingly, we define the estimator of non-NT as $\widehat{\NIT}_\nNT = \NIT - \widehat{\NIT}_\NT$. We use these estimates for the set of Lemma \ref{lem:401} instead of $\NIT_\NT$ and $\NIT_\nNT$. Next, we find an estimate of $\mathcal{D}$.  Let $\randomf_{1, (k)}$ be $k$-th order statistic of $\randomf_\NT (\bX_{ji} \con \widehat{\theta}_\NT)$ where cluster $j$ is assigned to treatment. The quantity $\mathcal{D}_1 := \randomf_{1, (\NIT_{1,\nNT}+1)} - \randomf_{1,(\NIT_{1,\nNT} )}$ is the difference of two two adjacent order statistics based only on $\NIT_1$ observations. Since $\mathcal{D}_1$ is based on fewer observation than $\mathcal{D}$, $\mathcal{D}_1$ tends to be larger than $\mathcal{D}$. Therefore, we again multiply the proportion $\NIT_1/\NIT$ to account for the difference of the number of observations. Therefore, we use $\widehat{\mathcal{D}} : = \NIT_1 \mathcal{D}_1/\NIT$ for the set of Lemma \ref{lem:401} instead of $\mathcal{D}$. As a result, we choose $(c,h)$ from $\mathcal{J}(\widehat{\mathcal{D}}, \widehat{N}_\NT, \widehat{N}_\nNT) $; i.e. we use estimates of $\{\mathcal{D}, \NIT_\NT,\NIT_\nNT\}$. 
	Under the above $c$ and $h$, we obtain $(\widehat{\theta}_\NT, \widehat{q}_\NT)$, the solution to the equations $\widehat{\theta}_\NT =\arg\min_{\theta} \sumji \iv_j L ( 1- \trt_{ji} , f_\NT(\bX_{ji} \con \theta) )$ and $\sumji \iv_j (1 - \trt_{ji}) = \sumji \iv_j \mathcal{I}_{c,h} (\randomf_\NT (\bX_{ji} \con \widehat{\theta}_\NT) \geq \widehat{q}_\NT)$. Note that $\widehat{q}_\NT$ satisfies $	\sumji \iv_j \big( 1- \trt_{ji} \big)
	= 	\sumji \iv_j \ind \big\{ \randomf_\NT (\bX_{ji} \con \widehat{\theta}_\NT)\geq \widehat{q}_\NT \big\}$; i.e. the quantile-threshold classifier based on $r$ from Section \ref{sec:choice-r} and $(c,h)$ chosen from Section  \ref{sec:choice-ch} is strength-calibrated for individuals in treated cluster.

	\subsection{Details of Section 4.4 in the Main Paper}											\label{sec:detail44}
	
	\subsubsection{Details of Elastic Programming}						\label{sec:EPdetail}
	
	We extensively restate the linear program about $\tau_t$ in \eqref{eq-LP-Population} in the main papaer below.
	{\fontsize{10}{8} \selectfont
	\begin{align}												\label{eq:LP-original}
		&
		\text{Min/Max } \tau_t = \big\{ ( \potY{\TP_t}{1} + \potY{\FN_t}{1} ) - (\potY{\TP_t}{0} + \potY{\FN_t}{0}) \big\}/\NIT_t \text{ over } \potY{\TP_t}{\ivp} , \potY{\FP_t}{\ivp} , \potY{\FN_t}{\ivp}  \text{ subject to }
	\\
	& \potY{\TP_\NT}{1} + \potY{\FP_\NT}{1} + \potY{\TP_\AT}{1} + \potY{\FP_\AT}{1} + \potY{\TP_\CO}{1} + \potY{\FP_\CO}{1}  = \potY{S}{1}
	\label{eq-LP-restrict1-1}
\\
	&
	\potY{\TP_\NT}{0} + \potY{\FP_\NT}{0} + \potY{\TP_\AT}{0} + \potY{\FP_\AT}{0} + \potY{\TP_\CO}{0} + \potY{\FP_\CO}{0} = \potY{S}{0}
	\label{eq-LP-restrict1-2}
		\\
			&
	\potY{\TP_\NT}{1}+ \potY{\FP_\NT}{1} = \potY{S_\NT}{1}
	, \
	\potY{\TP_\AT}{0}+ \potY{\FP_\AT}{0} = \potY{S_\AT}{0}
	\label{supp-eq-LP-restrict2}
	\\
	&
	\potY{\TP_\NT}{0} + \potY{\FN_\NT}{0} = \potY{S_{\BrandomC,\NT}}{0}
	, \
	\potY{\TP_\AT}{0} + \potY{\FN_\AT}{0}  =\potY{S_{\BrandomC,\AT}}{0}
	, \
	\potY{\TP_\CO}{0} + \potY{\FN_\CO}{0} = \potY{S_{\BrandomC,\CO}}{0}
	\label{eq-LP-restrict3-1}
	\\
	&
	\potY{\TP_\NT}{1} + \potY{\FN_\NT}{1} = \potY{S_{\BrandomC,\NT}}{1}
	, \
	\potY{\TP_\AT}{1} + \potY{\FN_\AT}{1} =\potY{S_{\BrandomC,\AT}}{1}
	, \
	\potY{\TP_\CO}{1} + \potY{\FN_\CO}{1} = \potY{S_{\BrandomC,\CO}}{1}
	\label{eq-LP-restrict3-2}
	\\
	&
	\potY{\TP_\NT}{0}  \leq \potY{\TP_\NT}{1}, \ 
	\potY{\FP_\NT}{0} \leq  \potY{\FP_\NT}{1} , \  
	\potY{\FN_\NT}{0} \leq \potY{\FN_\NT}{1}
	\ , \
	\potY{\TP_\AT}{0} \leq \potY{\TP_\AT}{1}, \ 
	\potY{\FP_\AT}{0} \leq  \potY{\FP_\AT}{1} , \  
	\potY{\FN_\AT}{0} \leq \potY{\FN_\AT}{1}
	\label{eq-LP-restrict4-2}
	\\
	&
	\potY{\TP_\CO}{0} \leq \potY{\TP_\CO}{1}, \ 
	\potY{\FP_\CO}{0} \leq  \potY{\FP_\CO}{1} , \  
	\potY{\FN_\CO}{0} \leq \potY{\FN_\CO}{1}
	\label{eq-LP-restrict4-3}
	\\
	&
	\potY{\TP_\NT}{1} \leq \NIT_\NT - R_\NT
	, \
	\potY{\TP_\AT}{1} \leq \NIT_\AT - R_\AT
	, \
	\potY{\TP_\CO}{1} \leq \NIT_\CO - R_\CO
	\label{supp-eq-LP-restrict5}
	\\
	&
	\potY{\FP_\NT}{1}  \leq R_\NT ,\  
	\potY{\FN_\NT}{1}  \leq R_\NT
	, \
	\potY{\FP_\AT}{1}  \leq R_\AT , \ 
	\potY{\FN_\AT}{1}  \leq R_\AT
	. \
	\potY{\FP_\CO}{1} \leq R_\CO , \ 
	\potY{\FN_\CO}{1} \leq R_\CO 
	\label{supp-eq-LP-restrict6}
	\\
	&
	0 \leq \potY{\TP_t}{\ivp} , \potY{\FP_t}{\ivp} , \potY{\FN_t}{\ivp} \ .
	\nonumber
	\end{align}}%
	The objective function of the linear program \eqref{eq:LP-original} is from the aforementioned decomposition. Constraints \eqref{eq-LP-restrict1-1} and \eqref{eq-LP-restrict1-2} are from the fact that individual $ji$ is either a never-taker, complier, or a ,complier. Constraints \eqref{supp-eq-LP-restrict2}-\eqref{eq-LP-restrict3-2} are based the decompositions of $\potY{S_\NT}{1}$, $\potY{S_\AT}{0}$, and $\potY{S_{\BrandomC,t}}{\ivp}$. Constraints \eqref{eq-LP-restrict4-2} and \eqref{eq-LP-restrict4-3} are from the monotonicity of the outcome in Assumption \hyperlink{(A6)}{(A6)}. Finally, constraints \eqref{supp-eq-LP-restrict5} and  \eqref{supp-eq-LP-restrict6} are from the boundedness of the outcome and the definition of the misclassification rate $R_t$.

	The linear program is always feasible if the population-level terms were known. However, we need to replace the population-level terms with the estimated terms from the observed data, and the estimated linear program may not be feasible because the particular realization of the observed sample may violate some restrictions. As a remedy, we consider the following elastic program. 
	
	{\fontsize{10}{8} \selectfont
	\begin{align}												\label{eq:LP-elastic}
		&
		\left.
		\begin{matrix}
			\text{Min } \tau_t + B \sum_{\ell=1}^{28} a_\ell + B \sum_{\ell=1}^{10} b_\ell
			\\
			\text{Max } \tau_t - B \sum_{\ell=1}^{28} a_\ell - B \sum_{\ell=1}^{10} b_\ell
		\end{matrix} 
		\right\}
		\text{ over } \potY{\TP_t}{\ivp} , \potY{\FP_t}{\ivp} , \potY{\FN_t}{\ivp}, a_\ell, b_\ell  \text{ subject to }
	\\
	& \potY{\TP_\NT}{1} + \potY{\FP_\NT}{1} + \potY{\TP_\AT}{1} + \potY{\FP_\AT}{1} + \potY{\TP_\CO}{1} + \potY{\FP_\CO}{1} + a_1 - b_1 = \potY{S}{1}
	\nonumber
	\\
	&
	\potY{\TP_\NT}{0} + \potY{\FP_\NT}{0} + \potY{\TP_\AT}{0} + \potY{\FP_\AT}{0} + \potY{\TP_\CO}{0} + \potY{\FP_\CO}{0} + a_2 - b_2 = \potY{S}{0}
	\nonumber		
		\\
			&
	\potY{\TP_\NT}{1}+ \potY{\FP_\NT}{1} + a_3 - b_3= \potY{S_\NT}{1}
	, \
	\potY{\TP_\AT}{0}+ \potY{\FP_\AT}{0} + a_4 - b_4 = \potY{S_\AT}{0}
	\nonumber		
	\\
	&
	\potY{\TP_\NT}{0} + \potY{\FN_\NT}{0} + a_5 - b_5 = \potY{S_{\BrandomC,\NT}}{0}
	, \
	\potY{\TP_\AT}{0} + \potY{\FN_\AT}{0} + a_6 - b_6 =\potY{S_{\BrandomC,\AT}}{0}
	, \
	\potY{\TP_\CO}{0} + \potY{\FN_\CO}{0} + a_7 - b_7 = \potY{S_{\BrandomC,\CO}}{0}
	\nonumber
	\\
	&
	\potY{\TP_\NT}{1} + \potY{\FN_\NT}{1} + a_8 - b_8 = \potY{S_{\BrandomC,\NT}}{1}
	, \
	\potY{\TP_\AT}{1} + \potY{\FN_\AT}{1} + a_9 - b_9 =\potY{S_{\BrandomC,\AT}}{1}
	, \
	\potY{\TP_\CO}{1} + \potY{\FN_\CO}{1} + a_{10} - b_{10} = \potY{S_{\BrandomC,\CO}}{1}
	\nonumber
	\\
	&
	\potY{\TP_\NT}{0} + a_{11} \leq \potY{\TP_\NT}{1}, \ \potY{\FP_\NT}{0} + a_{12} \leq  \potY{\FP_\NT}{1} , \  \potY{\FN_\NT}{0} + a_{13} \leq \potY{\FN_\NT}{1}
	\nonumber
	\\
	&
	\potY{\TP_\AT}{0} + a_{14} \leq \potY{\TP_\AT}{1}, \ \potY{\FP_\AT}{0} + a_{15} \leq  \potY{\FP_\AT}{1} , \  \potY{\FN_\AT}{0} + a_{16} \leq \potY{\FN_\AT}{1}
	\nonumber
	\\
	&
	\potY{\TP_\CO}{0} + a_{17} \leq \potY{\TP_\CO}{1}, \ \potY{\FP_\CO}{0} + a_{18} \leq  \potY{\FP_\CO}{1} , \  \potY{\FN_\CO}{0} + a_{19} \leq \potY{\FN_\CO}{1}
	\nonumber
	\\
	&
	\potY{\TP_\NT}{1} + a_{20} \leq \NIT_\NT - R_\NT
	, \
	\potY{\TP_\AT}{1} + a_{21} \leq \NIT_\AT - R_\AT
	, \
	\potY{\TP_\CO}{1} + a_{22} \leq \NIT_\CO - R_\CO
	\nonumber
	\\
	&
	\potY{\FP_\NT}{1} + a_{23}  \leq R_\NT ,\  \potY{\FN_\NT}{1} + a_{24} \leq R_\NT
	, \
	\potY{\FP_\AT}{1} + a_{25}  \leq R_\AT , \ \potY{\FN_\AT}{1} + a_{26} \leq R_\AT
	\nonumber
	\\
	&
	\potY{\FP_\CO}{1} + a_{27}  \leq R_\CO , \ \potY{\FN_\CO}{1} + a_{28} \leq R_\CO 
	\nonumber		
	\ , \
	0 \leq  \potY{\TP_t}{\ivp} , \potY{\FP_t}{\ivp} , \potY{\FN_t}{\ivp}, a_\ell, b_\ell \ .
	\nonumber
	\end{align}}%
	Here $B$ is a very large constant, say $B=10^6$.  The key idea is to use elastic variables to stretch the constraints that make the original linear program infeasible. Specifically, we add non-negative variables $a$ for inequality constraints and we additionally subtract non-negative variables $b$ for equality constraints. 
	
	When the population-level linear program \eqref{eq:LP-original} is feasible (which always hold with the true population-level quantities), the solution to the elastic program \eqref{eq:LP-elastic} is the same as the original linear program \eqref{eq:LP-original} by taking all $a_\ell$ and $b_\ell$ as zero. To be more specific, let $\vartheta_O^* \in \R^{18}$ be the values of $\{ \potY{\TP_\NT}{1}, \ldots, \potY{\FN_\CO}{0}\}$ that solves a minimization or a maximization of the original linear program \eqref{eq:LP-original} and $\vartheta_E^* \in \R^{56}$ be the values of $\{ \potY{\TP_\NT}{1}, \ldots, \potY{\FN_\CO}{0}, a_1, \ldots, b_{10} \}$ that solves the corresponding elastic program \eqref{eq:LP-elastic}. Then, if $\vartheta_O^*$ exists (i.e. \eqref{eq:LP-original} is feasible), $\vartheta_E^* = (\vartheta_O^*, 0,\ldots,0)$ where the last 38 components correspond to $(a_1,\ldots,b_{10})$.
	
	The elastic program \eqref{eq:LP-elastic} is feasible even though the linear program \eqref{eq:LP-original} is infeasible due to the usage of the estimated quantities $\widehat{\NIT}_t$, $\potY{\widehat{S}}{\ivp}$, $\potY{\widehat{S}_t}{\ivp}$, $\potY{\widehat{S}_{\BrandomC,t}}{\ivp}$, $\widehat{R}_t$. Let $\widehat{\vartheta}_E = \{ \potY{\widehat{T}_\NT}{1}, \ldots, \potY{\widehat{V}_\CO}{0}, \widehat{a}_1, \ldots, \widehat{b}_{10} \} \in \R^{56}$ be the values that solves the minimization of the elastic program \eqref{eq:LP-elastic} about $\tau_t$ using the estimated quantities. Then, the estimated lower bound for $\tau_t$ is given as $\widehat{{\rm LB}}_{\BrandomC,t} = \{ \potY{\widehat{T}_t}{1}  + \potY{\widehat{V}_t}{1} - \potY{\widehat{T}_t}{0}  - \potY{\widehat{V}_t}{0} \}/\widehat{\NIT}_t$. Similarly, the estimated upper bound for $\tau_t$ is given as $\widehat{{\rm UB}}_{\BrandomC,t} = \{ \potY{\widehat{T}_t}{1}  + \potY{\widehat{V}_t}{1} - \potY{\widehat{T}_t}{0}  - \potY{\widehat{V}_t}{0} \}/\widehat{\NIT}_t$ when $\widehat{\vartheta}_E = \{ \potY{\widehat{T}_\NT}{1}, \ldots, \potY{\widehat{V}_\CO}{0}, \widehat{a}_1, \ldots, \widehat{b}_{10} \} \in \R^{56}$ solves the maximization of the elastic program. As stated in Theorem \ref{thm:403} in the main paper, $[\widehat{{\rm LB}}_{\BrandomC,t}, \widehat{{\rm UB}}_{\BrandomC,t}]$ is consistent for the bound obtained from the population-level linear program \eqref{eq:LP-original}; see Section \ref{sec:Ourconsistency} for the proof.
	
	\subsubsection{Details of the Construction of the Extended Bounds}  \label{sec:LH}

	In this section, we consider methods proposed by \citet{GM2008} and \citet{LongHudgens2013} to obtain bounds for compliance group effects using binary covariates. Let $\bW_{ji}$ be a subset of binary covariates that are chosen by investigators. For each level of $\bW_{ji}$, we define $\NIT_t (\bw) = \sumji t_{ji} \ind(\bW_{ji} = \bw)$, the number of NT/AT/COs within stratum $\{ \bW_{ji} = \bw \}$, and $\potY{S_t}{\ivp}(\bw) = \sumji \potY{Y_{ji}}{\ivp} t_{ji} \ind(\bW_{ji} = \bw)$, the total potential outcome of NT/AT/COs under $\iv_{ji}=\ivp$ within stratum $\{ \bW_{ji} = \bw \}$. Using these quantities, we define the following quantities following \citet{LongHudgens2013}.
	\begin{align}								\label{eq-LH-parameters}
		&
		\potY{\pi}{0}(\bw) 
		= \frac{ 
			\potY{S_\NT}{0}(\bw) + \potY{S_\CO}{0}(\bw)
		}{
		\NIT_\NT(\bw) + \NIT_\CO(\bw)
		}
		\ , \
		\potY{\pi}{1}(\bw) 
		= \frac{ 
			\potY{S_\NT}{1}(\bw)
		}{
		\NIT_\NT(\bw)
		}
		\ , \
		\gamma(\bw) 
		= \frac{ 
			\NIT_\NT(\bw) 
		}{
		\NIT_\NT(\bw) + \NIT_\CO(\bw)
		}
		\ , \
		\nonumber
		\\
		&
		\potY{\lambda}{0}(\bw)
		=
		\frac{\potY{S_\AT}{0}(\bw)}{\NIT_\AT(\bw)}
		\ , \
		\potY{\lambda}{1}(\bw)
		=
		\frac{\potY{S_\AT}{1}(\bw)+\potY{S_\CO}{1}(\bw)}{\NIT_\AT(\bw)+\NIT_\CO(\bw)}
		\ , \
		\delta(\bw)
		=
		\frac{\NIT_\AT(\bw)}{\NIT_\AT(\bw)+\NIT_\CO(\bx)} \ .
	\end{align}
	Using these quantities, $\potY{S_\NT}{0}(\bw)/\NIT_\NT(\bw)$, $\potY{S_\AT}{1}(\bw)/\NIT_\AT(\bw)$, and $\potY{S_\CO}{\ivp}(\bw)/\NIT_\CO(\bw)$ are
	\begin{align*}
		&
		\frac{\potY{S_\NT}{0} (\bw)}{\NIT_\NT(\bw)}
		=
		\frac{\potY{\pi}{0}(\bw)}{\gamma(\bw)}
		-
		\frac{1-\gamma(\bw)}{\gamma(\bw)} \frac{\potY{S_\CO}{0}(\bw)}{\NIT_\CO(\bw)}
		\ , \
		\frac{		\potY{S_\AT}{1} (\bw)}{\NIT_\AT(\bw)}
		=
		\frac{\potY{\lambda}{1}(\bw)}{\delta(\bw)}
		-
		\frac{1-\delta(\bw)}{\delta(\bw)} \frac{\potY{S_\CO}{1}(\bw)}{\NIT_\CO(\bw)}
		\ , \\
		&
		\frac{\potY{S_\CO}{0} (\bw)}{\NIT_\CO(\bw)}
		=
		\frac{\potY{\pi}{0}(\bw)}{1-\gamma(\bw)}
		-
		\frac{\gamma(\bw)}{1-\gamma(\bw)} \frac{\potY{S_\NT}{0}(\bw)}{\NIT_\NT(\bw)}
		\ , \
		\frac{\potY{S_\CO}{1} (\bw)}{\NIT_\CO(\bw)}
		=
		\frac{\potY{\lambda}{1}(\bw)}{1-\delta(\bw)}
		-
		\frac{\delta(\bw)}{1-\delta(\bw)} \frac{\potY{S_\AT}{1}(\bw)}{\NIT_\AT(\bw)}
		\ .
	\end{align*}	
	If $\gamma(\bw)=0$, $\potY{S_\NT}{0}(\bw)=0$ because $\NIT_\NT(\bw)=0$. Similarly, $\potY{S_\AT}{1}=0$ if $\delta(\bw)=0$ and $\potY{S_\CO}{\ivp}=0$ if $\gamma(\bw)=\delta(\bw)=1$. Again, the quantities above are not estimable using the observed data, so $\tau_t(\bw) = \big\{ \potY{S_t}{1}(\bw) -  \potY{S_t}{0}(\bw) \big\}/\NIT_t(\bw)$ is also not estimable. But, we can obtain bounds for $\tau_t(\bw)$ using the fact that $\tau_t(\bw)$ and $\potY{S_t}{\ivp}(\bw) /\NIT_t(\bw)$ are bounded between 0 and 1 under Assumption \hyperlink{(A6)}{(A6)} in the main paper. For example, the bound for $\tau_\NT(\bw)$ is given as follows.
	{\fontsize{10}{12} \selectfont
	\begin{align}								\label{eq-LH1}
		\tau_\NT(\bw)
		& =
		\frac{\potY{S_\NT}{1} (\bw) }{\NIT_\NT(\bw)} 
		-
		\frac{\potY{\pi}{0}(\bw)}{\gamma(\bw)}
		+
		\frac{1-\gamma(\bw)}{\gamma(\bw)} \frac{\potY{S_\CO}{0}(\bw)}{\NIT_\CO(\bw)}
		\in 
		\begin{bmatrix}
			\max \big\{ 0 , \frac{\potY{S_\NT}{1} (\bw) }{\NIT_\NT(\bw)}  - \frac{\potY{\pi}{0}(\bw)}{\gamma(\bw)} \big\}
			\\
				\min \big\{  \frac{\potY{S_\NT}{1} (\bw) }{\NIT_\NT(\bw)}, \frac{\potY{S_\NT}{1} (\bw) }{\NIT_\NT(\bw)}  + \frac{1-\gamma(\bw) -\potY{\pi}{0}(\bw)}{\gamma(\bw)} \big\}
		\end{bmatrix} \ .
	\end{align}}%
	Since $\tau_\NT = \sum_{\bw} \NIT_\NT(\bw) \tau_\NT(\bw) / \NIT_\NT$, we can obtain a bound for $\tau_\NT$, $\big[ {\rm LB}_{W,\NT}, {\rm UB}_{W,\NT} \big]$, by taking a weighted sum of the bounds in \eqref{eq-LH1}; i.e. 
	\begin{align}													\label{eq-LH2}
		&
		\begin{bmatrix}
		{\rm LB}_{W,\NT} \\ {\rm UB}_{W,\NT}
		\end{bmatrix}
		=
		\sum_{\bw} \frac{\NIT_\NT(\bw)}{\NIT_\NT}
		\begin{bmatrix}
			\max \big\{ 0 , \frac{\potY{S_\NT}{1} (\bw) }{\NIT_\NT(\bw)}  - \frac{\potY{\pi}{0}(\bw)}{\gamma(\bw)} \big\}
			\\
				\min \big\{  \frac{\potY{S_\NT}{1} (\bw) }{\NIT_\NT(\bw)}, \frac{\potY{S_\NT}{1} (\bw) }{\NIT_\NT(\bw)}  + \frac{1-\gamma(\bw) -\potY{\pi}{0}(\bw)}{\gamma(\bw)} \big\}
		\end{bmatrix} \ .
	\end{align}
	The bound for $\tau_\AT$ and $\tau_\CO$, $[ {\rm LB}_{W,\AT}, {\rm UB}_{W,\AT} ]$ and $[ {\rm LB}_{W,\CO}, {\rm UB}_{W,\CO} ]$, respectively, are
	\begin{align*}
		\begin{bmatrix}
			{\rm LB}_{W,\AT} \\
			{\rm UB}_{W,\AT}
		\end{bmatrix}
		& = 
		\sum_{\bw} \frac{\NIT_\AT(\bw)}{\NIT_\AT}
		\begin{bmatrix}
			\max \Big\{ 0 ,  \frac{\potY{\lambda}{1}(\bw) - 1 + \delta(\bw)}{\delta(\bw)} - \frac{\potY{S_\AT}{0} (\bw) }{\NIT_\AT(\bw)} \Big\}
			\\
			\min \Big\{ 1- \frac{\potY{S_\AT}{0} (\bw)}{\NIT_\AT(\bw)} ,  \frac{\potY{\lambda}{1}(\bw)}{\delta(\bw)}- \frac{\potY{S_\AT}{0} (\bw)}{\NIT_\AT(\bw)}  \Big\}
		\end{bmatrix}\ ,
		\\
		\begin{bmatrix}
			{\rm LB}_{W,\CO} \\
			{\rm UB}_{W,\CO}
		\end{bmatrix}
		&=
		\sum_{\bw} \frac{\NIT_\CO(\bw)}{\NIT_\CO}
		\begin{bmatrix}
			\max \Big\{ 0 ,  
		\frac{\potY{\lambda}{1}(\bw) - \delta(\bw)}{1-\delta(\bw)}
		-
		\frac{\potY{\pi}{0}(\bw)}{1-\gamma(\bw)}
		\Big\}
		\\
		\min \Big\{ 1,  
		\frac{\potY{\lambda}{1}(\bw)}{1-\delta(\bw)}
		+
		\frac{\gamma(\bw) - \potY{\pi}{0}(\bw)}{1-\gamma(\bw)}
		\Big\}
		\end{bmatrix}
		\ .
	\end{align*}
	
	Again, the bounds $[ {\rm LB}_{W,t}, {\rm UB}_{W,t} ]$ are not available with real data because it requires the population-level terms. Instead, we construct estimates for the bounds by plugging in estimates for the parameters that comprise the bounds. First, we construct ratio-type estimates for $\NIT_t(\bw)$ as $
		{\widehat{\NIT}_\NT (\bw)}/{\NIT(\bw) }
		 = {\sumji \iv_j \NT_{ji} \ind(\bW_{ji} = \bw) }/\{\sumji \iv_j \ind(\bW_{ji} = \bw)\}$, $
		\widehat{\NIT}_\AT (\bw)/{\NIT(\bw) } = \sumji (1-\iv_j) \AT_{ji} \ind(\bW_{ji} = \bw) /\{\sumji (1-\iv_j) \ind(\bW_{ji} = \bw)\}$, and ${\widehat{\NIT}_\CO (\bw)}/{\NIT(\bw)}
		 =  1 - \{ \widehat{\NIT}_\NT (\bw) + \widehat{\NIT}_\AT (\bw)\}/{\NIT(\bw)}$.
	Here $\NIT(\bw) = \sumji \ind(\bW_{ji} = \bw)$ is the number of total units in stratum $\{ \bW_{ji} = \bw \}$. The total estimated numbers of NT/AT/COs are defined by $\widehat{\NIT}_t = \sum_{\bw} \widehat{\NIT}_t(\bw)$.	Similarly, we define the ratio-type estimates for $\potY{\widehat{S}}{\ivp}(\bw)$, $\potY{\widehat{S}_\NT}{1}(\bw)$, and $\potY{\widehat{S}_\AT}{0}(\bw)$ as ${\potY{\widehat{S}}{\ivp}(\bw) }/{\NIT(\bw)} = {\sumji \ind(\iv_j = \ivp) Y_{ji}  \ind(\bW_{ji} = \bw) }/\{\sumji \ind(\iv_j = \ivp) \ind(\bW_{ji} = \bw)\}$ and $\potY{\widehat{S}_t}{\ivp}(\bw)/\NIT(\bw) = \sumji \ind( \iv_j = \ivp) Y_{ji} t_{ji}  \ind(\bW_{ji} = \bw) /\{\sumji \ind( \iv_j = \ivp) \ind(\bW_{ji} = \bw)\}$.
	Using these estimates, we obtain plug-in estimates $\widehat{\NIT_t}(\bw)/\widehat{\NIT}_t$, $\potY{\pi}{0}(\bw)$, $\gamma(\bw)$, $\potY{\lambda}{1}(\bw)$, and $\delta(\bw)$ from \eqref{eq-LH-parameters}:
	\begin{align}							\label{eq-LH-Est}
	&
			\frac{ \widehat{\NIT}_t(\bw) }{ \widehat{\NIT}_t }
		=
		\frac{ \widehat{\NIT}_t(\bw) }{\sum_{\bw} \widehat{\NIT}_t(\bw)}
		\ , 
		\
		\potY{\widehat{\pi}}{0}(\bw) 
	=
	\frac{\potY{\widehat{S}}{0}(\bw) - \potY{\widehat{S}_\AT}{0}(\bw)}{\widehat{\NIT}_\NT(\bw) + \widehat{\NIT}_\CO(\bw)} \ , \
	\\
	& 
	\widehat{\gamma}(\bw) 
	= \frac{\widehat{\NIT}_\NT(\bw)}{\widehat{\NIT}_\NT(\bw) + \widehat{\NIT}_\CO(\bw)}
 \ ,
	\
	 \potY{\widehat{\lambda}}{1} (\bw) 
	=
	\frac{ \widehat{\NIT}_\AT(\bw) + \widehat{\NIT}_\CO(\bw)}{\potY{\widehat{S}}{1}(\bw) - \potY{\widehat{S}_\NT}{1}(\bw)}
	\ ,	\
	\widehat{\delta}(\bw) 
	=
	 \frac{\widehat{\NIT}_\AT(\bw)}{\widehat{\NIT}_\AT(\bw) + \widehat{\NIT}_\CO(\bw)}\ .
	\nonumber
	\end{align}
	Plugging in the above estimators, we define $\widehat{{\rm LB}}_{W,t}$ and $\widehat{{\rm UB}}_{W,t}$, the estimators for ${\rm LB}_{W,t}$ and ${\rm UB}_{W,t}$, respectively. As in Theorem \ref{thm:consistencyLH} below, $[ \widehat{{\rm LB}}_{W,t} , \widehat{{\rm UB}}_{W,t}]$ is consistent for $[ {\rm LB}_{W,t}, {\rm UB}_{W,t} ]$ under mild conditions. 	The proof is in Section \ref{sec:LHproof}. 
	\begin{theorem}								\label{thm:consistencyLH}
		Suppose Assumption \hyperlink{(A1)}{(A1)}-\hyperlink{(A6)}{(A6)} in the main paper holds. Furthermore, suppose that (i) $\NIT(\bw)/\NIT$ and $\tNC/\NC$ converge to constants in $(0,1)$ as $\NC \rightarrow \infty$; (ii) $\potY{S_\NT}{1}(\bw)/\NIT(\bw)$ and $\potY{S_\AT}{0}(\bw)/\NIT(\bw)$ converges to a constant in $[0,1]$ as $\NC \rightarrow \infty$; and (iii) $\NI_j$ is bounded for any $j=1,\ldots,\NC$ as $\NC \rightarrow \infty$. Then, 	the estimated bounds are consistent, i.e. for any $ \epsilon > 0$, 
		$\lim_{\NC \rightarrow \infty} P \big\{ \big| \widehat{{\rm UB}}_{W,t} - {\rm UB}_{W,t} \big| > \epsilon \, \big| \, \F_\NC, \mathcal{\iv}_\NC \big\} = 0$ and 
		$\lim_{\NC \rightarrow \infty} P \big\{ \big| \widehat{{\rm LB}}_{W,t} - {\rm LB}_{W,t} \big| > \epsilon \, \big| \,  \F_\NC, \mathcal{\iv}_\NC \big\} = 0$.
	\end{theorem}
	
	The bounds $[ {\rm LB}_{W,t}, {\rm UB}_{W,t}]$ depend on the collection of discrete covariates $\bW_{ji}$. In general, a richer $\bW_{ji}$ results narrower bounds for $\tau_t$, but too rich $\bW_{ji}$ brings problems in inference for the bounds. For instance, some denominators in the estimates may be zero if some levels of $\bW_{ji}$ contains few observations. Therefore, we recommend to choose $\bW_{ji}$ so that all strata defined by $\bW_{ji}$ contain both control and treated clusters, i.e. $\sumji \ind(\iv_j=\ivp) \ind(\bW_{ji} = \bw)>0$ for all $\ivp=0,1$ and $\bw$. 

Both $[{\rm LB}_{\BrandomC,t}, {\rm UB}_{\BrandomC,t}]$ and $[{\rm LB}_{W,t}, {\rm UB}_{W,t}]$ contain $\tau_t$ and they may differ according to the realization of the compliance status across covariates. Therefore, we can obtain a narrower bound for $\tau_t$ by taking the intersection of two bounds. Specifically, let ${\rm LB}_t = \max\{ {\rm LB}_{\BrandomC,t}, {\rm LB}_{W,t} \}$ and ${\rm UB}_t = \min\{ {\rm UB}_{\BrandomC,t}, {\rm UB}_{W,t} \}$. Accordingly, we analogously define an estimate for $[{\rm LB}_t, {\rm UB}_t]$, i.e. $\widehat{{\rm LB}}_t = \max \{ \widehat{{\rm LB}}_{\BrandomC,t}, \widehat{{\rm LB}}_{W,t} \}$ and $\widehat{{\rm UB}}_t = \min \{ \widehat{{\rm UB}}_{\BrandomC,t}, \widehat{{\rm UB}}_{W,t} \}$. Lastly, all bounds and corresponding estimators are affine transformation invariant as shown in the next subsection.

	\subsubsection{Details of the Affine Transformation Invariance of the Bounds}  \label{sec:AFinvarianceBound}
	
	Suppose we transform the outcome as $\potY{Y_{ji}'}{\ivp} = c + d \potY{Y_{ji}}{\ivp}$ where $d>0$. Accordingly, we find $\potY{S'}{\ivp} = c\NIT + d \potY{S}{\ivp}$, $\potY{S_t'}{\ivp} = c\NIT_t + d \potY{S_t}{\ivp}$, $\potY{S_{\BrandomC,t}'}{\ivp} =  c\NIT_t + d \potY{S_{\BrandomC,t}}{\ivp}$, and $R_t' = R_t$ (because $R_t$ does not depend on $Y_{ji}$). We obtain $\potY{\TP_t'}{\ivp}
		=	c(\NIT_t-R_t) + d \potY{\TP_t}{\ivp}$, $\potY{\FP_t'}{\ivp}
		=
		cR_t + d \potY{\FP_t}{\ivp}$, and $\potY{\FN_t'}{\ivp}=
		cR_t + d \potY{\FN_t}{\ivp}$, respectively.	
	Therefore, we obtain $\tau_t' = \big\{ \potY{\TP_t'}{1} + \potY{\FN_t'}{1} - \potY{\TP_t'}{0} - \potY{\FN_t'}{0} \big\}/\NIT_\NT = d \tau_t$.\\

	We consider the following elastic program with respect to $\potY{Y_{ji}'}{\ivp}$.
	{\fontsize{10}{8} \selectfont
		\begin{align}												\label{eq:LP-elastic-affine}
		&
		\left.
		\begin{matrix}
			\text{Min } \tau_t' + B \sum_{\ell=1}^{28} a_\ell' + B \sum_{\ell=1}^{10} b_\ell'
			\\
			\text{Max } \tau_t' - B \sum_{\ell=1}^{28} a_\ell' - B \sum_{\ell=1}^{10} b_\ell'
		\end{matrix} 
		\right\}
		\text{ over } \potY{\TP_t'}{\ivp} , \potY{\FP_t'}{\ivp} , \potY{\FN_t'}{\ivp}, a_\ell', b_\ell' \text{ subject to }
	\\
	& \potY{\TP_\NT'}{1} + \potY{\FP_\NT'}{1} + \potY{\TP_\AT'}{1} + \potY{\FP_\AT'}{1} + \potY{\TP_\CO'}{1} + \potY{\FP_\CO'}{1} + a_1' - b_1' = \potY{S'}{1}
	\nonumber
	\\
	&
	\potY{\TP_\NT'}{0} + \potY{\FP_\NT'}{0} + \potY{\TP_\AT'}{0} + \potY{\FP_\AT'}{0} + \potY{\TP_\CO'}{0} + \potY{\FP_\CO'}{0} + a_2' - b_2' = \potY{S'}{0}
	\nonumber		
		\\
			&
	\potY{\TP_\NT'}{1}+ \potY{\FP_\NT'}{1} + a_3' - b_3'= \potY{S_\NT'}{1}
	, \
	\potY{\TP_\AT'}{0}+ \potY{\FP_\AT'}{0} + a_4' - b_4' = \potY{S_\AT'}{0}
	\nonumber		
	\\
	&
	\potY{\TP_\NT'}{0} + \potY{\FN_\NT'}{0} + a_5' - b_5' = \potY{S_{\BrandomC,\NT}'}{0}
	, \
	\potY{\TP_\AT'}{0} + \potY{\FN_\AT'}{0} + a_6' - b_6' =\potY{S_{\BrandomC,\AT}'}{0}
	, \
	\potY{\TP_\CO'}{0} + \potY{\FN_\CO'}{0} + a_7' - b_7' = \potY{S_{\BrandomC,\CO}'}{0}
	\nonumber
	\\
	&
	\potY{\TP_\NT'}{1} + \potY{\FN_\NT'}{1} + a_8' - b_8' = \potY{S_{\BrandomC,\NT}'}{1}
	, \
	\potY{\TP_\AT'}{1} + \potY{\FN_\AT'}{1} + a_9' - b_9' =\potY{S_{\BrandomC,\AT}'}{1}
	, \
	\potY{\TP_\CO'}{1} + \potY{\FN_\CO'}{1} + a_{10}' - b_{10}' = \potY{S_{\BrandomC,\CO}'}{1}
	\nonumber
	\\
	&
	\potY{\TP_\NT'}{0} + a_{11} \leq \potY{\TP_\NT'}{1}, \ \potY{\FP_\NT'}{0} + a_{12}' \leq  \potY{\FP_\NT'}{1} , \  \potY{\FN_\NT'}{0} + a_{13}' \leq \potY{\FN_\NT'}{1}
	\nonumber
	\\
	&
	\potY{\TP_\AT'}{0} + a_{14}' \leq \potY{\TP_\AT'}{1}, \ \potY{\FP_\AT'}{0} + a_{15}' \leq  \potY{\FP_\AT'}{1} , \  \potY{\FN_\AT'}{0} + a_{16}' \leq \potY{\FN_\AT'}{1}
	\nonumber
	\\
	&
	\potY{\TP_\CO'}{0} + a_{17}' \leq \potY{\TP_\CO'}{1}, \ \potY{\FP_\CO'}{0} + a_{18}' \leq  \potY{\FP_\CO'}{1} , \  \potY{\FN_\CO'}{0} + a_{19}' \leq \potY{\FN_\CO'}{1}
	\nonumber
	\\
	&
	\potY{\TP_\NT'}{1} + a_{20}' \leq (c+d)(\NIT_\NT - R_\NT) \ , \
	\potY{\TP_\AT'}{1} + a_{21}' \leq (c+d)(\NIT_\AT - R_\AT) \ , \
	\potY{\TP_\CO'}{1} + a_{22}' \leq (c+d)(\NIT_\CO - R_\CO)
	\nonumber
	\\
	&
	\potY{\FP_\NT'}{1} + a_{23}'  \leq (c+d) R_\NT ,\  \potY{\FN_\NT'}{1} + a_{24}' \leq (c+d) R_\NT
	\nonumber
	\ , \
	\potY{\FP_\AT'}{1} + a_{25}'  \leq (c+d) R_\AT , \ \potY{\FN_\AT'}{1} + a_{26}' \leq (c+d) R_\AT
	\nonumber
	\\
	&
	\potY{\FP_\CO'}{1} + a_{27}'  \leq (c+d) R_\CO , \ \potY{\FN_\CO'}{1} + a_{28}' \leq (c+d) R_\CO 
	\nonumber
	\ , \
	c (\NIT_t - R_t) \leq \potY{\TP_t'}{\ivp} , c R_t \leq  \potY{\FP_t'}{\ivp} , \potY{\FN_t'}{\ivp}, 0 \leq a_\ell', b_\ell'
	\ .
	\nonumber
	\end{align}}%
	We transform $\potY{\TP_t'}{\ivp} , \potY{\FP_t'}{\ivp} , \potY{\FN_t'}{\ivp}, a_\ell', b_\ell'$ as $\potY{\TP_t}{\ivp} , \potY{\FP_t}{\ivp} , \potY{\FN_t}{\ivp}, a_\ell, b_\ell$ using the relationship between $\{\potY{\TP_t'}{\ivp} , \potY{\FP_t'}{\ivp} , \potY{\FN_t'}{\ivp}\}$ and $\{\potY{\TP_t}{\ivp} , \potY{\FP_t}{\ivp} , \potY{\FN_t}{\ivp}\}$, $a_\ell' = d a_\ell$, and $b_\ell' =d b_\ell$. Then, we find the elastic program \eqref{eq:LP-elastic-affine} has the same restrictions as \eqref{eq:LP-elastic} and the objective function of  \eqref{eq:LP-elastic-affine} is the same as that of  \eqref{eq:LP-elastic} multiplied by a factor $d$. Thus, we find ${\rm LB}_{\BrandomC,t}'$, the lower bound of $\tau_t'$, is the same as $d {\rm LB}_{\BrandomC,t}'$, and ${\rm UB}_{\BrandomC,t}' = d {\rm UB}_{\BrandomC,t}$ by the same reason.
	
	To show the invariance of the bound estimator, it suffices to show that the ratio estimators satisfy the affine transformation invariance property. By a similar manner, we have $ \potY{\widehat{S}'}{\ivp} = c + d\potY{\widehat{S}}{\ivp}$, $\potY{\widehat{S}_t'}{\ivp} = c + d\potY{\widehat{S}_t}{\ivp}$, and $\potY{\widehat{S}_{\BrandomC,t}'}{\ivp} = c + d \potY{\widehat{S}_{\BrandomC,t}}{\ivp} $. 
	As a consequence, $[ \widehat{{\rm LB}}_{\BrandomC,t}' , \widehat{{\rm UB}}_{\BrandomC,t} ' ]$, the bound estimator under $Y_{ji}'$, is the same as $[ d \widehat{{\rm LB}}_{\BrandomC,t} , d \widehat{{\rm UB}}_{\BrandomC,t} ]$.
	
	To show the invariance property of the extended bounds of \citet{GM2008} and \citet{LongHudgens2013}, we first study the quantities in \eqref{eq-LH-parameters} under the transformed outcome. Since $\gamma(\bw)$ and $\delta(\bw)$ are independent of the outcome, only $\pi$ and $\lambda$ vary as $\potY{\pi'}{\ivp}(\bw) = c + d \potY{\pi}{\ivp}(\bw)$ $(\ivp=0,1)$, and $\potY{\lambda'}{\ivp}(\bw)
		=
		c + d
		\potY{\lambda}{\ivp}(\bw) $. 
	Using these quantities, we have $\potY{S_\NT'}{0}(\bw)/\NIT_\NT(\bw)= c + d \potY{S_\NT}{0}(\bw)/\NIT_\NT(\bw) $, $\potY{S_\AT'}{1}(\bw)/\NIT_\AT(\bw) = c + d \potY{S_\AT}{1}(\bw)/\NIT_\AT(\bw)$, and $\potY{S_\CO'}{\ivp}(\bw)/\NIT_\CO(\bw) = c + d \potY{S_\CO}{\ivp}(\bw)/\NIT_\CO(\bw)$.
	Note that the above quantities are bounded in the interval $[c,d]$. We find a bound for $\tau_\NT' (\bw)$ based on the same reason in equation \eqref{eq-LH1} and $\tau_\NT' \in [0,d]$, which is given below.
	\begin{align}						\label{eq-LH3}
		\tau_\NT'(\bw)
		&
		\in 
		\begin{bmatrix}
			{\rm LB}_{W,\NT}'
			\\
			{\rm UB}_{W,\NT}'
		\end{bmatrix}
		=
		\begin{bmatrix}
			\max \Big\{ 0 , \frac{\potY{S_\NT'}{1} (\bw) }{\NIT_\NT(\bw)}  
		+ \frac{c \{ 1-\gamma(\bw) \} - \potY{\pi'}{0}(\bw) }{\gamma(\bw)}  
		\Big\}	
		\\
		\min \Big\{ \frac{\potY{S_\NT'}{1} (\bw) }{\NIT_\NT(\bw)}  ,
		\frac{\potY{S_\NT'}{1} (\bw) }{\NIT_\NT(\bw)}  
		+ \frac{ (c+d) \{ 1-\gamma(\bw) \} - \potY{\pi'}{0}(\bw) }{\gamma(\bw)}  
		\Big\}
		\end{bmatrix} \ .
	\end{align}
From straightforward algebra, the bound in \eqref{eq-LH3} satisfies $[ 
			{\rm LB}_{W,\NT}'
			,
			{\rm UB}_{W,\NT}' ] = d [
			{\rm LB}_{W,\NT}'
			,
			{\rm UB}_{W,\NT}']$
	where $ \big[ {\rm LB}_{W,\NT}, {\rm UB}_{W,\NT} \big]$ is a bound for $\tau_t$ defined in \eqref{eq-LH2}. This shows that the population-level bound for $\tau_\NT$ is invariant to affine transformations. We also find $\big[ {\rm LB}_{W,\AT}', {\rm UB}_{W,\AT}' \big] = d \big[ {\rm LB}_{W,\AT}, {\rm UB}_{W,\AT} \big]$ and $\big[ {\rm LB}_{W,\CO}', {\rm UB}_{W,\CO}' \big] = d \big[ {\rm LB}_{W,\CO}, {\rm UB}_{W,\CO} \big]$, showing the same  property. 
	
	To show the invariance of the bound estimator, it suffices to show that the ration estimators satisfy the affine transformation invariance property. As shown below, the ratio estimators using $\potY{Y_{ji}'}{\ivp}$is an affine transformation of the ratio estimators using $\potY{Y_{ji}}{\ivp}$, i.e. $\potY{\widehat{S'}_\NT}{1}(\bw)=c + d
		\potY{\widehat{S}_\NT}{1}(\bw)$ and $\potY{\widehat{S'}_\AT}{0}(\bw)=c + d
		\potY{\widehat{S}_\AT}{0}(\bw)$.
	As a consequence, we find $[ \widehat{{\rm LB}}_{W,t}' , \widehat{{\rm UB}}_{W,t} ' ]$, the bound estimator under $Y_{ji}'$, is the same as $[ d \widehat{{\rm LB}}_{W,t} , d \widehat{{\rm UB}}_{W,t} ]$.

	\subsection{Details of Section \ref{sec:asymptotic} in the main Paper} 	\label{sec:detail45}
		
	\subsubsection{Conditions on $f$ and $L$ for Consistent Estimation of the Sharp Bound}								\label{sec:generalfL}
	
	For given $f$ and $L$, we define the estimating equations about NT/AT classifiers as follows. Let $\Psi_t(\eta_t \con \NC)$ $(t \in \{\NT , \AT\})$ be the following population estimating equation:
\begin{align}											\label{eq-PEE}
	\Psi_t (\eta_t \con \NC) 
	= \frac{1}{\NC} \sum_j \psi_{t,j} (\eta_t)
	\ , \
	\psi_{t,j} (\eta_t)
	=
	\sum_{i=1}^{\NI_j} 
	\Bigg[
	\begin{matrix}
		\nabla_\theta L \big( t_{ji} , f_t(\bX_{ji} \con \theta_t) \big)
		\\[-0.2cm]
		t_{ji} -  \SI \big( \randomf_t (\bX_{ji} \con \theta_t) - q_t \big)
	\end{matrix}
	\Bigg]
\end{align}
\noindent
where $\nabla_\theta L$ is the gradient of $L$ with respect to $\theta$. By only using clusters under treatment/control, we consider the sample estimating equation about NT/AT classifiers as below.
\begin{align}											\label{eq-SEE}
	\widehat{\Psi}_\NT (\eta_\NT \con \NC) 
	= \frac{1}{\tNC} \sum_j \iv_j \psi_{\NT,j} (\eta_\AT)
	\ ,  \
	\widehat{\Psi}_\AT (\eta_\AT \con \NC) 
	= \frac{1}{\NC - \tNC} \sum_j (1-\iv_j) \psi_{\AT,j} (\eta_\AT)
	\ .
\end{align}
When $L$ and $f$ have a form either \hyperlink{(Linear)}{(Linear)} or \hyperlink{(Logistic)}{(Logistic)} in the main paper, $\eta_\NT^*=(\theta_\NT \sT,q_\NT^*)\T$, the solution to equation \eqref{eq:margin_1_welldefined} in the main paper, can be represented as the solution to the population estimating equation \eqref{eq-PEE}. Accordingly, $\widehat{\eta}_\NT = (\widehat{\theta}_\NT\T, \widehat{q}_\NT) \T$ and $\widehat{\eta}_\AT = (\widehat{\theta}_\AT\T, \widehat{q}_\AT) \T$, the solution to equation \eqref{eq:margin_2} in the main paper, are the solutions to the sample estimating equations in \eqref{eq-SEE}. 

To construct the CO classifier, we define $q_\CO^*$ and $\widehat{q}_\CO$ as solutions to $\NIT_\CO = \sumji  \SI \big( \randomf_\CO (\bX_{ji} \con \theta_\CO^* ) - q_\CO^* \big) $ and $\widehat{\NIT}_\CO = \sumji  \SI \big( \randomf_\CO (\bX_{ji} \con \widehat{ \theta}_\CO ) - \widehat{q}_\CO \big)$, respectively; here $\widehat{\theta}_\CO = \big( \widehat{w}_\NT, \widehat{w}_\AT, \widehat{\theta}_\NT , \widehat{ \theta}_\AT \big)$ is the plug-in estimate for $\theta_\CO$ and $\widehat{\NIT}_\CO = \NIT - \widehat{\NIT}_\NT - \widehat{\NIT}_\AT$ is the estimated number of COs in the population where $\widehat{\NIT}_\NT = \NIT \cdot {\sumji \ind(\iv_j = 1) \NT_{ji} }/{ \{\sum_j  \ind(\iv_j=1) \NI_j \} }$ and $\widehat{\NIT}_\AT =  \NIT \cdot \sumji \ind(\iv_j=0) \AT_{ji} / \{ \sum_j \ind(\iv_j=0) \NI_j \} $ are the ratio estimates for the total number of NTs and ATs.


To establish consistency, we consider the following conditions on \hyperlink{(Linear)}{(Linear)} and \hyperlink{(Logistic)}{(Logistic)} classifiers.

\begin{assumption} Let $\mathcal{E} = \mathcal{E}_\theta \otimes \mathcal{E}_q  $ be a finite-dimensional Euclidean parameter space of $\eta_t$ and let $\mathcal{E}^\circ = \mathcal{E}_{\theta}^\circ \otimes \mathcal{E}_{q}^\circ$ be its interior. Suppose the following assumptions on $(f_t,L)$ 
hold.						
									\label{assp:4-1}
	\begin{enumerate}
	\item[\hypertarget{(i)}{(i)}] (\textit{Asymptotics of NT/AT/CO and $J$}): For $t \in \{\NT, \AT, \CO\}$, $ {\tNC}/{\NC}$ and $\NIT_t/\NIT$ converge to constants in $(0,1)$, and $\potY{S_t}{\ivp}/\NIT$, $\potY{S_{\BrandomC,t}}{\ivp}/\NIT$, and $R_t/\NIT$ converge to constants in $[0,1]$ as $\NC \rightarrow \infty$. Also, $\NI_i$ is bounded for every $\NC$.
			\item[\hypertarget{(ii)}{(ii)}] (\textit{Compact $\mathcal{E}$ and true parameter in $\mathcal{E}^\circ$}): 
			$\mathcal{E}_{\theta}$ and $\mathcal{E}_q$ are compact and 
			$\eta_t^* \in \mathcal{E}^\circ$.
		\item[\hypertarget{(iii)}{(iii)}]  (\textit{Full rank $\bX_{ji}$ and rate of $\|\bX_{ji}\|$}): $  \{ \sumji \bX_{ji} \bX_{ji}\T \} /\NIT$ and $\lim_{\NC \rightarrow \infty} \{ \sumji \bX_{ji} \bX_{ji}\T \} /\NIT$ are finite and full rank. Also, 
		$\max_{ji} \| \bX_{ji} \| = O(\log \NC)$. 
					  
		\item[\hypertarget{(iv)}{(iv)}] (\textit{Behavior of randomized $\randomf$}):
		For $t \in \{\NT,\AT,\CO\}$, let $G_t(q \con \NC,  \theta_t)$ be the empirical cumulative distribution function (CDF) of $\randomf_t(\bX_{ji} \con \theta_t)$, i.e. $G_t(q \con \NC,  \theta) =  \sumji \ind \big\{ \randomf_t(\bX_{ji} \con \theta) \leq q \big\} / \NIT$. Then, $G_t(q \con \NC,  \theta)$ satisfies the following conditions.
		\begin{itemize}
			\item[(a)] (\textit{Smoothness}): 
			Let $\mathcal{E}_{\theta, {\rm const}}^\circ = \{ \theta \in \mathcal{E}_\theta^\circ \cond \randomf_t( \bX_{ji} \con \theta) = \randomf_t( \bX_{ji}' \con \theta) \ \forall \bX_{ji} \neq \bX_{ji}' \}$ be the set of $\theta$ where $\randomf_t( \cdot \con \theta)$ is constant. For any fixed $\theta_t \in \mathcal{E}_\theta^\circ \setminus \mathcal{E}_{\theta, {\rm const}}^\circ$ and fixed $q_1, q_2 \in \mathcal{E}_q^\circ$, we have $\big| G_t(q_2 \con \NC,  \theta_t) - G_t(q_1 \con \NC,  \theta_t) \big| \leq \mu_t (\NC, \theta_t) \big| q_2 - q_1 \big|$ where $\mu_t(\NC, \theta_t)$ depends on $\NC$ and $\theta_t$ and is bounded by a constant for any $\NC$. 
			\item[(b)]  (\textit{Local identifiability}): 
			There exists an interval $[q_L, q_U]$ containing $q_t^*$ such that for any fixed interval $(q_1, q_2) \subset [q_L, q_U]$, we have $\big| G_t (q_2  \con \NC, \theta_t^*) - G_t (q_1  \con \NC, \theta_t^*) \big| \geq \kappa_t(\NC) \big| q_2 - q_1 \big|$ where $\kappa_t(\NC)$ only depends on $\NC$ and converges to a positive constant as $\NC \rightarrow \infty$. 
		\end{itemize}
	\end{enumerate}
\end{assumption}
\noindent Condition \hyperlink{(i)}{(i)} of Assumption \ref{assp:4-1} is stronger than the assumption in Theorem \ref{thm:302} in the main paper and is a notable limitation of our theory. 
Condition \hyperlink{(ii)}{(ii)} and the full rank portion of condition \hyperlink{(iii)}{(iii)} are common assumptions when proving theoretical properties of estimators based on the likelihood principle or estimating equations. 
The max norm portion of condition \hyperlink{(iii)}{(iii)} states that the range of the covariates grows at $\log \NC$ rate and is satisfied if covariates are realizations from a non-degenerate sub-exponential distribution. 
Part (a) of condition \hyperlink{(iv)}{(iv)} states that the derivative of the CDF of the randomized $\randomf_t( \cdot \con \theta)$ is finite for every $\NC$; it is violated if the CDF ``jumps'' or changes drastically as $\NC$ grows. Part (b) of condition \hyperlink{(iv)}{(iv)} states that the derivative of the CDF near the true $q_t^*$ is non-zero; it is violated if the CDF remains flat around $q_t^*$. Both parts of condition \hyperlink{(iv)}{(iv)} hold if all $\bX_{ji}$ are realizations from a discrete support or from a compact support with a bounded density. 

Under general classifiers, we consider the following conditions on the experimental design and the estimating equations.

	\begin{assumption} 							\label{assmp-General}
	Suppose the following assumptions hold with probability tending to 1.
		\begin{itemize}
		
			\item[\hypertarget{(GC1)}{(GC1)}] (\textit{Asymptotics of NT/AT/CO and $J$}): For $t \in \{\NT, \AT, \CO\}$, $\NIT_t/\NIT$, $\potY{S_t}{\ivp}/\NIT$, $\potY{S_{\BrandomC,t}}{\ivp}/\NIT$, and $R_t/\NIT$ converge to constants in $[0,1]$ as $\NC \rightarrow \infty$. Also, $\NI_i$ is bounded for every $\NC$ and $ {\tNC}/{\NC}  \rightarrow p_{\rm trt} \in (0,1)$ as $\NC \to \infty$.
			
			\item[\hypertarget{(GC2)}{(GC2)}] (\textit{Compact $\mathcal{E}$ and true parameter in $\mathcal{E}^\circ$}): The parameter spaces $\mathcal{E}_{\theta}$ and $\mathcal{E}_q$ are compact and the true parameter $\eta_t^*$ $(t \in \{\NT,\AT,\CO\})$ is in the interior of the parameter spaces $\mathcal{E}^\circ$.
			
			\item[\hypertarget{(GC3)}{(GC3)}] (\textit{Invertible $\nabla_\theta^2 L$}): For $t \in \{\NT,\AT\}$, $ \sumji \nabla_\theta^2 L \big(t_{ji}, f_t(\bX_{ji} \con \theta_t) \big)$ is invertible for all $\theta_t$.
			
			\item[\hypertarget{(GC4)}{(GC4)}] (\textit{Slow growing $\| \psi_{t,j} \|$ and average rate of change of $f_t$}): For $t \in \{\NT,\AT\}$, $j=1,\ldots, \NC$, and $\eta_t$, there exist a constant $v_1>0$ satisfying $\big\| \psi_{t,j} (\eta_t) \big\| = O\big( (\log \NC)^{v_1} \big)$. Moreover, for $t \in \{\NT,\AT,\CO\} $ and for any $\theta_{t,1}$, $\theta_{t,2}$, and $\bX_{ji}$, we have $\| f_t (\bX_{ji} \con \theta_{t,1}) - f_t (\bX_{ji} \con \theta_{t,2}) \| / \| \theta_{t,1}  - \theta_{t,2} \| = O(\log\NC) $.
			
			\item[\hypertarget{(GC5)}{(GC5)}] (\textit{Smooth estimating equation}): For $t \in \{ \NT,\AT \}$, and any sequence $\overline{\eta}_{t,\NC} = O(\NC^{-1/2})$, we have the following result for some constant $v_2>0$.
			\begin{align*}
				\sup_{\eta_t \in \mathcal{E}^\circ } \frac{1}{\NC} \sum_j \Big\| \psi_{t,j} (\eta_t) - \psi_{t,j} (\eta_t+\overline{\eta}_{t,\NC}) \Big\| = O\big( \NC^{-1/2} \cdot \big( \log \NC \big)^{v_2}  \big) \ .
			\end{align*}
			
		\item[\hypertarget{(GC6)}{(GC6)}] (\textit{Identifiable estimating equation}): For $t \in \{\NT,\AT\}$, and any $\eta_t$ such that $\| \eta_t - \eta_t^* \| = \epsilon$ with $\epsilon = \Theta(\NC^{-r})$ for some $0<r<1/4$, we have the following result for some constant $v_3>0$.
		\begin{align*}
			\min_{\eta_t:\|\eta_t-\eta_t^*\|=\epsilon} \| \Psi_t(\eta_t \con \NC) \| \geq B(\NC, \epsilon) \big\{ 1 + o(1) \big\} \ , \ B(\NC, \epsilon) = \Theta\big( \NC^{-r} \cdot (\log \NC)^{-v_3} \big) \ .
			\end{align*}
			
		\item[\hypertarget{(GC7)}{(GC7)}] (\textit{Smooth accumulation of $\randomf$}): For $t \in \{\NT,\AT,\CO\}$, let $\mathcal{E}' = \{ ( q_t, \theta_t) \cond \randomf_\theta (\bx \con \theta_t) - q_t = 0 \}$. For any sequence $d_\NC$ such that $| d_\NC | = o(1)$, we have
		\begin{align*}
			& \sup_{\eta_t \in \mathcal{E}^\circ \setminus \mathcal{E}' } \frac{1}{\NC} \sumji \ind \big\{ | \randomf_t (\bX_{ji} \con \theta_t) - q_t | < |d_\NC| \big\} 
			= O\big( |d_\NC|) \ .
		\end{align*}

		\item[\hypertarget{(GC8)}{(GC8)}] (\textit{Identifiable accumulation of $\randomf$}): Let $e_\NC= O(\NC^{-1})$ and $e_\NC'=e'+O(\NC^{-1})$ be sequences satisfying with $0 < e'$ with $e'=\Theta(\NC^{-r})$ for some $0<r<1/4$. For $t \in \{\NT,\AT,\CO\}$, we have
		\begin{align*}
			& \frac{1}{\NC} \sumji \ind \big\{ e_\NC \leq \randomf_t (\bX_{ji} \con \theta_t^*) - q_t^* \leq  e_\NC'  \big\} 
			= \Theta(\NC^{-r}) \ , \
			\frac{1}{\NC} \sumji \ind \big\{ - e_\NC' \leq \randomf_t (\bX_{ji} \con \theta_t^*) - q_t^* \leq  - e_\NC  \big\} 
			= \Theta(\NC^{-r}) \ .
		\end{align*}

	\end{itemize}

	\end{assumption}
	Conditions \hyperlink{(GC1)}{(GC1)} and \hyperlink{(GC2)}{(GC2)} are the same as conditions (i) and (ii) of Assumption \ref{assp:4-1}. 
	Condition \hyperlink{(GC3)}{(GC3)} holds if $L$ is strictly convex in $\theta$.
	Condition \hyperlink{(GC4)}{(GC4)} means that the norm of the estimating equation is a slowly growing function of $\NC$ and the transformation function $f$ does not change dramatically across $\theta$ for a fixed $\bX_{ji}$.
	Condition \hyperlink{(GC5)}{(GC5)} implies that the estimating equation does not vary too much in the neighborhood of the given parameter $\eta_t$. 
	Condition \hyperlink{(GC6)}{(GC6)} implies that the estimating equation is curved to some degree at the neighborhood of the true parameter $\eta_t^*$. 
	Conditions \hyperlink{(GC7)}{(GC7)} and \hyperlink{(GC8)}{(GC8)} are similar to condition (iii) and (iv)  of Assumption \ref{assp:4-1}. 
	
	Lemma \ref{lem-EE12toGeneral} shows that Assumption \ref{assp:4-1} implies Assumption \ref{assmp-General} when $f$ and $L$ are chosen as either \hyperlink{(Linear)}{(Linear)} or \hyperlink{(Logistic)}{(Logistic)} in the main paper; the proof of Lemma \ref{lem-EE12toGeneral} is in Section \ref{sec:C4}.
	\begin{lemma}						\label{lem-EE12toGeneral}
		Suppose Assumptions \hyperlink{(A1)}{(A1)}-\hyperlink{(A6)}{(A6)} in the main paper and Assumption \ref{assp:4-1} hold. Moreover, $f$ and $L$ are chosen as either \hyperlink{(Linear)}{(Linear)} or \hyperlink{(Logistic)}{(Logistic)} in the main paper. Then, the conditions of Assumption \ref{assmp-General} hold.
	\end{lemma}

	\subsubsection{Consistency of $\widehat{\eta}_t$}

Let $\widetilde{\mathcal{F}}_\NC = \mathcal{F}_\NC \cup \{ e_{tji} \cond t\in\{\NT,\AT, \CO\},  j=1,\ldots,\NC, i = 1,\ldots,\NI_j \}$ be the extended set of $\mathcal{F}$ that includes the randomization term $e_{tji}$ from $\randomf_t(\cdot \con \theta_t)$. We treat $e_{tji}$ as fixed after being randomly generated once and condition on $\tilde{\mathcal{F}}_\NC$ in our asymptotic arguments; this allows the randomness in the study to still be from the treatment assignment $\IV$ only and is in alignment with finite-sample/randomization inference framework. Lemma \ref{lmm:401} establishes consistency of $\widehat{\eta}_t$  as well as its rate of convergence. 
	\begin{lemma}									\label{lmm:401}
		Suppose Assumptions \hyperlink{(A1)}{(A1)}-\hyperlink{(A6)}{(A6)} in the main paper and Assumption \ref{assmp-General} hold. Then, for any $\epsilon$ such that $\epsilon = \Theta(\NC^{-r})$ with $0<r<1/4$, the estimators  $\widehat{\eta}_\NT$ and $\widehat{\eta}_\AT$ from \eqref{eq:margin_2} in the main paper satisfy $\lim_{\NC \rightarrow \infty} P \big\{ \| \widehat{\eta}_t - \eta_t^* \| > \epsilon \, \big| \, \tilde{\mathcal{F}}_\NC , \mathcal{\iv}_\NC \big\} = 0$ for $t \in \{\NT,\AT\}$. Furthermore, suppose that $\widehat{w}_\NT$ and $\widehat{w}_\AT$ in $\widehat{\theta}_\CO$ satisfy $\widehat{w}_\NT = w_\NT^* + O(\NC^{-1/2})$ and $\widehat{w}_\AT = w_\AT^* + O(\NC^{-1/2})$, respectively. Then, we have $\lim_{\NC \rightarrow \infty} P \big\{ \| \widehat{\eta}_\CO - \eta_\CO^* \| > \epsilon \, \big| \, \tilde{\mathcal{F}}_\NC , \mathcal{\iv}_\NC \big\} = 0$ where $\epsilon = \Theta \big( \NC^{-r} \cdot ( \log\NC ) \big)$.
	\end{lemma}
\noindent The proof of Lemma \ref{lmm:401} is in Section \ref{sec:proofoflmm:401}. Note that the results of Lemma \ref{lmm:401} hold under Assumption \ref{assp:4-1} because of Lemma \ref{lem-EE12toGeneral}. The term $\epsilon$ in Lemma \ref{lmm:401} is allowed to decrease as $\NC$ increases, but the rate of $\epsilon$ must be slower than $\NC^{-1/4}$. Also, Lemma \ref{lmm:401} implies that with probability 1, $\NC^{r} \| \widehat{\eta}_t - \eta_t^* \|$ converges to zero as $\NC \rightarrow \infty$ for $0 < r< 1/4$. These are slower rates of convergence than typical parametric rates of $\NC^{-1/2}$, in part because of the surrogate indicator function in equation \eqref{eq:margin_2} in the main paper. Lastly, we choose $w_t^* = \NIT_t/\NIT$ and $\widehat{w}_t = \widehat{\NIT}_t /\NIT$ where $\widehat{\NIT}_t$ are ratio estimates. Then, the convergence rate about $\widehat{w}_t$ is satisfied from the finite population central limit theorem.

	\subsubsection{Visual Illustration of Assumption \ref{assp:4-1}}					
	
\begin{figure}[!htb]
	\centering
	\includegraphics[width=1\textwidth]{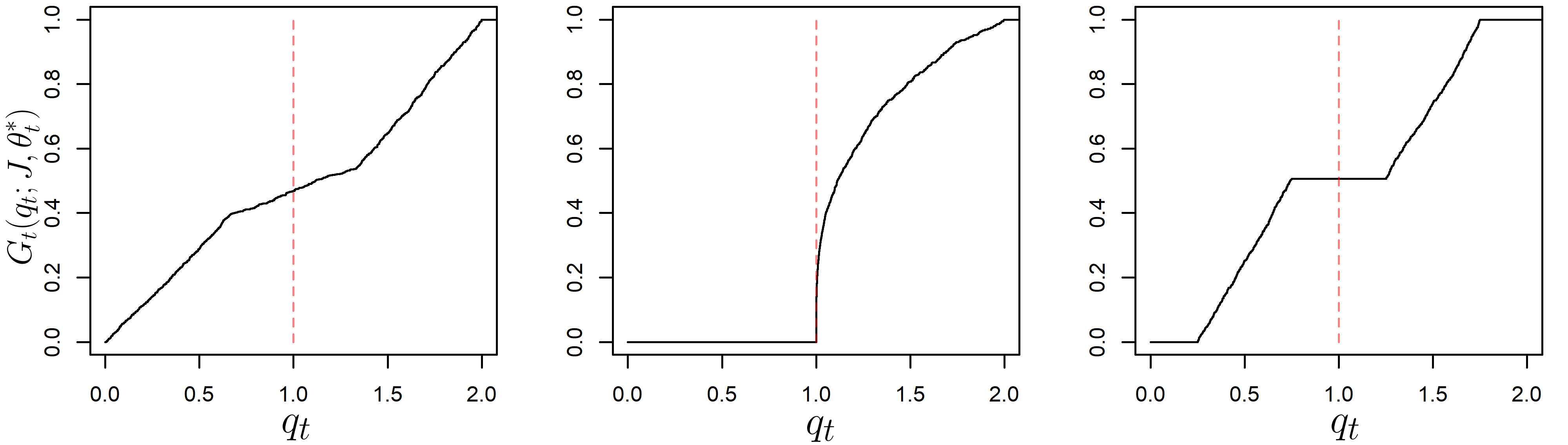}
	  \vspace*{-0.75cm}
   \caption{\footnotesize Visual Illustration of Condition {\protect\hyperlink{(iv)}{(iv)}} in Assumption \ref{assp:4-1}. }
   \label{Fig:violation}
\end{figure}

In Figure \ref{Fig:violation}, we plot out the value of the cumulative distribution function $G_t(q_t\con \NC, \theta_t^*)$ (y-axis) as a function of $q_t$ (x-axis). The red dashed line visually guides the true threshold $q_t^*=1$. The left plot shows an example of $G_t(q_t\con \NC, \theta_t^*)$ satisfying condition {\protect\hyperlink{(iv)}{(iv)}}. The middle plot shows an example of $G_t(q_t\con \NC, \theta_t^*)$ violating the first condition of {\protect\hyperlink{(iv)}{(iv)}} where the average rate of the derivative of $G_t(q_t\con \NC, \theta_t^*)$ near $q_t^*=1$ diverges as $\NC \rightarrow \infty$. Finally, the right plot shows an example of $G_t(q_t\con \NC, \theta_t^*)$ violating the second condition of {\protect\hyperlink{(iv)}{(iv)}} where the average rate of the derivative of $G_t(q_t\con \NC, \theta_t^*)$ near a neighborhood of $q_t^*=1$ converges to zero as $\NC \rightarrow \infty$.

	\subsubsection{Details of the Construction of Confidence Sets}				\label{sec:ConfidenceSet}

In general, deriving inferential properties for bounds is a difficult, sometimes impossible, task; see \citet{ImbensManski2004}, \citet{Romano2008}, \citet{Romano2010}, \citet{HiranoPorter2012}, \citet{Chernozhukov2013}, and citations within. In particular, to obtain uniformly valid confidence sets for bounds requires some understanding about the (asymptotic) distributions of the inputs to the linear program in Section \ref{sec:lp} in the main paper, uniformly over $\tau_t$ in ${\rm LB}_t$ and ${\rm UB}_t$. Given that we are using a non-trivial classifier that has been randomized and includes a surrogate indicator function, we defer a complete solution to this problem to a future paper.
For now, we propose a practical approach to construct confidence sets for the bounds of $\tau_t$ based on popular resampling procedures, such as the bootstrap \citep{Efron1979} or subsampling  \citep{Politis1994,Subsampling}. Here we use cluster-level bootstrap percentiles \citep{Efron1979, Bootstrap1993} to construct confidence intervals. Specifically, we resample $\tNC$ clusters among $\tNC$ treated clusters and $\NC - \tNC$ clusters among $\NC - \tNC$ control clusters with replacement for multiple times, say $B=10^3$. For each realized bootstrap cluster, we obtain the extended bounds of \citet{GM2008} and \citet{LongHudgens2013} that are denoted as $[ \widehat{{\rm LB}}_{\BrandomC,t}^{(b)}, \widehat{{\rm UB}}_{\BrandomC,t}^{(b)}]$ and $[ \widehat{{\rm LB}}_{W,t}^{(b)}, \widehat{{\rm UB}}_{W,t}^{(b)}]$ $(b=1,\ldots,B)$. Let $p_{\alpha} (\widehat{{\rm LB}}_{\BrandomC,t})$ be the $100\alpha$-th percentile of $\{ \widehat{{\rm LB}}_{\BrandomC,t}^{(1)} , \ldots, \widehat{{\rm LB}}_{\BrandomC,t}^{(B)} \}$. We define $p_{\alpha} (\widehat{{\rm UB}}_{\BrandomC,t})$, $p_{\alpha} (\widehat{{\rm LB}}_{W,t})$, and $p_{\alpha} (\widehat{{\rm UB}}_{W,t})$ in a similar manner. If the lower bound estimate is chosen as $\widehat{{\rm LB}}_{\BrandomC,t}$ (i.e. $\widehat{{\rm LB}}_{W,t} \leq \widehat{{\rm LB}}_{\BrandomC,t}$), we choose $q_{\alpha/2} (\widehat{{\rm LB}}_{\BrandomC,t})$ as $\widehat{{\rm LB}}_{\alpha/2, t}$, the lower end of a $100(1-\alpha)$ confidence interval for $[ {\rm LB}_t, {\rm UB}_t ]$. On the other hand, if the lower bound estimate is chosen as $\widehat{{\rm LB}}_{W,t}$ (i.e. $\widehat{{\rm LB}}_{\BrandomC,t} \leq \widehat{{\rm LB}}_{W,t}$), we choose $q_{\alpha/2} (\widehat{{\rm LB}}_{W,t})$ as $\widehat{{\rm LB}}_{\alpha/2, t}$; we similarly define $\widehat{{\rm UB}}_{1-\alpha/2, t}$, the upper end of a $100(1-\alpha)$ confidence interval for $[ {\rm LB}_t, {\rm UB}_t ]$: $
	\widehat{{\rm LB}}_{\alpha/2, t}
	=
	q_{\alpha/2} (\widehat{{\rm LB}}_{\BrandomC,t}) \ind \big\{ \widehat{{\rm LB}}_{W,t} \leq \widehat{{\rm LB}}_{\BrandomC,t} \big\}
	+
	q_{\alpha/2} (\widehat{{\rm LB}}_{W,t}) \ind \big\{ \widehat{{\rm LB}}_{W,t} \geq \widehat{{\rm LB}}_{\BrandomC,t} \big\}$ and $
	\widehat{{\rm UB}}_{1-\alpha/2, t}
	=
	q_{1-\alpha/2} (\widehat{{\rm UB}}_{\BrandomC,t}) \ind \big\{ \widehat{{\rm UB}}_{W,t} \geq \widehat{{\rm UB}}_{\BrandomC,t} \big\}
	+
	q_{1-\alpha/2} (\widehat{{\rm UB}}_{W,t}) \ind \big\{ \widehat{{\rm UB}}_{W,t} \leq \widehat{{\rm UB}}_{\BrandomC,t} \big\}$. 

\subsection{Additional Results of Section 5 and 6 in the Main Paper}						\label{sec:addsim}

\begin{figure}[!htb]
	\centering
	\includegraphics[width=1\textwidth]{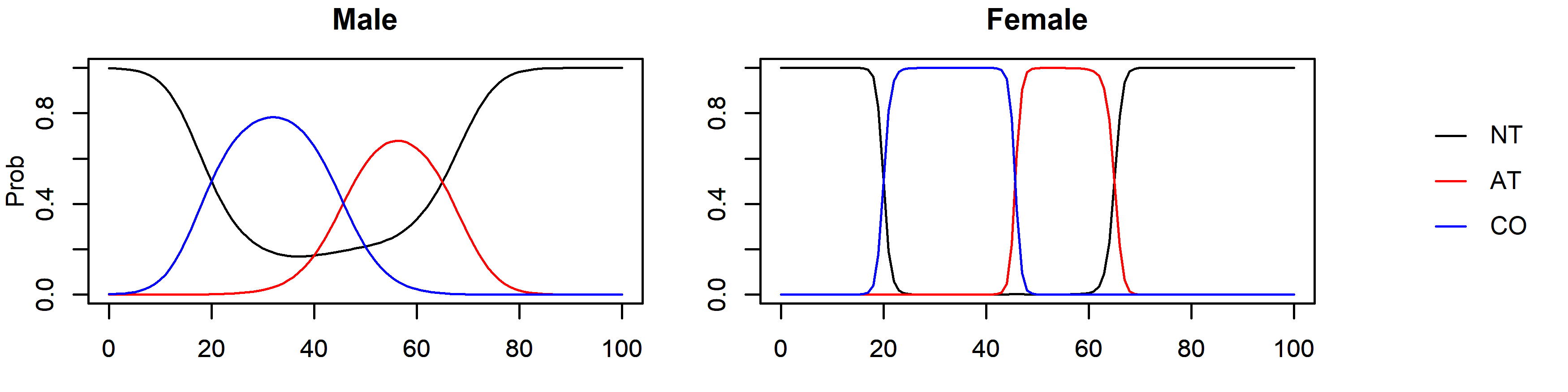}
	\vspace*{-1cm}
	\caption{\footnotesize Probability of Each Compliance Type across $X_{ji,{\rm male}}$ and $X_{ji, {\rm age}}$.}
	\label{Fig:Compliance}
\end{figure}

In this subsection, we present additional results of simulation and data analysis. 
Figure \ref{Fig:Compliance} shows the probability of each compliance type across $X_{ji,{\rm male}}$ and $X_{ji, {\rm age}}$.

\newpage

\begin{figure}[!htb]
	\centering
		\vspace*{-0.25cm}
	\includegraphics[width=1\textwidth]{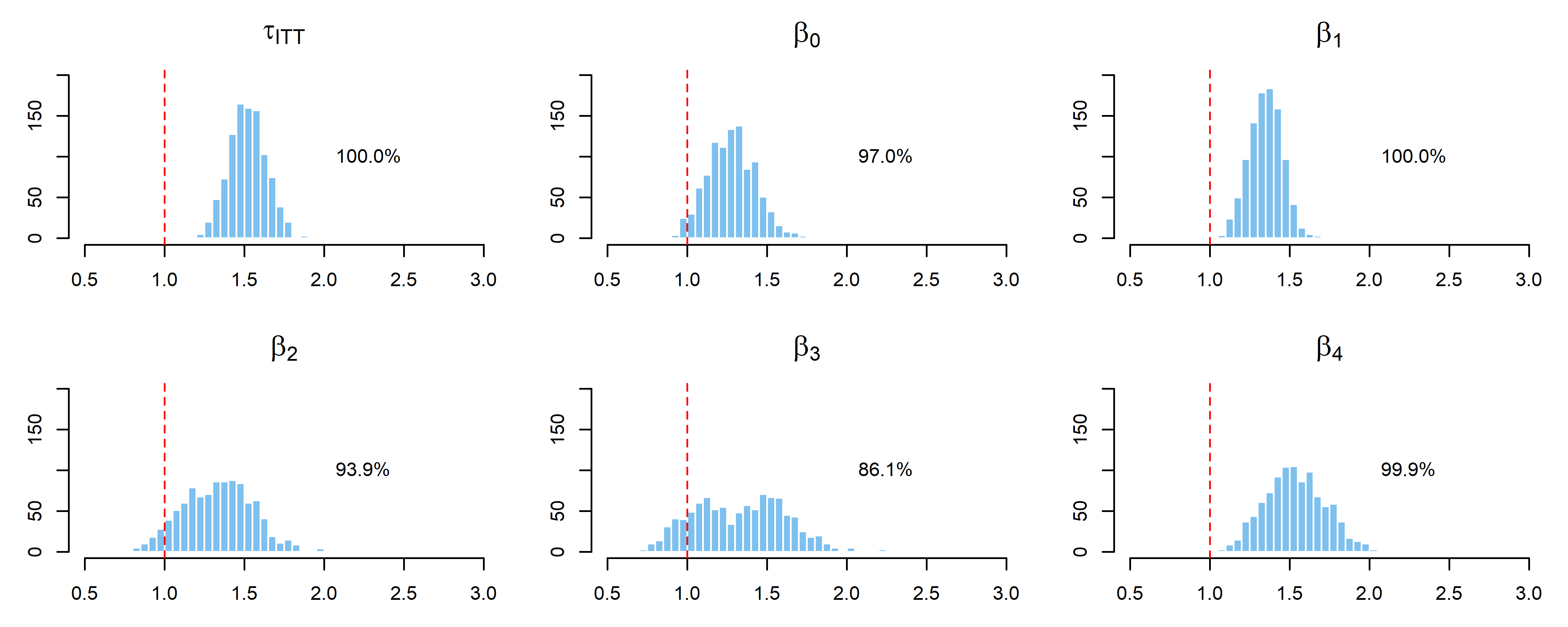}
	\caption{\footnotesize Histogram of the Ratio of the Variance Estimates to the Empirical Variance of the Estimates. The number in each figure shows the proportion of the ratio that is greater than 1.}
	\label{Fig:VarianceRatio}
\end{figure}

We present Figure \ref{Fig:VarianceRatio} that shows the histograms of the ratio of the variance estimators $\widehat{\sigma}_\ITT^2$ and $\widehat{\Sigma}_{\B}$ to the empirical variance of $\widehat{\tau}_\ITT$ and $\widehat{\B}$. The histograms shows the variance estimators are conservative for the true variance of the overall and heterogeneous ITT effects.

\begin{figure}[!htb]
	\centering
	\includegraphics[width=1\textwidth]{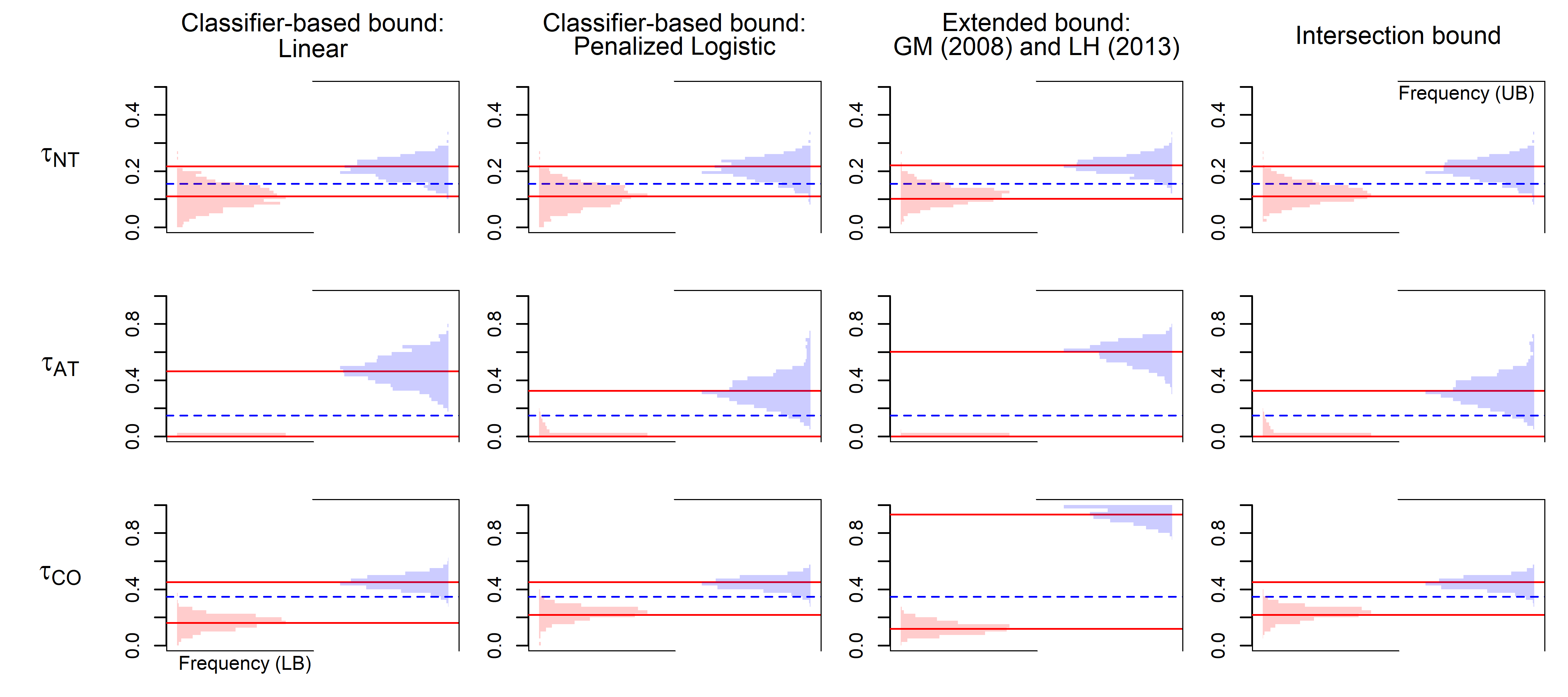}
	\vspace*{-0.25cm}
   \caption{\footnotesize Histograms of the Lower  ({\color{red1}$\blacksquare$}) and Upper ({\color{blue1}$\blacksquare$}) Limits of Bound Estimates. Each row corresponds to the compliance group effect. Each column corresponds to the classifier-based bounds, the extended bounds of \citet{GM2008} and \citet{LongHudgens2013}, and the intersection bounds created by taking the intersection of our bounds and the extended bounds. The $y$-axis across all plots represents effect size. The left column shows histograms of the bound estimates and the $x$-axes show the relative frequencies.  The red solid lines ({\color{red}\rule[0.5ex]{0.5cm}{0.5pt}}) represent the true sharp bounds under each method. The blue dashed lines ({\color{blue}\rule[0.5ex]{0.1cm}{0.5pt}\,\rule[0.5ex]{0.1cm}{0.5pt}\,\rule[0.5ex]{0.1cm}{0.5pt}}) show the true compliance group effects $\tau_\NT=0.155$, $\tau_\AT=0.148$, and $\tau_\CO=0.347$.}
   \label{Fig:HongKong3-1}
   		\vspace*{-2cm}
\end{figure}

We present Figure \ref{Fig:HongKong3-1} that shows the histogram of the bound estimates. The histograms visually show that the bounds contain the true compliance group effect and the bound estimates are consistent for the bounds.

\newpage

Table \ref{tab:HK-Balance} shows the result of assessing covariate balance in the Hong Kong study.

\begin{table}[!htp]
		\renewcommand{\arraystretch}{1.1} \centering
		\footnotesize
\begin{tabular}{|c|c|c|c|c|}
\hline
                                                    &  Treated  &   Control  &  Absolute value of $t$-statistic \\ \hline
                                Number of clusters  &         45  &        51  &                            -  \\ \hline
                             Number of individuals  &        129  &       161  &                            -  \\ \hline
 Number of individuals who actually took treatment  &         71  &         3  &                            -  \\ \hline
                               Average of outcomes  &     0.9535  &    0.8758  &                            -  \\ \hline
                              Average cluster size  &     2.8667  &    3.1569  &                       $1.5417$  \\ \hline
                                Proportion of male  &     0.3824  &    0.3701  &                      $0.2136$  \\ \hline
                                    Average of age  &    34.4118  &   36.2727  &                      $0.9357$  \\ \hline
              Proportion of vaccinated individuals  &     0.1397  &    0.1234  &                      $0.4086$  \\ \hline
                     Proportion of male index individuals  &        0.5000	  &    0.4481  &                      $0.8821$  \\ \hline
                      Average of age of index individuals  &    11.7059  &   12.1364  &                        $0.2870$  \\ \hline
               Proportion of vaccinated index individuals  &     0.1544  &    0.1299  &                      $0.5942$  \\ \hline
                                        House size  &   817.8603  &  831.9221  &                       $0.2689$  \\ \hline
\end{tabular}		
\caption{\footnotesize Result of Covariate Balance Assessment. $t$-statistics are from the $t$-test comparing treated and control groups.}
\label{tab:HK-Balance}
\vspace*{-0.5cm}
\end{table}

To incoporate the cluster size in the analysis, we use background information by \citet{Bartlett1957} and \citet[Chapter 6]{Keeling2011} on mathematical models of infectious disease within small clusters to encode each cluster size as a fixed effect where $\NI_i =2$ is specified as the base level and three indicators $\ind(\NI_j=3)$, $\ind(\NI_j=4)$, and $\ind(\NI_j \geq 5)$ are used as dummy variables.  In total, with a constant intercept term, $\bX_{ji}$ consist of the following 13 components: $
	\big[
		1 
		,
		X_{ ji, {\rm male} } = \ind( ji \text{ is male})
		,
		X_{ ji, {\rm age} }
		,
		X_{ ji, {\rm age} }^2
		,
		X_{ ji, {\rm vaccine} } = \ind( ji \text{ is vaccinated in 2008})
		, \\
		X_{ ji, {\rm ind.male} } = \ind( \text{cluster $j$'s index individual is male})
		,
		X_{ ji, {\rm ind.age} }
		,
		X_{ ji, {\rm ind.age} }^2
		, 
		X_{ ji, {\rm ind.vaccine} } = \\
		\ind(\text{cluster $j$'s index individual is vaccinated in 2008})
		,
		\ind(\NI_j=3)
		,
		\ind(\NI_j=4)
		,
		\ind(\NI_j \geq 5)
		,
		X_{ji , {\rm housesize} } \big]$.

Lastly, Table \ref{tab:HK-NTSE2} presents the  results including the estimated bounds of $\tau_\AT$.

\begin{table}[!htp]
		\renewcommand{\arraystretch}{1.1} \centering
		\footnotesize
\begin{tabular}{|c|c|c|c|c|c|}
\hline
\multirow{3}{*}{Estimand}   & \multirow{3}{*}{Statistic} & \multicolumn{2}{c|}{\multirow{2}{*}{Classifier-based Bound}} & \multirow{3}{*}{\begin{tabular}[c]{@{}c@{}}Extended Bound of\\ \citet{GM2008} and \\ \citet{LongHudgens2013}\end{tabular}} & \multirow{3}{*}{Intersection Bound} \\
                            &                             & \multicolumn{2}{c|}{}                            &                                                                      &                                     \\ \cline{3-4}
                            &                             &              Linear            &    Penalized Logistic                 &                                                                      &                                     \\ \hline
\multirow{2}{*}{$\tau_\NT$}                  & Bound                      & $ [0.000 , 0.173]$      & $[0.000 , 0.173]$      & $[0.054 , 0.254]$                                                    & $[0.054 , 0.173]$                   \\ \cline{2-6}
                            & 95\% CI                     & $ [0.000 , 0.374]$      & $[0.000 , 0.375]$      & $[0.000 , 0.395]$                                                    & $[0.000 , 0.374]$                   \\ \hline
 \multirow{2}{*}{$\tau_\AT$} & Bounds & $ [0.000 , 0.000]$ & $[0.000 , 0.000]$ &  $[0.000 , 0.000]$ & $[0.000 , 0.000]$ \\ \cline{2-6}
                                   & 95\% CI & $ [0.000 , 0.000]$ & $[0.000 , 0.000]$ & $[0.000 , 0.000]$ & $[0.000 , 0.000]$ \\ \hline                            
\multirow{2}{*}{$\tau_\CO$} & Bound                      & $ [0.000 , 0.146]$      & $[0.000 , 0.146]$      & $[0.000 , 0.186]$                                                    & $[0.000 , 0.146]$                   \\ \cline{2-6} 
                            & 95\% CI                     & $ [0.000 , 0.299]$      & $[0.000 , 0.288]$      & $[0.000 , 0.297]$                                                    & $[0.000 , 0.288]$                   \\ \hline
\end{tabular}		
\caption{\footnotesize Network Effects Among Compliance Types in the Hong Kong Study. Each row corresponds to the compliance group effects and relevant statistics.  Each column shows the classifiers used in our bounds, the extended bound of \citet{GM2008} and \citet{LongHudgens2013}, and the intersection bound created by taking the intersection of our bounds and the extended bound.}
\label{tab:HK-NTSE2}
\vspace*{-1cm}
\end{table}

\newpage

\section{Useful Lemmas}								\label{sec:Lemma}

In this section, we introduce lemmas that facilitates the proofs in Section \ref{sec:B} and \ref{sec:C}.

	\begin{lemma}								\label{lem:301}
	Suppose that Assumption \hyperlink{(A1)}{(A1)} and \hyperlink{(A2)}{(A2)} in the main paper hold. Then, the expectation and the variance of $\IV$ are given by $\EXP \big( \IV \cond \mathcal{F}_\NC ,\mathcal{\iv}_\NC \big)
	=
	\tNC\bm{1}_\NC/\NC$ and $
	\VAR \big( \IV \cond \mathcal{F}_\NC , \mathcal{\iv}_\NC \big)
	=
	 \tNC ( \NC - \tNC)\Pi_\NC/\{\NC ( \NC - 1 )	\}$ where $\Pi_\NC =I_\NC-\bm{1}_\NC \bm{1}_\NC\T/\NC$ is the centering matrix.
\end{lemma}

\begin{lemma}							\label{lem:302}
	Suppose that Assumption \hyperlink{(A1)}{(A1)} and \hyperlink{(A2)}{(A2)} in the main paper hold. Then, for $\ivp=0,1$, we have $ 
		\EXP \big\{ \widehat{\mo}_\ivp (\bm{S}) \cond \mathcal{F}_\NC , \mathcal{\iv}_\NC \big\}
		=
		\mo(\bm{S}) $ and $
		\EXP \big\{ \widehat{\vo}_\ivp (\bm{S},\bm{T}) \cond \mathcal{F}_\NC , \mathcal{\iv}_\NC \big\}
		=
		\vo(\bm{S},\bm{T}) $.
\end{lemma}

\begin{lemma}			\label{lemma:WLLN}
	Let $\bW_i$ be the discrete covariates that are used to define $[{\rm LB}_{W,t}, {\rm UB}_{W,t}]$. 	For stratum $\{\bW_i = \bw \}$, let $\mathcal{S}_j(\bw) = \sum_{i=1}^{\NI_j} S_{ji} \ind (\bW_{ji} = \bw)$. Suppose Assumption \hyperlink{(A1)}{(A1)} and \hyperlink{(A2)}{(A2)} in the main paper hold.  Furthermore, suppose that (i) $\max_{ji} | S_{ji} |$ and $\NI_j$ are bounded for all $\NC$; (ii) $\sum_j \mathcal{S}_j(\bw) / \NIT$ converges to a constant; and (iii) $\tNC/\NC$ converges to a constant in $(0,1)$.  Then, we have $\{\NC  \sumji \ind(\iv_j = \ivp) \mathcal{S}_j(\bw) \}/\{\NIT \sum_j \ind(\iv_j=\ivp) \} \stackrel{P}{\rightarrow} {\sumji \mathcal{S}_j(\bw)}/{\NIT}$; i.e. for $\ivp \in \{0,1\}$ and any constant $\epsilon>0$, 
	\begin{align*}
		\displaystyle{\lim_{\NC \rightarrow \infty}	P \bigg\{ \bigg|  
		\frac{\NC  \sumji \ind(\iv_j = \ivp) \mathcal{S}_j(\bw) }{\NIT \sum_j \ind(\iv_j=\ivp) }
		-
		\frac{\sumji \mathcal{S}_j(\bw) }{\NIT} \bigg| > \epsilon \, \bigg| \, \F_\NC , \mathcal{\iv}_\NC \bigg\} = 0} \ .
	\end{align*}
\end{lemma}

\noindent The proofs of Lemma \ref{lem:301} and \ref{lem:302} are trivial. The proof of \ref{lemma:WLLN} is 
\ref{sec:C6}.

\section{Proof of Lemmas and Theorems in the Main Paper}							\label{sec:B}

\subsection{Proof of Theorem \ref{thm:302} in the main Paper and Theorem \ref{Supp-thm:302}}							\label{sec:proof-thm1}

	We introduce Theorem \ref{thm:Ding5} of \citet{LiDing2017} which is formally stated as follows.
	\begin{manualtheorem}{5}\citep{LiDing2017}				\label{thm:Ding5}
		Suppose that $N$ units are completely randomized into $Q$ treatment groups of size $(n_1, \ldots , n_Q)$. Let $L_i$ be the group number,
where $L_i = q$ if unit i belongs to group $q$. Let $\bm{Y}_i(q) \in \R^p$ be unit $i$’s potential outcome under treatment $q$ and $\overline{\bm{Y}} = \sum_{i=1}^N \bm{Y}_i(q) / N$ be the average of $\bm{Y}_i(q)$s. Let the average causal effect $\bm{\tau}(\bm{A})$ and its estimator $\widehat{\bm{\tau}}(\bm{A})$ be $\bm{\tau}(\bm{A})
			=
			\sum_{q=1}^Q \bm{A}_q \overline{\bm{Y}}(q)$, $	\widehat{ \bm{\tau} }(\bm{A})
			=
			\sum_{q=1}^Q \bm{A}_q \widehat{\overline{\bm{Y}}}(q) $, $
			\widehat{\overline{\bm{Y}}}(q) 
			=
			\sum_{i : L_i = q} \bm{Y}_i(q)/n_q$. Let $\bm{S}_{qr}$ $(q,r=1,\ldots, Q, q \neq r)$ be $\bm{S}_{qr}
	=
	\sum_{i=1}^N \big\{ \bm{Y}_i(q) - \overline{\bm{Y}}(q) \big\} \big\{ \bm{Y}_i(r) - \overline{\bm{Y}}(r) \big\}\T/(N-1) $ and $	\bm{S}_q^2 = \bm{S}_{qq}$. Suppose that, for any $1 \leq q \neq r \leq Q$, (a) $\bm{S}_q^2$ and $\bm{S}_{qr}$ have limiting values, (b) $n_q/N$has positive limiting value, and (c) $\max_{1\leq q \leq Q} \max_{1 \leq i \leq N} \| \bm{Y}_i(q) - \overline{\bm{Y}}(q) \|_2^2/N\rightarrow 0$. Then, $N \VAR \big\{ \widehat{\bm{\tau}}(\bm{A}) \big\} $ has a limiting value and $
			\sqrt{N} \big\{ \widehat{\bm{\tau}}(\bm{A}) - \bm{\tau} (\bm{A}) \big\} 
			\stackrel{D}{\rightarrow } 
			N \big( 0 , \lim_{N \rightarrow \infty} N \VAR \big\{ \widehat{\bm{\tau}}(\bm{A}) \big\} \big)$.
	\end{manualtheorem}		
	Our framework corresponds to the case of $Q=2$ with indices $q=1,2$ where $q=1$ and $q=2$ means treatment $(\ivp=1)$ and control $(\ivp=0)$, respectively.  Thus, $n_1$, $n_2$, and $N$ in Theorem \ref{thm:Ding5} of \citet{LiDing2017} correspond to our $\tNC$, $\NC - \tNC$, and $\NC$, respectively, and condition (b) of Theorem \ref{thm:Ding5} is satisfied from Assumption \ref{assp:3-1}-(i).
	
	We only prove the result about $\widehat{\B}$ because $\widehat{\tau}_\ITT$ is a special case of $\widehat{\B}$ where $\bX_{ji}$ is chosen as $1$ leading $\X_j = (\NC/\NIT) \sum_{i=1}^{\NI_j} \bX_{ji} \bX_{ji}\T =  (\NC/\NIT) \NI_j = \mathcal{N}_j $. Let $\B_\ivp^*
	=
	\big\{ \mo \big( \bm{\X} \big) \big\}^{-1} \mo \big\{ \potY{ \bm{\Y}_X }{\ivp} \big\}$.
From Assumption \ref{assp:3-1}-(ii) and (iii), $\| \B_1^*\|$ and $\|\B_0^*\|$ are finite. 
Also, $\B_1^*$ and $\B_0^*$ solve the following equations: $
	\sum_j \big( \potY{ \Y_{X, j} }{\ivp} - \X_j \B_\ivp^* \big)
	=
	\sum_j \potY{ \mathcal{R}_{X, j} }{\ivp}  = 0$. We take $\bm{Y}_i(1)$ and $\bm{Y}_i(2)$ in Theorem \ref{thm:Ding5} of \citet{LiDing2017} as $\potY{ \mathcal{R}_{X,j}}{1}$ and 0, respectively. As a consequence, $\big\{ \overline{\bm{Y}}(1), \overline{\bm{Y}}(2), \bm{S}_1^2, \bm{S}_2^2, \bm{S}_{12} \big\} $ correspond to $\big\{0, 0, \vo \big\{ \potY{\bm{ \mathcal{R} }_X}{1} \big\}, 0, 0 \big\}$ and conditions (b) and (c) of Theorem \ref{thm:Ding5} of \citet{LiDing2017} are satisfied under Assumption \ref{assp:3-1}. As a result, ${	\sqrt{\NC} \big\{ \sum_j \iv_j \potY{ \mathcal{R}_{X,j}  }{1} / \tNC \big\}
	\stackrel{D}{\rightarrow}
	N \big( 0 , \lim_{\NC \rightarrow \infty} \NC \vo\big\{ \potY{\bm{ \mathcal{R} }_X}{1} \big\} \big)}$. That is, $\tNC^{-1} \sum_j \iv_j \potY{ \mathcal{R}_{X,j}  }{1} = O_P(\NC^{-1/2})$ and, similarly, $(\NC - \tNC)^{-1} \sum_j (1-\iv_j) \potY{ \mathcal{R}_{X,j}  }{0} = O_P(\NC^{-1/2})$.

Next, we take $\bm{Y}_i(1)$ and $\bm{Y}_i(2)$ in Theorem \ref{thm:Ding5} of \citet{LiDing2017} as $\textsf{vec}(\X_j)$ and 0, respectively. As a consequence, $\big\{ \overline{\bm{Y}}(1), \overline{\bm{Y}}(2) ,\bm{S}_1^2, \bm{S}_2^2 , \bm{S}_{12} \big\}$ correspond to $\big\{ \textsf{vec}\big\{ \mo ( \bm{\X}) \big\} , 0 , \vo\big\{ \potY{\bm{ \mathcal{R} }_X}{1} \big\}, 0 , 0 \big\}$ and conditions (b) and (c) of Theorem \ref{thm:Ding5} of \citet{LiDing2017} are satisfied under Assumption \ref{assp:3-1}. As a result, $ \sqrt{\NC}
	\big[  \sum_j \iv_j \textsf{vec}(\X_j) / \tNC - \textsf{vec}\big\{ \mo ( \bm{\X}) \big\}
	\big]
	\stackrel{D}{\rightarrow}
	N \big( 0 , \lim_{\NC \rightarrow \infty} \NC \vo ( \bm{\X} ) \big)$. 
This implies $
	\widehat{\mo}_1( \bm{\X} ) =  \sum_j \iv_j \X_j / \tNC  = \mo(\bm{\X})+O_P(\NC^{-1/2})$. Since $\widehat{\mo}_1( \bm{\X} )$ is used in the denominator of $\widehat{\B}$, we study the gap between $\widehat{\mo}_1 ( \bm{\X} )$ and $\mo( \bm{\X} )$ which is given as follows.
\begin{align*}	
	\Big\| \Big\{ \widehat{\mo}_1( \bm{\X} ) \Big\}^{-1} - \Big\{ \mo(\bm{\X}) \Big\}^{-1} \Big\|_2
	&
	\leq \Big\| \Big\{ \widehat{\mo}_1( \bm{\X} ) \Big\}^{-1} \Big\|_2
	\Big\|  \widehat{\mo}_1( \bm{\X} )  - \mo(\bm{\X}) \Big\|_2
	\Big\| \Big\{ \mo(\bm{\X}) \Big\}^{-1}  \Big\|_2
	\\
	& 
	\leq \bigg[ \frac{1}{ \lambda_{\min} \big\{  \mo(\bm{\X}) \big\} }+ o(1) \bigg] \cdot O_P(\NC^{-1/2}) \bigg[ \frac{1}{\lambda_{\min} \big\{  \mo(\bm{\X}) \big\} }  \bigg]
	= O_P(\NC^{-1/2})\ .
\end{align*}
The first inequality is from the property of the matrix norm. The second equality is from $\widehat{\mo}_1( \bm{\X} ) - \mo(\bm{\X}) = O_P(\NC^{-1/2})$ and the full-rank $\mo(\bm{\X})$ where $\lambda_{\min} \big( \bm{A} \big)$ is the minimum eigenvalue of matrix $\bm{A}$. The last equality is trivial from $1/\lambda_{\min} \big\{  \mo(\bm{\X}) \big\} < \infty$. Thus, we obtain $\big\{ \widehat{\mo}_\ivp( \bm{\X} ) \big\}^{-1} - \big\{ \mo(\bm{\X}) \big\}^{-1} = O_P(\NC^{-1/2})$. Combining the above results, we obtain $\widehat{\B} - \B^* =
	\big\{ \mo( \bm{\X} ) \big\}^{-1}
	\big[
		\widehat{\mo}_1 \big\{  \potY{ \bm{\mathcal{R}}_X }{1} \big\} 
	-	\widehat{\mo}_0 \big\{  \potY{ \bm{\mathcal{R}}_X }{0} \big\} 
	\big] + O_P(\NC^{-1}) $.

Lastly, we take $\bm{Y}_i(1)$ and $\bm{Y}_i(2)$ in Theorem \ref{thm:Ding5} of \citet{LiDing2017} as $\potY{ \mathcal{R}_{X,j}}{1}$ and $\potY{ \mathcal{R}_{X,j}}{0}$, respectively, and $ \big\{ \overline{\bm{Y}}(1), \overline{\bm{Y}}(2), \bm{S}_1^2, \bm{S}_2^2, \bm{S}_{12} \big\}$ correspond to $\big\{ 0, 0, \vo\big\{ \potY{\bm{ \mathcal{R} }_X}{1} \big\},\vo\big\{ \potY{\bm{ \mathcal{R} }_X}{0} \big\}, \vo\big\{ \potY{\bm{ \mathcal{R} }_X}{1} , \potY{\bm{ \mathcal{R} }_X}{0} \big\} \big\}$ and conditions (b) and (c) of Theorem \ref{thm:Ding5} of \citet{LiDing2017} are satisfied under Assumption \ref{assp:3-1}. As a result,  $\sqrt{\NC}
	\big[
		\widehat{\mo}_1 \big\{  \potY{ \bm{\mathcal{R}}_X }{1} \big\} 
	-	\widehat{\mo}_0 \big\{  \potY{ \bm{\mathcal{R}}_X }{0} \big\} 
	\big]
	\stackrel{D}{\rightarrow}
	N \big( 0 , \lim_{\NC \rightarrow \infty}\NC \cdot \VAR \big[ \widehat{\mo}_1 \big\{  \potY{ \bm{\mathcal{R}}_X }{1} \big\} 
	-	\widehat{\mo}_0 \big\{  \potY{ \bm{\mathcal{R}}_X }{0} \big\} \, \big| \, \mathcal{F}_\NC, \mathcal{\iv}_\NC \big] \big) $.
Using Lemma \ref{lem:301} and the relationship between the covariance operator $\mathcal{V}$ and the centering matrix $\Pi_\NC$, the variance of $\widehat{\mo}_1 \big\{  \potY{ \bm{\mathcal{R}}_X }{1} \big\} 
	-	\widehat{\mo}_0 \big\{  \potY{ \bm{\mathcal{R}}_X }{0} \big\}$ is $\VAR \big[ \widehat{\mo}_1  \big\{  \potY{ \bm{\mathcal{R}}_X }{1} \big\} 
	-	\widehat{\mo}_0 \big\{  \potY{ \bm{\mathcal{R}}_X }{0} \big\} \, \big| \, \mathcal{F}_\NC, \mathcal{\iv}_\NC \big] = \mathcal{V} \big\{ \potY{\bm{\mathcal{R}}_X}{1} \big\} / \tNC
	+
		 \mathcal{V} \big\{ \potY{\bm{\mathcal{R}}_X}{0} \big\} /   (\NC - \tNC)
	-
	\mathcal{V} \big\{ \potY{\bm{\mathcal{R}}_X}{1} - \potY{\bm{\mathcal{R}}_X}{0} \big\}/\NC$.
As a result, the asymptotic Normality of $\widehat{\B}$ is obtained from the Slutsky's theorem:
\begin{align*}
	\sqrt{\NC} \big( \widehat{\B} - \B^* \big)
		\stackrel{D}{\rightarrow}
		N \Big( 0 , \lim_{\NC \rightarrow \infty} \NC V_{\B} \Big)
		\ , \
		V_{\B}
	 &=
	\Big\{ \mo \big( \bm{\X} \big)  \Big\}^{-1} 
	\bigg[
	 \frac{\mathcal{V} \big\{ \potY{\bm{\mathcal{R}}_X}{1} \big\}}{\tNC} 
	 +
	 \frac{ \mathcal{V} \big\{ \potY{\bm{\mathcal{R}}_X}{0} \big\}}{\NC - \tNC}
	 -
	 \frac{\mathcal{V}  \big\{ \potY{\bm{\mathcal{R}}_X}{1} - \potY{\bm{\mathcal{R}}_X}{0} \big\}}{\NC} 
	 \bigg] 
	 \Big\{ \mo \big( \bm{\X} \big)  \Big\}^{-1} \ .
\end{align*}

To show that the variance estimator $\Sigma_\beta$ is conservative, we use a result from Proposition 3 of \citet{LiDing2017} which is formally stated as follows.

\begin{manualprop}{3}\citep{LiDing2017}				\label{thm:Ding3}
	Let $\bm{s}_q^2$ be $
		\bm{s}_q^2
		=  \sum_{i: L_i = q} 
		\big\{ \bm{Y}_i - 	\widehat{\overline{\bm{Y}}}(q) \big\} \big\{ \bm{Y}_i - \widehat{\overline{\bm{Y}}}(q) \big\} \T / (n_q - 1)$. Under the regularity conditions in Theorem \ref{thm:Ding5}, $\bm{s}_q^2 - \bm{S}_q^2 \stackrel{P}{\rightarrow} 0$ for each $1 \leq q \leq Q$.
\end{manualprop}		
Under our notation, we have $\bm{s}_\ivp =  \widehat{\mathcal{V}}_\ivp(  \bm{ \mathcal{R}}_X )$. Therefore, from Proposition \ref{thm:Ding3}, we have $\bm{s}_\ivp^2 - \bm{S}_\ivp^2 = \widehat{\mathcal{V}}_\ivp(  \bm{ \mathcal{R}}_X ) - \vo\big\{ \potY{\bm{ \mathcal{R} }_X}{1} \big\} = o_P(1)$. Combining these results with $\big\{ \widehat{\mo}_\ivp( \bm{\X} ) \big\}^{-1} - \big\{ \mo(\bm{\X}) \big\}^{-1} = O_P(\NC^{-1/2})$, we obtain $\widehat{\Sigma}_{\B}
	= \NC
	\big\{ \mo \big( \bm{\X} \big)  \big\}^{-1} 
	\big[ \tNC^{-1} \mathcal{V} \big\{ \potY{\bm{\mathcal{R}}_{X}}{1} \big\}
	 + (\NC - \tNC)^{-1}\mathcal{V} \big\{ \potY{\bm{\mathcal{R}}_{X}}{0} \big\}
	 \big] 
	 \big\{ \mo \big( \bm{\X} \big)  \big\}^{-1} + o_P(1)$ and $
	\widehat{\Sigma}_{\B} - \Sigma_{\B}
	=
	\big\{ \mo \big( \bm{\X} \big)  \big\}^{-1} \mathcal{V}  \big\{ \potY{\bm{\mathcal{R}}_{X}}{1} - \potY{\bm{\mathcal{R}}_{X}}{0} \big\}
	 \big\{ \mo \big( \bm{\X} \big)  \big\}^{-1} + o_P(1)$. The probability limit of the right hand side is positive semi-definite, i.e. $\widehat{\Sigma}_{\B}$ is conservative. The results related to $\widehat{\tau}_\ITT$ can be similarly shown by replacing $\{ \bX_{ji},\X_j,\bm{\X}\}$ with $\{1,\mathcal{N}_j ,1\}$, respectively.

\subsection{Proof of Theorem \ref{thm:402} in the main Paper}

For $t \in \{ \NT,\AT,\CO\}$, there is no additional constraints about $\{ \potY{\TP_\NT}{1}, \ldots, \potY{\FN_\CO}{0} \}$ other than the constraints in the linear program in \eqref{eq:LP-original}. Let $v$ be a value that is smaller than ${\rm LB}_{\BrandomC,t}$, the solution to the minimization problem of $\tau_t$. Suppose that there exist $\{ \potY{\TP_\NT}{1}, \ldots, \potY{\FN_\CO}{0} \}$ that satisfy $v = \{ \potY{\TP_\NT}{1} + \potY{\FN_\NT}{1} - \potY{\TP_\NT}{0} - \potY{\FN_\NT}{0} \} / \NIT_t$ and satisfy all restrictions in the linear program in \eqref{eq:LP-original}. Then, ${\rm LB}_{\BrandomC,t}$ is not the solution to the minimization problem of $\tau_t$, which is a contradiction. Moreover, since ${\rm LB}_{\BrandomC,t}$ is the solution to the minimization problem of $\tau_t$, there exist $\{ \potY{\TP_\NT}{1}, \ldots, \potY{\FN_\CO}{0} \}$ that satisfy ${\rm LB}_{\BrandomC,t} = \{ \potY{\TP_\NT}{1} + \potY{\FN_\NT}{1} - \potY{\TP_\NT}{0} - \potY{\FN_\NT}{0} \} / \NIT_t$ and satisfy all restrictions in the linear program in \eqref{eq:LP-original}. That is, any value larger than ${\rm LB}_{\BrandomC,t}$ cannot be a lower bound for $\tau_t$ because ${\rm LB}_{\BrandomC,t}$ is feasible for some $\{ \potY{\TP_\NT}{1}, \ldots, \potY{\FN_\CO}{0} \}$. This shows the sharpness of ${\rm LB}_{\BrandomC,t}$. The sharpness of ${\rm UB}_{\BrandomC,t}$ can be shown in a similar manner.

	\subsection{Proof of Theorem \ref{thm:403} in the main Paper}						\label{sec:Ourconsistency}

We first establish consistency of $\widehat{\NIT}_t$, $\potY{\widehat{S}}{\ivp}$, $\potY{\widehat{S}_\NT}{1}$, $\potY{\widehat{S}_\AT}{0}$, $\potY{\widehat{S}_{\BrandomC, t}}{\ivp}$, and $\widehat{R}_t$. From Lemma \ref{lemma:WLLN} and the continuous mapping theorem, consistency of the estimators that does not use the classifiers is established. For example, we obtain
\begin{align*}
\frac{\widehat{\NIT}_\NT}{\NIT}
= \frac{ \frac{\NC}{\NIT \tNC} \sumji \ind(\iv_j = 1) \NT_{ji} }{ \frac{\NC}{\NIT \tNC}\sum_j  \ind(\iv_j=1) \NI_j  }
\stackrel{P}{\rightarrow}
\frac{  \sumji  \NT_{ji} }{ \sum_j  \NI_j  }
=
\frac{\NIT_\NT}{\NIT} \ .
\end{align*}
We find the following consistency for other estimators by a similar manner.
\begin{align}										\label{eq-consistency1}
	\frac{\widehat{\NIT}_t}{\NIT} \stackrel{P}{\rightarrow} \frac{\NIT_t}{\NIT}
	\quad , \quad
	\frac{\potY{\widehat{S}}{\ivp}}{\NIT} \stackrel{P}{\rightarrow} \frac{\potY{S}{\ivp}}{\NIT}
	\quad , \quad
	\frac{\potY{\widehat{S}_\NT}{1}}{\NIT} \stackrel{P}{\rightarrow} \frac{\potY{S_\NT}{1}}{\NIT} 
	\quad , \quad
	\frac{\potY{\widehat{S}_\AT}{0}}{\NIT} \stackrel{P}{\rightarrow} \frac{\potY{S_\AT}{0}}{\NIT} \ .
\end{align}
Next we study consistency of the classifier-based estimators. Since consistency of other estimators can be shown in a similar manner, we only study ${\potY{ \widehat{S}_{\BrandomC,\NT}}{\ivp}}/{ \NIT }$ that is represented as
\begin{align}								\label{eq-420}
	\frac{\potY{ \widehat{S}_{\BrandomC,\NT}}{\ivp}}{ \NIT }
	& =
	\frac{\widehat{\NIT}_\NT}{\NIT}
	\frac{\sumji \ind(\iv_j=1) Y_{ji} \ind \big\{ \randomf_\NT (\bX_{ji} \con \widehat{\theta}_\NT) \geq \widehat{q}_\NT \big\}  }{ \sumji \ind (\iv_j = 1)  \ind \big\{ \randomf_\NT (\bX_{ji} \con \widehat{\theta}_\NT) \geq \widehat{q}_\NT \big\}  }
 \ .
\end{align}
The numerator of \eqref{eq-420} is
\begin{align}											\label{eq-423}
	 \sumji \iv_j \potY{Y_{ji}}{1} \ind \big\{ \randomf_\NT (\bX_{ji} \con \widehat{\theta}_\NT) \geq \widehat{q}_\NT \big\} 
	 & =
	  \frac{\NC}{\NIT \tNC} \sumji \iv_j \potY{Y_{ji} }{1} \Big[ \ind \big\{ \randomf_\NT (\bX_{ji} \con \widehat{\theta}_\NT) \geq \widehat{q}_\NT \big\}   
	  - \ind \big\{ \randomf_\NT (\bX_{ji} \con \theta_\NT^*) \geq q_\NT^* \big\}  \Big]
	  \nonumber
	  \\
	  & \hspace*{2cm}
	  +
	  \frac{\NC}{\NIT \tNC} \sumji \iv_j \potY{Y_{ji} }{1} \ind \big\{ \randomf_\NT (\bX_{ji} \con \theta_\NT^*) \geq q_\NT^* \big\} 
	   \ .
\end{align}
We study the right-hand side of \eqref{eq-423}. An upper bound of the first term of  \eqref{eq-423} is
	\begin{align}									\label{eq-424}
		\bigg|  \frac{\NC}{\NIT \tNC} \sumji & \iv_j \potY{Y_{ji} }{1} \Big[ \ind \big\{ \randomf_\NT (\bX_{ji} \con \widehat{\theta}_\NT) \geq \widehat{q}_\NT \big\}   -  \ind \big\{ \randomf_\NT (\bX_{ji} \con \theta_\NT^*) \geq q_\NT^* \big\}   \Big] \bigg| 
		\nonumber
		\\
		& \leq
		\frac{\NC}{\NIT \tNC}  \sumji \Big| \ind \big\{ \randomf_\NT (\bX_{ji} \con \widehat{\theta}_\NT) \geq \widehat{q}_\NT \big\}   -  \ind \big\{ \randomf_\NT (\bX_{ji} \con \theta_\NT^*) \geq q_\NT^* \big\}   \Big| \ .
	\end{align}
	For a fixed constant $0<r<1/4$, we decompose the summand of \eqref{eq-424} as follows.
	{\fontsize{10}{12} \selectfont
	\begin{align}							\label{eq-425}
	& \Big|    \ind \big\{ \randomf_\NT (\bX_{ji} \con \widehat{\theta}_\NT) \geq \widehat{q}_\NT \big\}    -  \ind \big\{ \randomf_\NT (\bX_{ji} \con \theta_\NT^*) \geq q_\NT^* \big\}  \Big|
	\nonumber
	\\
	& \leq 
	\ind \Big[
		\big|  \randomf_\NT (\bX_{ji} \con \theta_\NT^*) - q_\NT^*    \big| 
		\leq
		\big|	 \big\{  \randomf_\NT (\bX_{ji} \con \theta_\NT^*) - q_\NT^*  \big\} - \big\{ \randomf_\NT (\bX_{ji} \con \widehat{\theta}_\NT) - \widehat{q}_\NT \big\}	\big|
	\Big]
	\nonumber
	\\
	& =
	\ind \Big[
		\big| \randomf_\NT (\bX_{ji} \con \theta_\NT^*) - q_\NT^* \big| 
		\leq
		\big|	 \big\{ f_\NT (\bX_{ji} \con \theta_\NT^*) - q_\NT^*  \big\} - \big\{ \randomf_\NT (\bX_{ji} \con \widehat{\theta}_\NT) - \widehat{q}_\NT \big\}	\big|
	\Big] 
	\nonumber
	\\
	& \hspace*{3cm}
	\times \ind \Big[
		\big|	 \big\{ f_\NT (\bX_{ji} \con \theta_\NT^*) - q_\NT^*  \big\} - \big\{ \randomf_\NT (\bX_{ji} \con \widehat{\theta}_\NT) - \widehat{q}_\NT \big\}	\big|
		\leq
		\NC^{-r/2}
	\Big] 
	\nonumber
	\\
	& \hspace*{1cm} +
	\ind \Big[
		\big| \randomf_\NT (\bX_{ij} \con \theta_\NT^*) - q_\NT^* \big| 
		\leq
		\big|	 \big\{ f_\NT (\bX_{ji} \con \theta_\NT^*) - q_\NT^*  \big\} - \big\{ \randomf_\NT (\bX_{ji} \con \widehat{\theta}_\NT) - \widehat{q}_\NT \big\}	\big|
	\Big] 
	\nonumber
	\\
	& \hspace*{3cm}
	\times \ind \Big[
		\big|	 \big\{ f_\NT (\bX_{ji} \con \theta_\NT^*) - q_\NT^*  \big\} - \big\{ \randomf_\NT (\bX_{ji} \con \widehat{\theta}_\NT) - \widehat{q}_\NT \big\}	\big|
		>
		\NC^{-r/2} \log \NC
	\Big] 
	\nonumber
	\\
	& \leq 
	\ind \Big[
		\big| \randomf_\NT (\bX_{ji} \con \theta_\NT^*) - q_\NT^* \big| 
		\leq
		\NC^{-r/2} \log \NC
	\Big]
	+
	\ind \Big[
		( K \log \NC + 1) \| \widehat{\eta}_\NT - \eta_\NT^*\|
		>
		\NC^{-r/2} \log \NC
	\Big] \ .
	\end{align}}%
	The inequality in the second line is from Lemma 1 of \citet{Kennedy2020}; for any $a$ and $b$ in $\R$, we have $ | \ind (a \geq 0 ) - \ind (b \geq 0) | \leq \ind \{ |a| \leq |a-b| \}$. 
	The equality in the third line is from $\randomf_\NT (\bX_{ji} \con \theta_\NT) = f_\NT (\bX_{ji} \con \theta) + e_{\NT ji}$ and $\ind(E) = \ind(E) \ind(F) + \ind(E) \ind(F^c)$ for two events $E$ and $F$. 
	The last inequality is from $\ind(E) \leq \ind(F)$ if $E \subset F$ and $\ind(E) \leq 1$ for all $E$, and  $
		\big|	 \big\{ f_\NT (\bX_{ji} \con \theta_\NT^*) - q_\NT^*  \big\} - \big\{ \randomf_\NT (\bX_{ji} \con \widehat{\theta}_\NT) - \widehat{q} \big\}	\big|		\leq
		( K \log \NC + 1) \| \widehat{\eta}_\NT - \eta_\NT^*\|$ obtained from \hyperlink{(GC4)}{(GC4)} with a generic constant $K$. Combining \eqref{eq-424} and \eqref{eq-425}, we obtain an upper bound of \eqref{eq-424}:
	\begin{align}													\label{eq-426}
		&
		\frac{\NC}{\NIT \tNC}  \sumji \Big|  \ind \big\{ \randomf_\NT (\bX_{ji} \con \widehat{\theta}_\NT) \geq \widehat{q}_\NT \big\}    -  \ind \big\{ \randomf_\NT (\bX_{ji} \con \theta_\NT^*) \geq q_\NT^* \big\}  \Big|
		\\
		&
		\leq
		\frac{\NC}{\NIT \tNC}  \sumji 
		\ind \Big[
		\big| \randomf_\NT (\bX_{ji} \con \theta_\NT^*) - q_\NT^* \big| 
		\leq
		\NC^{-r/2} \log \NC
	\Big]
	+
	\frac{\NC}{\NIT \tNC}  \sumji 
	\ind \Big[
		( K \log \NC + 1) \| \widehat{\eta}_t - \eta_t^*\|
		>
		\NC^{-r/2} \log \NC
	\Big] \ .
	\nonumber
	\end{align}
	The first term in \eqref{eq-426} is $O\big( \NC^{-r/2}  \log \NC \big)=o(1)$ from \hyperlink{(GC7)}{(GC7)}.
	To study the second term in \eqref{eq-426}, we find the rate of $\mu = \EXP \big[ \ind \big\{ ( K \log \NC + 1) \| \widehat{\eta}_\NT - \eta_\NT^*\|	>	\NC^{-r/2} \log \NC \big\} \big]$:
	\begin{align*}
		\mu 
	& =
	P \big[ 
		( K \log \NC + 1) \| \widehat{\eta}_\NT - \eta_\NT^*\|
		>
		\NC^{-r/2} \log \NC
		 \, \big| \, \extendF_\NC , \mathcal{\iv}_\NC \big]
	\leq 
	P \big[ 
		\| \widehat{\eta}_\NT  - \eta_\NT^*\|
		>
		\epsilon' \NC^{-r}
		 \, \big| \, \extendF_\NC , \mathcal{\iv}_\NC \big] 
	= o(1)
	\end{align*}
	where $\displaystyle{\epsilon' = \inf_{\NC =1,2,\ldots } \big\{ \NC^{r/2} \log \NC/(K \log \NC + 1) \big\} > 0 } $. Therefore, the inequality in the third line is straightforward from the definition of $\epsilon'$. The asymptotic result in the last line is from Lemma \ref{lmm:401}. Moreover, the variance of $\ind \big[
		( K \log \NC + 1) \| \widehat{\eta}_\NT - \eta_\NT^*\|	>	\NC^{-r/2} \log\NC \big] $ is bounded above by 2. Thus, by the law of large number, the second term  in \eqref{eq-426} is $o_P(1)$. 
	\begin{align*}
	\frac{\NC}{\tNC} \frac{1}{\NIT}  \sumji 
		\ind \Big[
		( K \log \NC + 1) \| \widehat{\eta}_\NT - \eta_\NT^*\|
		>
		\NC^{-r/2} \log \NC
	\Big] 
	= O(1) \big( \mu + o_P(1) \big)
	= o_P(1) \ .
	\end{align*}
	This concludes  \eqref{eq-426} is $o_P(1)$ and, as a result, the first term of \eqref{eq-423} is $o_P(1)$ from \eqref{eq-424}.

	From Lemma \ref{lemma:WLLN} with $\mathcal{S}_j  = \sum_{i=1}^{\NI_j} \potY{Y_{ji}}{1}  \ind \big\{ \randomf_\NT (\bX_{ji} \con \theta_\NT^*) \geq q_\NT^* \big\}$, we have
	\begin{align*}
		\frac{\NC}{\NIT \tNC} \sumji \iv_j \potY{Y_{ji} }{1} \ind \big\{ \randomf_\NT (\bX_{ji} \con \theta_\NT^*) \geq q_\NT^* \big\} 
		\stackrel{P}{\rightarrow} 
	 \frac{1}{\NIT} \sumji \potY{Y_{ji}}{1} \ind \big\{ \randomf_\NT (\bX_{ji} \con \theta_\NT^*) \geq q_\NT^* \big\}  \ .
	\end{align*}
	Combining the above results, the  limits of \eqref{eq-423} and the denominator of \eqref{eq-420} are
	\begin{align}												
		&
		 \frac{\NC}{\NIT \tNC} \sumji\iv_j \potY{Y_{ji}}{1}\ind \big\{ \randomf_\NT (\bX_{ji} \con \widehat{\theta}_\NT) \geq \widehat{q}_\NT \big\}  
	\stackrel{P}{\rightarrow}
	\frac{1}{\NIT} \sumji \potY{Y_{ji}}{1} \ind \big\{ \randomf_\NT (\bX_{ji} \con \theta_\NT^*) \geq q_\NT^* \big\} \ , \label{eq-427}
	\\
	&
	\frac{\NC}{\NIT \tNC} \sumji\iv_j \ind \big\{ \randomf_\NT (\bX_{ji} \con \widehat{\theta}_\NT) \geq \widehat{q}_\NT \big\}  
	\stackrel{P}{\rightarrow}
	\frac{1}{\NIT} \sumji \ind \big\{ \randomf_\NT (\bX_{ji} \con \theta_\NT^*) \geq q_\NT^* \big\} \ .
	\label{eq-428}
	\end{align}
	Therefore, from the continuous mapping theorem, \eqref{eq-427}, and \eqref{eq-428}, we find  \eqref{eq-420} converges to $\potY{ S_{\BrandomC,\NT} }{1}/\NIT_\NT$ and the similar results for other quantities:
	\begin{align}										\label{eq-consistency2}
			\frac{\potY{ \widehat{S}_{\BrandomC,\NT}}{1}}{ \widehat{\NIT}_\NT }
	\stackrel{P}{\rightarrow}
	\frac{\NIT_\NT}{\NIT}
	\frac{  \sumji \potY{Y_{ji}}{1} \ind \big\{ \randomf_\NT (\bX_{ji} \con \theta_\NT^*) \geq q_\NT^* \big\}  }{  \sumji \ind \big\{ \randomf_\NT (\bX_{ji} \con \theta_\NT^*) \geq q_\NT^* \big\}  }
	=
	\frac{ \potY{ S_{\BrandomC,\NT} }{1} }{\NIT}
	\ , \
		\frac{\potY{ \widehat{S}_{\BrandomC,t}}{\ivp}}{ \widehat{\NIT}_t }
	\stackrel{P}{\rightarrow}
	\frac{ \potY{ S_{\BrandomC,t} }{\ivp} }{\NIT} \ .
	\end{align}
	Consistency of the classification error rate estimators (${\widehat{R}_t }/{\NIT} \stackrel{P}{\rightarrow} {R_\NT }/{\NIT}$) is similarly shown.
	
	Next we consider the elastic program in \eqref{eq:LP-elastic}. We only discuss the maximization of $\tau_\NT$ to discuss consistency of $\widehat{{\rm UB}}_{\BrandomC,\NT}$, but consistency of other bound estimators can be shown in a similar manner. Let $\psi \in \R^{16}$ and $\vartheta_E \in \R^{74}$ be 
	\begin{align*}
		& \psi
		= ( \NIT_\NT, \NIT_\AT, \NIT_\CO, \potY{S}{1}, \potY{S}{0}, \potY{S_\NT}{1}, \potY{S_\AT}{0}, \potY{S_{C,\NT}}{1}, \potY{S_{C,\AT}}{1}, \potY{S_{C,\CO}}{1}, \potY{S_{C,\NT}}{0}, \potY{S_{C,\AT}}{0}, \potY{S_{C,\CO}}{0}, R_\NT, R_\AT, R_\CO) \T \ , \\
		&
		\vartheta_E = (\potY{\TP_\NT}{0}, \ldots, \potY{\FN_\CO}{1}, s_1,\ldots,s_{18}, a_{1},\ldots,a_{28}, b_{1}, \ldots, b_{10} ) \T \ .
	\end{align*}
	Here $s_j$s are non-negative slack variables that convert the inequality constraints to the equality constraints in \eqref{eq:LP-elastic}, e.g., $\potY{S_\NT}{1} + s_1 - \potY{S_\NT}{0} - a_{11} = 0$. The objective function are written as linear combinations as $\{ \potY{\TP_\NT}{1} + \potY{\FN_\NT}{1} - \potY{\TP_\NT}{0} - \potY{\FN_\NT}{0} - M \sum_{\ell=1}^{28} a_\ell- M \sum_{\ell=1}^{10} b_\ell \}/\NIT_\NT = {\mathfrak{c}_E \T \vartheta_E }/{\NIT_\NT} $.
	The linear constraints of the elastic programming is represented as $ \mathfrak{A}_E \vartheta_E = \mathfrak{b}$; here $\mathfrak{A}_E \in \R^{28 \times 74}$ and $\mathfrak{b} \in \R^{28}$ where $\mathfrak{A}_E$ is a full-rank constant matrix having $(0, \pm 1)$ and $\mathfrak{b}$ is a linear function of $\psi$, i.e. $\mathfrak{b} = \mathfrak{D} \psi$ for some matrix $\mathfrak{D} \in \R^{28 \times 16}$. Let $\widehat{\psi} = (\widehat{\NIT}_\NT, \ldots, \widehat{R}_\CO)\T$. From \eqref{eq-consistency1},  \eqref{eq-consistency2}, and consistency of ${\widehat{R}_\NT }/{\NIT}$, we have $	{\widehat{\psi}}/{\NIT} \stackrel{P}{\rightarrow} {\psi}/{\NIT}$.

	Let $\texttt{POS}(\mathfrak{A}) = \big\{ 	\mathfrak{A}\bx \cond \bx \geq 0 \big\}$, $\texttt{HOM}(\mathfrak{A}) = \big\{ 	\bx \cond \mathfrak{A}\bx=0 , \bx \geq 0 \big\}$, and $\texttt{HOM}(\mathfrak{A})^* = \big\{ 	\by \cond \by\T \bx \leq 0 , ^\forall \bx \in \texttt{HOM}(\mathfrak{A})  \big\}$. Then, we find $\mathfrak{c}_E \in \texttt{HOM}(\mathfrak{A}_E)^*$. For any $\bx \in \texttt{HOM}(\mathfrak{A}_E)$,  we have $ \potY{\TP_\NT}{0} + \potY{\FN_\NT}{0}= - a_3 + b_3$ and $ \potY{\TP_\NT}{1} + \potY{\FN_\NT}{1}= -a_6 + b_6$. Thus, $\mathfrak{c}_E\T \bx = (-M+1) a_3 - a_6 - M \sum_{t \neq 3} a_t - b_3 + (-M+1) b_6 - M \sum_{t\neq6} b_t$. Since $M$ is sufficiently large, we find $\mathfrak{c}_E\T \bx \leq 0$, which implies $\mathfrak{c}_E \in \texttt{HOM}(\mathfrak{A}_E)^*$. 
	
	Combining the aforementioned results, Theorem 4 and 5 of \citet{Ward1990} can be applied. Let $\phi_E(\mathfrak{b})$ be the maximum of $\mathfrak{c}_E\T \vartheta_E $ subject to $\mathfrak{A}_E \vartheta_E = \mathfrak{b}$ and $\vartheta_E \geq 0$.  $\phi_E(\mathfrak{b})$ is a piecewise linear, continuous, and concave function in $\mathfrak{b} \in \texttt{POS}(\mathfrak{A}_E) = \big\{ 	\mathfrak{A}_E\bx \cond \bx \geq 0 \big\}$. Moreover, $\texttt{POS}(\mathfrak{A}_E)$ is partitioned into closed polyhedral cones $\{ \mathcal{B}_1,\ldots,\mathcal{B}_K \}$ of which boundaries overlap but interiors do not overlap. For $\mathfrak{b} \in \mathcal{B}_k$, $\phi_E$ is a linear function in $\mathfrak{b}$ as $\phi_E(\mathfrak{b}) = \mathfrak{c}_{E,k} \T B_{E,k}^{-1} \mathfrak{b}$ for some $I_k$ where $I_k = \{ i_1,\ldots,i_{28} \cond i_j \in \{1,\ldots,74\} \}$ is a collection of indices that takes linearly independent 28 columns of $\mathfrak{A}_E$, $B_{E,k} \in \R^{28 \times 28}$ is a submatrix of $\mathfrak{A}_E$ of which 28 columns are chosen from $I_k$, and  $\mathfrak{c}_{E,k} \in \R^{28}$ is a subvector of $\mathfrak{c}_E$ chosen from $I_k$; see \citet{Ward1990} for details. This implies
	\begin{align}										\label{eq-ElastLP}
		\begin{matrix}
			\max \  \mathfrak{c}_E \T \vartheta_E/\NIT_\NT
			\text{ subject to } \Bigg\{
			\begin{matrix}
				\mathfrak{A}_E \vartheta_E = \mathfrak{b}\\[-0.2cm]
				\vartheta_E \geq 0	
			\end{matrix} 		
		\end{matrix}
		= \frac{\phi_E(\mathfrak{b}) }{\NIT_\NT} 
		=  \frac{\phi_E(\mathfrak{b}) /\NIT }{\NIT_\NT/\NIT}
		=  \frac{\phi_E(\mathfrak{b}/\NIT)  }{\NIT_\NT/\NIT}
		=  \frac{\phi_E(\mathfrak{D} \psi /\NIT)  }{\NIT_\NT/\NIT} \ .
	\end{align}
	The third equality holds from piecewise linear form of $\phi_E$. 
	
	Next we consider the original linear program where $\mathfrak{A}_O\in \R^{28 \times 36}$, $\vartheta_O, \mathfrak{c}_O \in \R^{36}$, $\mathfrak{c}_{O,k} \in \R^{28}$, $B_{O,k} \in \R^{28 \times 28}$, and $\phi_O$ are similarly defined without elastic variables $a$ and $b$s. We also find
	\begin{align}									\label{eq-OriginalLP}
		\begin{matrix}
			\max \  \mathfrak{c}_O \T \vartheta_O/\NIT_\NT
			\text{ subject to }\Bigg\{ \begin{matrix}
			 \mathfrak{A}_O \vartheta_O = \mathfrak{b} \\[-0.2cm] \vartheta_O \geq 0
			\end{matrix} 
		\end{matrix}
		= \frac{\phi_O(\mathfrak{b}) }{\NIT_\NT} 
		=  \frac{\phi_O(\mathfrak{b}) /\NIT }{\NIT_\NT/\NIT}
		=  \frac{\phi_O(\mathfrak{b}/\NIT)  }{\NIT_\NT/\NIT}
		=  \frac{\phi_O(\mathfrak{D} \psi /\NIT)  }{\NIT_\NT/\NIT} \ .
	\end{align}
	At true $\psi$ and $\mathfrak{b}$, we find the solutions to \eqref{eq-ElastLP} and \eqref{eq-OriginalLP} are the same because the original linear program \eqref{eq-OriginalLP} is feasible and $a_\ell$ and $b_\ell$s are zero in the elastic program in \eqref{eq-ElastLP}. Therefore, this implies $\phi_E(\mathfrak{b}) = \phi_O(\mathfrak{b})$ for $\mathfrak{b}$ that makes the original linear program feasible. If $\mathfrak{b}$ makes the original linear program infeasible, $I_k$ includes a subset of $\{ 37,\ldots,74 \}$, the indices corresponding to $\{a_1,\ldots,b_{10}\}$.

	Let $\widehat{\vartheta}_E$ be the solution vector that solves the elastic program using $\widehat{\mathfrak{b}}$, i.e. $\mathfrak{c}_E \T \widehat{\vartheta}_E = \max_{\vartheta_E} \  \mathfrak{c}_E \T \vartheta_E$
			$\text{ subject to } \mathfrak{A}_E \vartheta_E = \widehat{ \mathfrak{b} } \text{ and }  \vartheta_E \geq 0$. Then we find $\phi_E(\widehat{\mathfrak{b}}) = \mathfrak{c}_E \T \widehat{\vartheta}_E$.
	The approximated upper bound for $\tau_\NT$ is defined as
	\begin{align*}
		\widehat{{\rm UB}}_{\BrandomC, \NT}
		& = \frac{1}{\widehat{\NIT}_\NT} \bigg( 	\mathfrak{c}_E \T \widehat{\theta}_E 
		-
		M \sum_{\ell=1}^{28} \widehat{a}_\ell
		-
		M \sum_{\ell=1}^{10} \widehat{b}_\ell
		\bigg) 
		=
		\frac{1}{\widehat{\NIT}_\NT} \bigg( 	
		\mathfrak{c}_{E,k}\T B_{E,k}^{-1} \widehat{\mathfrak{b}}
		-
		\mathfrak{m}_{E,k}\T B_{E,k}^{-1} \widehat{\mathfrak{b}}
		\bigg)
		\text{ for } \mathfrak{b} \in \mathcal{B}_k
		\ .
	\end{align*}
	Here $\mathfrak{m}_{E,k,j} = 
			M \mathfrak{c}_{E,k,j}  \text{ if } j \text{ is associated with elastic variables}$ and $\mathfrak{m}_{E,k,j} = 
			0 \text{ otherwise}$. 	If $\widehat{a}_\ell$ and/or $\widehat{b}_\ell$ is zero, $\mathfrak{m}_{E,k}\T=0$ because all components of $\mathfrak{c}_{E,k}$ are not associated with elastic variables. Let $\nu(\mathfrak{b}): \texttt{POS}(\mathfrak{A}_E) \rightarrow \R$ be the function that satisfies  $\nu(\mathfrak{b}) = \mathfrak{m}_{E,k}\T B_{E,k}^{-1} \mathfrak{b}$ for $\mathfrak{b} \in \mathcal{B}_k$. Due to the construction, $\nu(\mathfrak{b})$ is piecewise linear in $\mathfrak{b}$. Moreover, since $\mathcal{B}_k$s are closed sets and $\nu(\mathfrak{b}$ is linear in $\mathfrak{b}$ over $\mathcal{B}_k$, $\nu(\mathfrak{b})$ is continuous in $\mathfrak{b}$ over $\texttt{POS}(\mathfrak{A}_E) =\bigcup_{k=1}^K \mathcal{B}_k$. For $\mathfrak{b} $ making the original linear program \eqref{eq-OriginalLP} feasible, we find $\mathfrak{m}_{E,k}=0$, i.e. $\nu(\mathfrak{b})=0$.
	
	Combining the above result, $\widehat{{\rm UB}}_{\BrandomC, \NT} = \big\{ \phi_E(\widehat{\mathfrak{b}}) - \nu(\widehat{\mathfrak{b}}) \big\}/\widehat{\NIT}_\NT$ and, from the continuous mapping theorem,
	\begin{align*}
		\widehat{{\rm UB}}_{\BrandomC, \NT}
		=  
		\frac{\phi_E(\mathfrak{D}\widehat{\psi}/\NIT) - \nu(\mathfrak{D}\widehat{\psi}/\NIT)}{\widehat{\NIT}_\NT/\NIT}
		\stackrel{P}{\rightarrow} 
		\frac{\phi_E(\mathfrak{D}\psi/\NIT) - \nu(\mathfrak{D}\psi/\NIT)}{\widehat{\NIT}_\NT/\NIT}
		=  \frac{\phi_O(\mathfrak{D} \psi /\NIT)  }{\NIT_\NT/\NIT}
		=
		{\rm UB}_{\BrandomC, \NT} \ .
	\end{align*}
	The last equality holds because $\mathfrak{D}\psi$ belongs to a feasible region of the original LP \eqref{eq-OriginalLP}.

\subsection{Proof of Theorem \ref{thm:consistencyLH}}								\label{sec:LHproof}

Taking $S_{ji} \in \{ 1 , \NT_{ji} , \AT_{ji} , \potY{Y_{ji}}{\ivp}, \potY{Y_{ji}}{1} \NT_{ji} , \potY{Y_{ji}}{0} \AT_{ji}\}$ in Lemma \ref{lemma:WLLN}, we obtain
{\fontsize{10}{12} \selectfont
\begin{align*}
	&
	\frac{\NC \sumji \ind(\iv_j =\ivp) \ind(\bW_{ji} = \bw)   }{\NIT \sum_j \ind(\iv_j=\ivp)} \stackrel{P}{\rightarrow}
\frac{\NIT(\bw) }{\NIT} \ , 
&&
\frac{\NC  \sumji \ind(\iv_j =\ivp)  \potY{Y_{ji}}{\ivp} \ind(\bW_{ji} = \bw)  }{\NIT \sum_j \ind(\iv_j=\ivp)} \stackrel{P}{\rightarrow}
\frac{\potY{S}{\ivp} (\bw) }{\NIT} \ , 
\\
	&
	\frac{\NC  \sumji \iv_j  \NT_{ji} \ind(\bW_{ji} = \bw) }{\NIT \tNC}  \stackrel{P}{\rightarrow}
\frac{\NIT_\NT (\bw) }{\NIT} \ , &&
	\frac{\NC \sumji (1-\iv_j)  \AT_{ji} \ind(\bW_{ji} = \bw) }{\NIT (\NC -\tNC)}   \stackrel{P}{\rightarrow}
\frac{\NIT_\AT (\bw) }{\NIT} \ , 
\\
	&
	\frac{\NC  \sumji \iv_j \potY{Y_{ji}}{1} \NT_{ji} \ind(\bW_{ji} = \bw)}{\NIT \tNC}  \stackrel{P}{\rightarrow}
\frac{\potY{S_\NT}{1} (\bw) }{\NIT} \ , 
&&
	\frac{\NC  \sumji \ind(1-\iv_j) \potY{Y_{ji}}{0} \AT_{ji} \ind(\bW_{ji} = \bw) }{\NIT (\NC -\tNC)}  \stackrel{P}{\rightarrow}
\frac{\potY{S_\AT}{0} (\bw) }{\NIT} \ .
\end{align*}	}%
Using the above result, we find $\widehat{\NIT}_t(\bw)/\NIT(\bw)$ are consistent (i.e. $\widehat{\NIT}_t(\bw)/\NIT(\bw) \stackrel{P}{\rightarrow} \NIT_t(\bw)/\NIT(\bw)$). 
Similarly, $\potY{S}{\ivp}(\bw)/ \NIT(\bw)$, $\potY{S_\NT}{1}(\bw)/ \NIT(\bw)$, and $\potY{S_\AT}{0}(\bw)/ \NIT(\bw)$ are consistent.
Therefore, from the continuous mapping theorem, estimators in \eqref{eq-LH-Est} are consistent and, as a consequence, the plug-in estimators $\widehat{{\rm LB}}_{W,t}$ and $\widehat{{\rm UB}}_{W,t}$ are consistent for ${\rm LB}_{W,t}$ and ${\rm UB}_{W,t}$, respectively.

\section{Proof of Lemmas in the Supplementary Material}							\label{sec:C}

\subsection{Proof of Lemma \ref{lem:401}} \label{sec:C-3}

 Let $F_1 = \{ ji \cond \randomf_\NT (\bX_{ji} \con \theta_\NT^*) \geq \randomf_{( \NIT_\text{nNT}+1)} \}$ and $F_2 = \{ ji \cond \randomf_\NT (\bX_{ji} \con \theta_\NT^*) \leq f_{( \NIT_\text{nNT})} \}$. Let $ H(q_\NT) =	\sumji \SI \big( \randomf_\NT (\bX_{ji} \con \theta_\NT^*) - q_\NT \big) $ which is continuous and strictly decreasing function in $q_\NT \in \R$.
 
First, we study the value $H(\randomf_{(\NIT_\text{nNT})} )$. The summand of $H(\randomf_{(\NIT_\text{nNT})} )$ for  $ji \in F_1$ is
\begin{align*}
	\SI \big( \randomf_\NT (\bX_{ji} \con \theta_\NT^*) - \randomf_{(\NIT_\text{nNT})}\big )
	& \geq
	1 - c \cdot \exp \bigg[ - \frac{1-2c}{2ch} \Big\{ \randomf_{(\NIT_\text{nNT}+1)} - \randomf_{(\NIT_\text{nNT})} - h \Big\}	\bigg]
\end{align*}
where we use the monotonicity of $\SI$ and the definition of $\SI$ and $\randomf_{(\NIT_\text{nNT}+1)} - \randomf_{(\NIT_\text{nNT})} - h \geq 2h - h =h $.  We find that the summand of $H(\randomf_{(\NIT_\text{nNT})} )$ for individual $ji$ in $F_2$ is $	\SI \big( \randomf_\NT (\bX_{ji} \con \theta_\NT^*) - \randomf_{(\NIT_\text{nNT})} \big)
	\geq
	c \cdot \ind\{ \randomf_\NT(\bX_{ji} \con \theta_\NT^*) = \randomf_{(\NIT_\text{nNT})} \} $ 
	where we use the positivity of $\SI$ and $\SI(-h) = c$.  Combining the above results, we can derive a lower bound of $H(\randomf_{(\NIT_\text{nNT})} )$ as follows.
\begin{align*}
	H(\randomf_{(\NIT_\text{nNT})} )
	& = \sum_{ji \in F_1} \SI \big( \randomf_\NT (\bX_{ji} \con \theta_\NT^*) - \randomf_{(\NIT_\text{nNT})} \big) + \sum_{ji \in F_2} \SI \big( \randomf_\NT(\bX_{ji} \con \theta_\NT^*) - \randomf_{(\NIT_\text{nNT})} \big)
	\\
	&\geq \sum_{ji \in F_1} \bigg[ 1 - c \exp \bigg[ - \frac{1-2c}{2ch} \Big\{ \randomf_{(\NIT_\text{nNT}+1)} - \randomf_{(\NIT_\text{nNT})} - h \Big\}	\bigg] \bigg] + \sum_{ji \in F_2} c  \ind\{ \randomf_\NT (\bX_{ji} \con \theta_\NT^*) = \randomf_{(\NIT_\text{nNT})} \} 
	\\
	& = \NIT_\NT - c \NIT_\NT\exp \bigg[ - \frac{1-2c}{2ch}\Big\{ \randomf_{(\NIT_\text{nNT}+1)} - \randomf_{(\NIT_\text{nNT})} - h \Big\} \bigg] + c \ .
\end{align*}
The equality in the first line is straightforward from the definition of $F_1$ and $F_2$. The inequality in the second line is from the above results. The equality in the last line is from $|F_1| = \NIT_\NT$ and the uniqueness of the randomized learner $\randomf_\NT(\bX_{ji} \con \theta_\NT^*)$. Subtracting $\NIT_\NT$ both hand sides, we find that
\begin{align}										\label{proof-401}
	H \big(\randomf_{(\NIT_\text{nNT})} \big) - \NIT_\NT
	& \geq
	c \bigg[
	1- 	\NIT_\NT\exp \bigg[ - \frac{1-2c}{2ch} \Big\{ \randomf_{(\NIT_\text{nNT}+1)} - \randomf_{(\NIT_\text{nNT})} - h \Big\}	\bigg]
	\bigg] 
	> 0 \ .
\end{align}
The second inequality is from the specific choice of $c$ and $h$.

Next we study the value $H(\randomf_{(\NIT_\text{nNT}+1)} )$. The summand of $H(\randomf_{(\NIT_\text{nNT}+1)})$ in $F_1$ becomes
\begin{align*}
	\SI \big( \randomf_\NT(\bX_{ji} \con \theta_\NT^*) - \randomf_{(\NIT_\text{nNT}+1)} \big)
	& \leq
	\ind \{ \randomf_\NT (\bX_{ji} \con \theta_\NT^*) \geq \randomf_{(\NIT_\text{nNT}+1)} \} - c \cdot\ind\{ \randomf_\NT(\bX_{ji} \con \theta_\NT^*)=\randomf_{(\NIT_\text{nNT}+1)} \} 
\end{align*}
where we use $\SI (t) \leq 1$ for all $t$ and $\SI(h) =1-c$.  The summand of $H(\randomf_{(\NIT_\text{nNT}+1)} )$ in $F_2$ is
\begin{align*}
	\SI \big( \randomf_\NT (\bX_{ji} \con \theta_\NT^*)- \randomf_{(\NIT_\text{nNT}+1)} \big)
	& \leq
	\SI \big( \randomf_{(\NIT_\text{nNT})} - \randomf_{(\NIT_\text{nNT}+1)} \big)
	= c \cdot \exp \bigg[ \frac{1-2c}{2ch} \Big\{ \randomf_{(\NIT_\text{nNT})} - \randomf_{(\NIT_\text{nNT}+1)} + h \Big\} 	\bigg] 
\end{align*}
from the monotonicity of $\SI$, the definition of $\SI$, and $ \randomf_{(\NIT_\text{nNT})} - \randomf_{(\NIT_\text{nNT}+1)} + h \leq  - h$.

Therefore, $H(\randomf_{(\NIT_\text{nNT}+1)} )$ is upper-bounded as follows by the similar reasons in the construction of the lower bound of $H(\randomf_{(\NIT_\text{nNT})} )$.
\begin{align*}
H( \randomf_{(\NIT_\text{nNT}+1)} )
	\leq \NIT_\NT-c +c \NIT_\nNT\exp \bigg[ \frac{1-2c}{2ch} \Big\{ \randomf_{(\NIT_\text{nNT})} - \randomf_{(\NIT_\text{nNT}+1)} + h \Big\}	\bigg] \ .
\end{align*}
Subtracting $\NIT_\NT$ both hand sides, we find that
\begin{align}										\label{proof-402}
	H \big(\randomf_{(\NIT_\text{nNT}+1)}\big) - \NIT_\NT 
	& \leq 
	-c \bigg[
	1- 	\NIT_\nNT \exp \bigg[ \frac{1-2c}{2ch} \Big\{ \randomf_{(\NIT_\text{nNT})} - \randomf_{(\NIT_\text{nNT}+1)} + h \Big\}	\bigg]
	\bigg] 
		< 0 \ .
\end{align}
The second inequality is from the specific choice of $c$ and $h$. 

Since $H(q_\NT)$ is continuous and strictly decreasing on $\R$, \eqref{proof-401} and \eqref{proof-402} imply that the unique solution to $H(q_\NT) - \NIT_\NT$ exists on the interval $( \randomf_{(\NIT_\text{nNT})} , \randomf_{(\NIT_\text{nNT}+1)} )$. 

\subsection{Proof of Lemma \ref{lem-EE12toGeneral}}								\label{sec:C4}

	Conditions \hyperlink{(GC1)}{(GC1)} and \hyperlink{(GC2)}{(GC2)} are exactly  the same as condition (i) and (ii) of Assumption \ref{assp:4-1}. Therefore, it is sufficient to show that conditions \hyperlink{(GC3)}{(GC3)}-\hyperlink{(GC8)}{(GC8)} hold which are given in the following Section \ref{sec:GC3}-\ref{sec:GC8}.
	
\subsubsection{Proof of Condition \protect\hyperlink{(GC3)}{(GC3)}}								\label{sec:GC3}
Under \hyperlink{(Linear)}{(Linear)} and  \hyperlink{(Logistic)}{(Logistic)}, we find the both quantities are invertible so long as $\lambda >0$.
\begin{align*}
	&
	 \sumji \nabla_\theta^2 L \big(t_{ji}, f_t(\bX_{ji} \con \theta_t) \big)
	 =
	\sumji \bX_{ji} \bX_{ji}\T
	=
	\frac{\NIT}{\NC} \mo(\bm{\bm{\X}})
	\ , \
	\\
	&
	 \sumji \nabla_\theta^2 L \big(t_{ji}, f_t(\bX_{ji} \con \theta_t) \big)
	 =
	 \sumji 
	 \frac{1+ \exp ( \bX_{ji}\T \theta_t )}{\{ 1+ \exp ( \bX_{ji}\T \theta_t ) \}^{2}} \big( \bX_{ji}\bX_{ji}\T \big) + \NIT \lambda I_p
\end{align*}

\subsubsection{Proof of Condition \protect\hyperlink{(GC4)}{(GC4)}}								\label{sec:GC4}
Under \hyperlink{(Linear)}{(Linear)}, we have upper bounds for $\| \psi_{t,j} (\eta) \|^2 = \Big\|
				\sum_{i=1}^{\NI_j} 
				\big\{
				\big( \bX_{ji}\T \theta_t - t_{ji} \big) \bX_{ji} 
				\big\}\big\|^2
				+
				\big[ \sum_{i=1}^{\NI_j} \big\{ t_{ji} - \SI \big( \randomf_t (\bX_{ji} \con \theta_t) - q_t \big) \big\} \big]^2		$ as $ \big\| \psi_{t,j} (\eta) \big\|^2
				 \leq 
				M^2 \big\{ (K \log \NC)^2 \|\theta_t\| + (K \log \NC) \big\}^2
				+
				(2M)^2  $.
	From Assumption \ref{assp:4-1}, $\| \bX_{ji} \| \leq K \log \NC$ for some constant $K$. Similarly, under \hyperlink{(Logistic)}{(Logistic)}, we have  $\big\| \psi_{t,j} (\eta) \big\|^2 \leq 
				M^2 \big\{ 2 (K \log \NC) + \lambda  \|\theta_t\| \big\}^2
				+
				(2M)^2 $.
		This implies that $\| \psi_{t,j} (\eta_k) \| / (\log \NC)^2 < \infty$ for all $j=1,\ldots, \NC$ for both \hyperlink{(Linear)}{(Linear)} and \hyperlink{(Logistic)}{(Logistic)}, i.e. the first result of \hyperlink{(GC4)}{(GC4)} hold with $v_1=2$. 
		
		Next we find the second result holds for $t \in \{\NT, \AT\}$ under \hyperlink{(Linear)}{(Linear)} as follows.
		\begin{align}									\label{eq-averagerateLinear}
		\big\| f_t(\bX_{ji} \con \theta_1) -  f_t(\bX_{ji} \con \theta_2) \big\| 
		=
		\big\|
			\bX_{ji} ( \theta_1 - \theta_2) 
		\big\|
		\leq \big\| \bX_{ji} \big\| \big\| \theta_1 - \theta_2 \big\| 
		\leq K \log \NC \big\| \theta_1 - \theta_2 \big\| \ .
		\end{align}
		We also find the second result holds for $t \in \{\NT, \AT\}$ under \hyperlink{(Logistic)}{(Logistic)} as follows.
\begin{align}									\label{eq-averagerateLogistic}
	\big\| f_t(\bX_{ji} \con \theta_1) -  f_t(\bX_{ji} \con \theta_2) \big\| 
	\leq
	\frac{ \exp \big\{ | \bX_{ji}\T ( \theta_1 - \theta_2 ) | / 2  \big\} -1 }{\exp \big\{ | \bX_{ji}\T (\theta_1 - \theta_2 ) | / 2 \big\} + 1}
	\leq K \log \NC \| \theta_1 - \theta_2 \|\ .
\end{align}
The first inequality is from the definition of $f_t(\bX_{ji} \con \theta_t)$ and the following inequality:
\begin{align*}
	\frac{\exp(x+t)}{1+\exp(x+t)} - \frac{\exp(x)}{1+\exp(x)} 
	\leq \frac{\exp(t/2)}{1+ \exp(t/2)} - \frac{\exp(-t/2)}{1+ \exp(-t/2)}
	= \frac{\exp(t/2) - 1}{\exp(t/2) + 1} \leq t \ , \ ^\forall x \in \R , ^\forall t >0 \ . 
\end{align*}
In the second inequality, we use that $\{ \exp(t/2) - 1 \}/ \{ \exp(t/2) + 1 \}$ is increasing in $t$ and \eqref{eq-averagerateLinear}. 

For $t = \CO$ and under  \hyperlink{(Linear)}{(Linear)}, we find 
\begin{align*}
	\big\|
	 f_\CO(\bX_{ji} \con \theta_1) -  f_\CO(\bX_{ji} \con \theta_2) 
	 \big\|
	 & =
	 \big\|
	- w_{\NT,1} \bX_{ji}\T \theta_{\NT,1} - w_{\AT,1} \bX_{ji}\T \theta_{\AT,1}
	+ w_{\NT,2} \bX_{ji}\T \theta_{\NT,2} + w_{\AT,2} \bX_{ji}\T \theta_{\AT,2}
	\big\|
	\\
	& \leq
	\frac{1}{2}
	\sum_{t\in \{\NT,\AT\} }
	\Big\{
		\big\| w_{t,2} - w_{t,1} \big\| \big\|  \bX_{ji} \big\| \big\| \theta_{t,1} + \theta_{t,2} \big\|
		+
		\big\| w_{t,2} + w_{t,1} \big\| \big\| \bX_{ji} \big\| \big\|\theta_{t,1} - \theta_{t,2} \big\|
	\Big\}
	\\
	& \leq
	2 D \big\| \theta_1 - \theta_2 \big\|
		\big\|  \bX_{ji} \big\| 
		\leq
	2 D K \log\NC \big\| \theta_1 - \theta_2 \big\| 
	\quad 
		\Leftarrow
		\quad 
		D := \max_{\theta_\CO \in \mathcal{E}_\theta } \big\| \theta_\CO \big\|
		\ .
\end{align*}
Since $\mathcal{E}_\theta$ is compact, $D$ is finite.

For $t = \CO$ and \hyperlink{(Logistic)}{(Logistic)}, let $L_1=- w_{\NT,1} \bX_{ji}\T \theta_{\NT,1} - w_{\AT,1} \bX_{ji}\T \theta_{\AT,1}$ and $L_2 = - w_{\NT,2} \bX_{ji}\T \theta_{\NT,2} - w_{\AT,2} \bX_{ji}\T \theta_{\AT,2}$. From the same procedure in \eqref{eq-averagerateLogistic}, we have
\begin{align*}
	\big\|
	 f_\CO(\bX_{ji} \con \theta_1) -  f_\CO(\bX_{ji} \con \theta_2) 
	 \big\|
	&
	=
	 \bigg\|
	\frac{\exp(L_1)}{1+\exp(L_1)}
	-
	\frac{\exp(L_2)}{1+\exp(L_2)}
	\bigg\|
	\leq  \big\| L_1 - L_2\big\| 
	\leq 2 D K \log\NC \big\| \theta_1 - \theta_2 \big\| \ .
\end{align*}

\subsubsection{Proof of Condition \protect\hyperlink{(GC5)}{(GC5)} under \protect\hyperlink{(Linear)}{(Linear)}}								\label{sec:GC5-1}

If $\overline{\eta}_{t,\NC} = 0$, condition \hyperlink{(GC5)}{(GC5)} trivially holds. Therefore, we consider non-zero $\overline{\eta}_{t,\NC}$.

	We split the cases where $\overline{\eta}_{t,\NC} \in \mathcal{E}' $ and $\overline{\eta}_{t,\NC} \in \mathcal{E}^\circ \setminus \mathcal{E}'$. If the noise $e_{tji}$ is generated from a uniform distribution, $\mathcal{E}'$ is empty. Therefore, non-empty $\mathcal{E}'$ implies $e_{tji}=0$ for all $ji$; i.e. the randomizer is not used. As a result, when $ \overline{\eta}_{t,\NC}$ belongs to non-empty $ \mathcal{E}'$, we have $\bX_{ji}\T \overline{\theta}_t = \overline{q}_t$ and $\psi_{ji} (\eta + \overline{\eta}_{t,\NC} )-\psi_{ji} (\eta_t)$ is given by
	{\fontsize{10}{12} \selectfont
	\begin{align*}
		\psi_{ji} (\eta_t + \overline{\eta}_{t,\NC} )
		-
		\psi_{ji} (\eta_t )
		=
		\begin{bmatrix}
			\big\{ \bX_{ji}\T ( \theta_t + \overline{\theta}_t)   - t_{ji}   \big\} \bX_{ji}
			\\
			t_{ji} - \SI \big( \bX_{ji}\T ( \theta_t + \overline{\theta}_t)  - q_t - \overline{q}_t \big) 
		\end{bmatrix}
		-
		\begin{bmatrix}
			\big\{ \bX_{ji}\T\theta_t   - t_{ji}   \big\} \bX_{ji}
			\\
			t_{ji} - \SI \big( \bX_{ji}\T \theta_t  - q_t \big) 
		\end{bmatrix}
		=
		\begin{bmatrix}
			\bX_{ji}\T \overline{\theta}_t
			\\
			0
		\end{bmatrix} \ .
	\end{align*}}%
	Thus, we find that $\| \psi_{ji} (\eta_t + \overline{\eta}_{t,\NC} )
		-
		\psi_{ji} (\eta_t ) \| \leq \| \bX_{ji} \| | \overline{\theta}_t| \leq K \log \NC  \| \overline{\eta}_{t,\NC}\| $ and 
		{\fontsize{10}{12} \selectfont
	\begin{align*}
		\frac{1}{\NC} \sum_j \Big\| \psi_j(\eta_t) - \psi_j(\eta_t+\overline{\eta}_{t,\NC}) \Big\| 
		\leq 
		\frac{1}{\NC} \sumji K \log \NC \| \overline{\eta}_{t,\NC} \| 
		= M K  \log \NC  \| \overline{\eta}_{t,\NC} \| 
		=
		O\big( \NC^{-1/2}  \big(  \log \NC \big)  \big) 
		=
		o(1) \ .
	\end{align*}}%
	
	 Next, we consider the case where $ \overline{\eta}_{t,\NC} \in \mathcal{E}^\circ \setminus \mathcal{E}'$.	
	We consider the derivative of $\psi_j(\eta_t)$.
	{\fontsize{10}{12} \selectfont
	\begin{align}									\label{lem-402-2-psideriv}
		\frac{\partial  \psi_j(\eta_t)}{\partial \eta_t}
		=
		\sum_{i=1}^{\NI_j}  \psi_{ji}' (\eta_t)
		\ , \
		\psi_{ji} ' (\eta_t) 
		= \begin{bmatrix}
			\bX_{ji}\bX_{ji}\T
			& 0 
			\\ 
			- \SI' ( \bX_{ji}\T \theta_t  + e_{tji} - q_t )  \bX_{ji}\T
			&
			\SI' ( \bX_{ji}\T \theta_t + e_{tji}  - q_t ) 
		\end{bmatrix}
	\end{align}}%
	where $\SI'  $ is the derivative of $\SI$ which is given in \eqref{eq-SIdiff}.	Note that $\partial \psi_j(\eta_t) / \partial \eta$ is continuous function for any $\eta_t$. Therefore, we have 
	{\fontsize{10}{12} \selectfont
	\begin{align}										\label{lem-402-2-000}
		\big\| \psi_j(\eta_t + \overline{\eta}_{t,\NC}) - 
		\psi_j(\eta_t) \big\| 
		=
		\bigg\| \sum_{i=1}^{\NI_j}  \frac{\partial}{\partial \eta} \psi_{ji} (\eta_m) \cdot \overline{\eta}_{t,\NC} \bigg\|
		\leq
		\sum_{i=1}^{\NI_j} \bigg\| \frac{\partial}{\partial \eta} \psi_{ji}' (\eta_m) \bigg\| \cdot \| \overline{\eta}_{t,\NC}\|
	\end{align}}
	where $\eta_m = (q_m , \theta_{m}\T)\T$ is the intermediate value between $\eta_t$ and $\eta_t + \overline{\eta}_{t,\NC}$ that satisfies the mean value theorem condition on $\psi_{ji}$; i.e. $\displaystyle{		 \psi_{ji}(\eta_t + \overline{\eta}_{t,\NC}) =
		\psi_{ji}(\eta_t) +\frac{\partial}{\partial \eta} \psi_{ji}(\eta_m) \cdot \overline{\eta}_{t,\NC} }$.

We first consider the case of $\eta_m \in \mathcal{E}'$. Then, $\{ \partial \psi_{ji}(\eta_m) / \partial \eta\} \cdot \overline{\eta}_{t,\NC}$ for $\eta_m \in \mathcal{E}'$ is given by
{\fontsize{10}{12} \selectfont
	\begin{align}										\label{lem-402-2-001}
		\frac{\partial}{\partial \eta} \psi_{ji}(\eta_m) \cdot \overline{\eta}_{t,\NC}
		= 
		\begin{bmatrix}
			\bX_{ji} \bX_{ji}\T
			& 
			0
			\\
			- \frac{1-2c}{2h}  \bX_{ji}\T
			& 
			\frac{1-2c}{2h}
		\end{bmatrix}		
		\begin{bmatrix}
			\overline{\theta}_t
			\\
			\overline{q} _t
		\end{bmatrix}
		=
		\begin{bmatrix}
			\bX_{ji} \bX_{ji}\T \overline{\theta} _t
			\\
			\frac{1-2c}{2h} \overline{q} _t  - \frac{1-2c}{2h}  \bX_{ji}\T \overline{\theta} _t
		\end{bmatrix} \ .
	\end{align}}%
	Non-empty $\mathcal{E}'$ implies $e_{tji}= 0$ for all $ji$. Also, based on the definition of $\mathcal{E}'$, we obtain $\bX_{ji}\T \theta_m + e_{tji} - q_m = \bX_{ji}\T \theta_m - q_m = 0$. Also, since $\eta_m$ is the internal division point of $\eta_t+\overline{\eta}_{t,\NC}$ and $\eta_t$, we obtain $q _t- q_m = - k \overline{q}_t$ and $q_t + \overline{q}_t - q_m = (1- k) \overline{q}_t$. Similarly, we obtain $\theta_t - \theta_{m} = -k \overline{\theta}_t$ and $\theta_t + \overline{\theta}_t - \theta_{m} = (1-k) \overline{\theta}_t$. Therefore, we have $q_t + k \overline{q}_t  = q_m = \bX_{ji}\T \theta_{m} = \bX_{ji}\T \theta_t + k \bX_{ji}\T \overline{\theta}_t$ and, as a result, $\bX_{ji}\T \theta_t - q_t = -k(\bX_{ji}\T \overline{\theta}_t - \overline{q}_t)$. This gives
	{\fontsize{10}{12} \selectfont
	\begin{align}								\label{lem-402-2-002}
		\psi_{ji} (\eta_t + \overline{\eta}_{t,\NC} ) - \psi_{ji} (\eta_t )
		& = \begin{bmatrix}
				\bX_{ji} \bX_{ji}\T \overline{\theta}_t
				\\
				\SI \big\{  -k ( \bX_{ji} \T \overline{\theta}_t - \overline{q}_t ) \big\}
				- \SI \big\{ (1-k) ( \bX_{ji} \T \overline{\theta}_t - \overline{q}_t ) \big\}
		\end{bmatrix} \ .
	\end{align}}%
	Since \eqref{lem-402-2-001} and \eqref{lem-402-2-002} are equivalent, this implies that $ | (1-k) ( \bX_{ji} \T \overline{\theta}_t - \overline{q}_t)  | \leq h$ and $ | k( \bX_{ji} \T \overline{\theta}_t - \overline{q}_t) | \leq h$. Thus, we obtain $| \bX_{ji}\T \overline{\theta}_t - \overline{q}_t  | = | \bX_{ji}\T \overline{\theta}_t - \overline{q}_t +e_{tji} | \leq 2h$; note that the equality holds because $\eta_m \in \mathcal{E}'$ implies $e_{tji} = 0$ for all $ji$. Thus, $\eta_m \in \mathcal{E}'$ implies $| \bX_{ji}\T \overline{\theta}_t - \overline{q}_t +e_{tji} | \leq 2h$ where $\overline{\eta}_{t,\NC}$ is assumed to belong in $\mathcal{E}^\circ \setminus \mathcal{E}'$; i.e.
	\begin{align}														\label{lem-402-2-002-2}
		\ind \big\{ \eta_m \in \mathcal{E}'  \big\}
		\leq 
		 \ind \big\{
			| \bX_{ji}\T \overline{\theta}_t + e_{tji} - \overline{q}_t  | \leq 2h , \overline{\eta}_{t,\NC} \in  \mathcal{E}^\circ \setminus \mathcal{E}'
		\big\}  \ .
	\end{align}
	
	We separate the case $\{\eta_m \in \mathcal{E}^\circ\}$ into three following cases: $\{\eta_m \in \mathcal{E}'\}$, $\{\eta_m \in \mathcal{E}^\circ \setminus \mathcal{E}' \} \cap \{	| \bX_{ji}\T \theta_m + e_{tji} - q_m  | \leq 2h \}$, and $\{\eta_m \in \mathcal{E}^\circ \setminus \mathcal{E}' \} \cap \{	| \bX_{ji}\T \theta_m + e_{tji} - q_m  | > 2h \}$. We find the following result using \eqref{lem-402-2-000} and \eqref{lem-402-2-002-2}
	{\fontsize{10}{12} \selectfont
	\begin{align}											\label{lem-402-2-003}
		\frac{1}{\NC} \sum_j 
		\big\| \psi_j(\eta_t  + \overline{\eta}_{t,\NC}) - 
		\psi_j(\eta_t) \big\| 
		& \leq
		\frac{1}{\NC} \sumji 
		\Big\| \psi_{ji} ' (\eta_m) \Big\|  \| \overline{\eta}_{t,\NC} \|
		 \ind \big\{
			| \bX_{ji}\T \overline{\theta}_t + e_{tji} - \overline{q}_t  | \leq 2h , \overline{\eta}_{t,\NC} \in  \mathcal{E}^\circ \setminus \mathcal{E}'
		\big\} 
		\nonumber
		\\
		&+
		\frac{1}{\NC} \sumji 
		\Big\| \psi_{ji} ' (\eta_m) \Big\|  \| \overline{\eta}_{t,\NC} \|
		 \ind \big\{
			| \bX_{ji}\T \theta_m + e_{tji} - q_m  | \leq 2h , \eta_m \in \mathcal{E}^\circ \setminus \mathcal{E}'
		\big\}
		\nonumber
		\\ 
		& + \frac{1}{\NC} \sumji 
		\Big\| \psi_{ji} ' (\eta_m) \Big\|  \| \overline{\eta}_{t,\NC}\|
		 \ind \big\{
			| \bX_{ji}\T \theta_m + e_{tji} - q_m  | > 2h , \eta_m \in \mathcal{E}^\circ \setminus \mathcal{E}'
		\big\}   \ .
	\end{align}}%
	The first inequality is from \eqref{lem-402-2-000}. The second inequality is from \eqref{lem-402-2-002-2}. 
	
	Note that we have the following result for $\eta_t$ satisfying $|\bX_{ji} \T \theta_t + e_{tji} - q_t| \leq 2h$ from the straightforward calculation involving $\psi_{ji} ' (\eta_t)$ in \eqref{lem-402-2-psideriv}.
	\begin{align}				\label{lem-402-2-005}
		& \Big\| \psi_{ji} ' (\eta_t) \Big\| 
		\leq  (p+1) \max \bigg\{ \frac{1-2c}{2h} , \frac{1-2c}{2h} \big( K \log \NC \big)  , \big( K \log \NC \big)^2 \bigg\} 
		= O \big( \NC \big( \log \NC \big) \big) \ . 
		\end{align}
		Here $p={\rm dim}(\bX_{ji})$. Moreover, from condition  \hyperlink{(GC7)}{(GC7)} (which is proven later), we obtain	
		\begin{align}				\label{lem-402-2-004}
		& \sup_{\eta_t \in \mathcal{E}^\circ \setminus \mathcal{E}'} \frac{1}{\NC} \sumji 
		 \ind \big\{
			| \bX_{ji}\T \theta_t  + e_{tji}  - q_t | \leq 2h \big\}
		=
		O(h) 
		= 
		O ( \NC^{-1} ) \ .
	\end{align}
	Thus, the first term of \eqref{lem-402-2-003} is bounded above by
	\begin{align*}
		& \frac{1}{\NC} \sumji 
		\Big\| \psi_{ji} ' (\eta_m) \Big\| \cdot \| \overline{\eta}_{t,\NC} \|
		\cdot \ind \big\{
			| \bX_{ji}\T \overline{\theta}_t  + e_{tji}  - \overline{q}_t | \leq 2h , \overline{\eta}_{t,\NC} \in  \mathcal{E}^\circ \setminus \mathcal{E}'
		\big\} 
		\\
		& \leq 
		(p+1) \max \bigg\{ \frac{1-2c}{2h} , \frac{1-2c}{2h} \big( K \log \NC \big)  ,  \big( K \log \NC \big)^2 \bigg\} 
		\|\overline{\eta}_{t,\NC} \| 
		\bigg[  \sup_{\eta_t \in \mathcal{E}^\circ \setminus \mathcal{E}' } 
		 \frac{1}{\NC} \sumji 
		 \ind \big\{
			| \bX_{ji}\T \theta_t + e_{tji}  - q_t | \leq 2h \big\}
			\bigg]
			\\
		& = O \big( \NC^{-1/2}  \big( \log \NC \big) \big) \ .
	\end{align*}
	The inequality in the second line  is from \eqref{lem-402-2-005}. The asymptotic representation in the third line is from \eqref{lem-402-2-004}. Similarly, the second term of \eqref{lem-402-2-003} is bounded above by a term that is $O \big( \NC^{-1/2}  \big( \log \NC \big) \big)$.
	If $| \bX_{ji}\T \theta_m + e_{tji}  - q_m | > 2h $, we have
	{\fontsize{10}{12} \selectfont
	\begin{align}									\label{lem-402-2-006}
		& \Big\| \psi_{ji} ' (\eta_m) \Big\|
		\leq  (p+1) \max \Bigg\{ \frac{1-2c}{2h} e^{ -\frac{1-2c}{2c} } , \frac{1-2c}{2h} e^{ -\frac{1-2c}{2c} } \big( K \log \NC \big)  ,  \big( K \log \NC \big)^2 \Bigg\} 
		=
		O \big( \big( \log \NC \big)^2  \big) \ .
	\end{align}}%
	
	From \eqref{lem-402-2-004} and \eqref{lem-402-2-006}, the last term of \eqref{lem-402-2-003} is bounded above by
	\begin{align*}
		& \frac{1}{\NC} \sumji 
		\Big\| \psi_{ji} ' (\eta_m) \Big\| \cdot \| \overline{\eta}_{t,\NC} \|
		\cdot \ind \big\{
			| \bX_{ji}\T \theta_m  + e_{tji}  - q_m | > 2h  , \eta_m \in \mathcal{E}^\circ \setminus \mathcal{E}'
		\big\}
		\\
		& \leq 
		 (p+1) \max \bigg\{ \frac{1-2c}{2h} e^{ -\frac{1-2c}{2c} } , \frac{1-2c}{2h} e^{ -\frac{1-2c}{2c} } \big( K \log \NC \big)  ,  \big( K \log \NC \big)^2 \bigg\} 
		 \\
		 & \hspace*{2cm}  \times \|\overline{\eta}_{t,\NC} \|  \cdot
		 \sup_{\eta_t \in \mathcal{E}^\circ \setminus \mathcal{E}' } 
		 \frac{1}{\NC} \sumji 
		 \ind \big\{
			| \bX_{ji}\T \theta_t + e_{tji} - q_t | > 2h \big\}
			= O\big( \NC^{-1/2} \big( \log \NC \big)^2  \big) \ .
	\end{align*}
	
	Combining the above results with \eqref{lem-402-2-003}, we obtain the convergence to zero as $\NC \rightarrow \infty$. 
	\begin{align*}
	\frac{1}{\NC} \sum_j 
		\big\| \psi_j(\eta_t + \overline{\eta}_{t,\NC}) - 
		\psi_j(\eta_t) \big\| 
		& \leq O\big( \NC^{-1/2} \big( \log \NC \big)^2  \big) \ .
	\end{align*}
	The result holds for any $\eta_t \in \mathcal{E}^\circ $. Thus, condition \hyperlink{(GC5)}{(GC5)} holds under \hyperlink{(Linear)}{(Linear)} with $v_2 = 2$.
	
	\subsubsection{Proof of Condition \protect\hyperlink{(GC5)}{(GC5)} under \protect\hyperlink{(Logistic)}{(Logistic)}}							\label{sec:GC5-2}
	
	The result holds with $v_2 = 2$ where the detail is similar to the proof in Section \ref{sec:GC5-1}.

\subsubsection{Proof of \protect\hyperlink{(GC6)}{(GC6)} under \protect\hyperlink{(Linear)}{(Linear)}}							\label{sec:GC6-1}

We define $\Phi(\eta_t)$ and two disjoint sets $E_1$ and $E_2$ as follows.
		\begin{align*}
			&
			\Phi(\eta_t)
			=
			\| \Psi_t(\eta_t) \|^2
			= 
			\underbrace{ \frac{1}{\NC^2}\bigg\| \sumji  \Big( \bX_{ji}\T \theta_t - t_{ji} \Big) \bX_{ji} \bigg\|^2 }_{ \Phi_1 (\theta_t) }
			+
			\underbrace{ \frac{1}{\NC^2} 
			\bigg[ \sumji  \Big\{ t_{ji} -  \SI ( \bX_{ji}\T \theta_t + e_{tji} - q_t ) \Big\} \bigg]^2  }_{ \Phi_2 (\theta_t,q_t) } 
			\ , \\
	&
		E_1
		=
		\Big\{ \eta \, \Big| \, \|\theta - \theta_t^* \| \geq \frac{\epsilon}{10  (K \log \NC +1) } , \| \eta - \eta_t^* \| = \epsilon \Big\}
		\ , \
		E_2 
		= 
		\Big\{ \eta \, \Big| \, \|\theta - \theta_t^* \| < \frac{\epsilon}{10 (K\log \NC +1) } , \| \eta - \eta_t^* \| = \epsilon \Big\} \ .
\end{align*}
	where $K$ is a constant satisfying $\| \bX_{ji} \| / \log \NC \leq K$. Note that $\{ \eta \cond \| \eta - \eta_t^* \| = \epsilon\} = E_1 \cup E_2$. 
	
	We split the cases of $\eta_t \in E_1$ and $\eta_t \in E_2$. If $\eta_t \in E_1$, we find that $\Phi_1(\theta_t)$ lower-bounded by a constant. 
	\begin{align}	\label{lem-402-1-001}
		\Phi_1(\theta_t)
		& = \frac{1}{\NC^2}  \bigg\| \sumji \bX_{ji} \bX_{ji}\T \big(  \theta_t -  \theta_t^*  \big)   \bigg\|^2
		\geq \frac{\epsilon^2  \sigma_{\min}^2 \big( \frac{1}{\NIT} \sumji \bX_{ji} \bX_{ji}\T \big)  }{100  (K \log \NC+1) ^2}
		= M'(\NC, \epsilon) \big\{ 1 + o(1) \big\} \ .
	\end{align}
	The first equality is from $\sum_j \sum_i \bX_{ji} t_{ji} = \sum_j \sum_i \bX_{ji} \bX_{ji}\T \theta_t^*$. The second equality is from the definition of the norm of a vector. The inequality is from $\|\theta_t - \theta_t^*\| \geq \epsilon/\{ 10  (K \log \NC+1) \}$ where $\sigma_{\min}(A)$ is the minimum of the singular value of a matrix $A$. From condition (iii) of Assumption \ref{assp:4-1}, $\sigma_{\min}^2 \big( \NIT^{-1}  \sum_{j,i} \bX_{ji} \bX_{ji}\T \big) $ is strictly positive for all $\NC$. Thus, we have the asymptotic result with  some $M'(\NC, \epsilon) =O(  \NC^{-2r} (\log\NC)^{-2} )$.

Next we consider the case of $\eta \in E_2$.  From \eqref{eq-averagerateLinear} in the proof of \hyperlink{(GC4)}{(GC4)}, we find $\big|
		f_t ( \bX_{ji} \con \theta_t ) - f_t ( \bX_{ji} \con \theta_t^* )
	\big|
	\leq
	K \log\NC \big\| \theta_t - \theta_t^* \big\|$. Thus, from the monotonicity of $\SI$ and $\randomf_t (\bX_{ji} \con \theta_t ) = f_t (\bX_{ji} \con \theta_t) + e_{tji}$, we observe the following inequality:
	\begin{align}							\label{lem-402-1-002}
	\SI \big( \bX_{ji}\T \theta_t^* + e_{tji} - K \log\NC \| \theta_t - \theta_t^*\| - q_t \big)
	\leq
	\SI \big( \bX_{ji}\T \theta_t^* + e_{tji}+ K \log\NC \| \theta_t - \theta_t^*\| - q_t \big)
	\end{align}
Since $\|\eta_t - \eta_t^*\| = \sqrt{ (q_t-q_t^*)^2 + \|\theta_t - \theta_t^* \|^2 } = \epsilon$, we find that $q_t = q_t^* \pm \sqrt{ \epsilon^2 - \| \theta_t - \theta_t^* \|^2 }$. First, for $q_t = q_t^* + \sqrt{ \epsilon^2 - \| \theta_t - \theta_t^* \|^2 }$, we get $q_t \geq q_t^* +  \epsilon - \| \theta_t - \theta_t^* \|$. Thus, the right term in \eqref{lem-402-1-002} is upper-bounded by  $
	\SI \big( \bX_{ji}\T \theta_t^* + e_{tji}  + K \log\NC \| \theta_t - \theta_t^*\| - q_t \big)
	\leq
	\SI \big( \bX_{ji}\T \theta_t^* + e_{tji} + K \log\NC \| \theta_t - \theta_t^*\| - q_t^* - \epsilon + \| \theta_t - \theta_t^* \| \big)$.
For $\eta \in E_2$, we find that $\| \theta_t - \theta_t^* \| < \epsilon/\{ 10  (K \log\NC +1) \}$. Hence, each $\SI \big( \bX_{ji}\T \theta_t - q_t \big)$ is upper-bounded by $
	\SI \big( \bX_{ji}\T \theta_t + e_{tji} - q_t \big)
	\leq
	\SI \big( \bX_{ji}\T \theta_t^* + e_{tji} + K \log\NC \| \theta_t - \theta_t^*\| - q_t \big)
	< 
	\SI \big( \bX_{ji}\T \theta_t^* + e_{tji} - q_t^* - 0.9 \epsilon \big)$. Therefore, for $\eta_t \in E_2$, we find that
\begin{align}						\label{lem-402-1-003}
	&
	\frac{1}{\NC} \sumji  \Big\{ t_{ji} -  \SI \Big( \bX_{ji}\T \theta_t + e_{tji} - q_t \Big) \Big\} 
	\geq
	\frac{1}{\NC} \sumji \Bigg\{
		\begin{matrix}
		\SI \big( \bX_{ji}\T \theta_t^*+ e_{tji}  - q_t^* \big) \hfill
		\\[-0.25cm]
		 \quad \quad \quad
		-  \SI \big( \bX_{ji}\T \theta_t^*+ e_{tji}  - q_t^* - 0.9 \epsilon \big)
		\end{matrix}
	\Bigg\}\ .
\end{align}

We consider a set of individuals $I_+$ where $ I_+
	=
	\big\{
		ji \, \big| \,  h < \bX_{ji}\T \theta_t^* + e_{tji} - q_t^* < 0.9\epsilon -  h \big\}$. Because of condition \hyperlink{(GC8)}{(GC8)} (which is proven later), $| I_+ | = O(\NC) C_+(\epsilon) = O(\NC^{1-r})$ because $C_+(\epsilon) = \Theta(\NC^{-r})$; i.e. $I_+$ is asymptotically non-empty. If individual $ji \in I_+$, we find that each summand of \eqref{lem-402-1-003} is lower-bounded by a constant related to the parameter $c$.
\begin{align*}
	\SI \Big( \bX_{ji}\T \theta_t^* + e_{tji} - q_t^* \Big) -  \SI \Big( \bX_{ji}\T \theta_t^* + e_{tji} - q_t^*  - 0.9 \epsilon \Big)
	\geq \SI (h) - \SI(-h) 
	=
	1 - 2c \ .
\end{align*}
If individual $ji \notin I_+$, we find that each summand of \eqref{lem-402-1-003} is lower-bounded by zero because of the monotonicity of $\SI$. 
Thus, the left hand side of \eqref{lem-402-1-003} is lower-bounded by a constant:
\begin{align*}
	& \frac{1}{\NC} \sumji  \Big\{ t_{ji} -  \SI \Big( \bX_{ji}\T \theta_t + e_{tji} - q_t \Big) \Big\} 
	\nonumber
		\\
	&
	\geq
	\frac{1}{\NC} \sum_{ji \in I_+}  \Big\{ \SI \Big( \bX_{ji}\T \theta_t^* + e_{tji} - q_t^* \Big) -  \SI \Big( \bX_{ji}\T \theta_t^* + e_{tji} - q_t^* - 0.9 \epsilon \Big) \Big\} 
	\nonumber
	\\[-0.2cm]
	& \hspace*{2cm}
	+
	\frac{1}{\NC} \sum_{ji \in I_+^c}  \Big\{ \SI \Big( \bX_{ji}\T \theta_t^*  + e_{tji} - q_t^* \Big) -  \SI \Big( \bX_{ji}\T \theta_t^* + e_{tji} - q_t^* - 0.9 \epsilon \Big) \Big\} 
	\nonumber
	\\[-0.2cm]
	& 
	\geq 
	\frac{1}{\NC} \sum_{ji \in I_+} (1-2c)
	=
	(1-2c) \frac{1}{\NC} \sumji \ind \Big\{ h < \bX_{ji}\T \theta_t^* + e_{tji} - q_t^* < 0.9\epsilon -  h \Big\} 
	=
	\big\{ 1- o(1) \big\}  C_+ (\epsilon) \ .
\end{align*}
The inequality in the second line is from \eqref{lem-402-1-003}. The last line from the above results and the definition of $I_+$. The asymptotic representation in the last line is from the rate of $c$ and condition \hyperlink{(GC8)}{(GC8)} where $C_+(\epsilon) = \Theta(\NC^{-r} )$ is the limit of $ \NC^{-1} \sumji \ind \big( h < \bX_{ji}\T \theta_t^* + e_{tji} - q_t^* < 0.9\epsilon -  h \big)$ which holds with probability tending to 1.

	Similarly, for $q_t = q_t^* - \sqrt{ \epsilon^2 - \| \theta_t - \theta_t^* \|^2 }$, we find that $q_t \leq q_t^* -  \epsilon + \| \theta_t - \theta_t^* \|$. Thus, the left term in \eqref{lem-402-1-002} is lower-bounded by  $	\SI \big( \bX_{ji}\T \theta_t^* + e_{tji} - K \log\NC \| \theta_t - \theta_t^*\| - q_t \big)
	\geq
	\SI \big( \bX_{ji}\T \theta_t^*+ e_{tji} - K \log\NC \| \theta_t - \theta_t^*\| - q_t^* + \epsilon - \| \theta_t - \theta_t^* \| \big)$. 
For $\eta_t \in E_2$, we find that $\| \theta_t - \theta_t^* \| < \epsilon/\{ 10  (K \log \NC +1) \}$. Hence, each $\SI \big( \bX_{ji}\T \theta_t + e_{tji} - q_t \big)$ is lower-bounded by $
	\SI \big( \bX_{ji}\T \theta_t + e_{tji} - q_t \big)
	\geq
	\SI \big( \bX_{ji}\T \theta_t^*+ e_{tji}  - K \log\NC \| \theta_t - \theta_t^*\| - q_t \big)
	>
	\SI \big( \bX_{ji}\T \theta_t^*+ e_{tji}  - q_t^* + 0.9 \epsilon \big)$.
For $\eta_t \in E_2$, we find that $ \sumji  \big\{ t_{ji} -  \SI \big( \bX_{ji}\T \theta_t + e_{tji} - q_t \big) \big\} 
	\leq \sumji  \big\{ \SI \big( \bX_{ji}\T \theta_t^* + e_{tji} - q_t^* \big) -  \SI \big( \bX_{ji}\T \theta_t^* + e_{tji} - q_t^* + 0.9 \epsilon \big) \big\}$.
We consider a set of individuals $I_-$ where $ I_-
	=
	\big\{
		ji \, \big| \,  -0.9\epsilon + h < \bX_{ji}\T \theta_t^* + e_{tji} - q_t^* < - h \big\}$. Similar to the case of $I_+$, we find $ \NC^{-1} \sumji  \big\{ t_{ji} -  \SI \big( \bX_{ji}\T \theta_t + e_{tji} - q_t \big) \big\} 
	=
	\big\{ -1+ o(1) \big\} C_-(\epsilon)$. 
As a result, with probability tending to 1, we obtain a lower bound of $\Phi_2(\theta_t,q_t)$ for $\eta_t \in E_2$ as follows.
\begin{align}												\label{lem-402-1-006}
	\Phi_2(\theta_t,q_t)
	=
	 \frac{1}{\NC^2} 
			\Big[ \sumji  \Big\{ t_{ji} -  \SI ( \bX_{ji}\T \theta_t + e_{tji} - q_t ) \Big\} \Big]^2
	\geq \{ 1+ o(1) \}\min \Big[  C_+(\epsilon)^2 , C_-(\epsilon)^2 \Big] \ .
\end{align}
Combining \eqref{lem-402-1-001} and \eqref{lem-402-1-006}, we obtain the lower bound of $\Phi(\eta_t)$ for all $\eta_t \in E_1 \cup E_2$ with probability tending to 1. 
\begin{align*}
	\Phi(\eta_t) 
	& =
	\Phi_1(\theta_t) + \Phi_2(\theta_t,q_t)
	=
	\ind \big\{ \eta_t \in E_1 \big\} \Big\{ \Phi_1(\theta_t)  + \Phi_2(\theta_t,q_t) \Big\}
	+
	\ind \big\{ \eta_t \in E_2 \big\} \Big\{ \Phi_1(\theta_t)  + \Phi_2(\theta_t,q_t)\Big\} 
	\\
	& \geq 
	\ind \big\{ \eta_t \in E_1 \big\} \Phi_1(\theta_t)
	+
	\ind \big\{ \eta_t \in E_2 \big\} \Phi_2(\theta_t,q_t) 
	\geq
	\big\{ 1 + o(1) \big\} \min \Big[  M'(\NC, \epsilon)   ,  C_+(\epsilon)^2 , C_-(\epsilon)^2 \Big] \ .
\end{align*}
Note that $M'(\NC, \epsilon)  \times  \NC^{2r} (\log \NC)^2$ converges to a finite constant while $C_+(\epsilon)^2 \times  \NC^{2r} (\log \NC)^2 $ and $ C_-(\epsilon)^2  \times \NC^{2r} (\log \NC)^2 $ diverges. This concludes \hyperlink{(GC6)}{(GC6)} under \hyperlink{(Linear)}{(Linear)}  with $v_3 = 1$.
	
\subsubsection{Proof of \protect\hyperlink{(GC6)}{(GC6)} under \protect\hyperlink{(Logistic)}{(Logistic)}}								\label{sec:GC6-2}
	
		The result holds with $v_3 = 1$ where the detail is similar to the proof in Section \ref{sec:GC6-1}.

	\subsubsection{Proof of \protect\hyperlink{(GC7)}{(GC7)}}							\label{sec:GC7}
	
		We prove condition \hyperlink{(GC7)}{(GC7)}. If $\eta_t \in \mathcal{E}^\circ \setminus \mathcal{E}'$, $\eta_t$ satisfies either $\theta_t \in \mathcal{E}_\theta^\circ \setminus \mathcal{E}_\theta'$ (i.e. $\randomf_t (\bX_{ji} \con \theta_t^*)$ is not constant across $ji$) or $\randomf_t(\bX_{ji} \con \theta_t) \neq q_t$ while $\randomf_t(\bX_{ji} \con \theta_t)$ is constant across $ji$. First, we consider the case of $\theta_t \in \mathcal{E}_\theta^\circ \setminus \mathcal{E}_\theta'$. Since $q_t$ is chosen as an interior points of $\mathcal{E}_q^\circ$, we have $[q_t-|d_\NC|, q_t+|d_\NC| ] \subset \mathcal{E}_q$ for sufficiently large $\NC$. Therefore, the quantity in condition \hyperlink{(GC7)}{(GC7)} is
		\begin{align*}
			 \frac{1}{\NC} \sumji \ind \big\{ | \randomf_t (\bX_{ji} \con \theta_t) - q_t | < |d_\NC| \big\}
			 & \leq \frac{\NIT}{\NC} \Big\{ G_t \big(  q_t +  |d_\NC| \con \NC, \theta_t \big)  - G_t\big(  q_t -  |d_\NC| \con \NC , \theta_t \big)  \Big\}
			 \leq \frac{\NIT}{\NC}  2 \mu_t(\NC, \theta_t) | d_\NC |
			 = O \big(  | d_\NC | \big) \ .
		\end{align*}
		The first inequality is from $ \NIT^{-1} \sumji  \ind \big\{ \randomf_t (\bX_{ji} \con \theta_t) - q_t < v \big\} 
		\leq \NIT^{-1} \sumji  \ind \big\{ \randomf_t (\bX_{ji} \con \theta_t) - q_t \leq v \big\} 
		= G_t(q_t+v \con \NC, \theta_t)$. The second inequality is from condition (iv)-(a) of Assumption \ref{assp:4-1}. The last asymptotic representation is from $\NIT/\NC = O(1)$ and $\mu_t(\NC,\theta_t) = O(1)$. 
		
	Second, we consider the case of $\randomf_t (\bX_{ji} \con \theta_t) \neq q_t$ while $\randomf_t (\bX_{ji} \con \theta_t)$ is constant across $ji$ as $\randomf_t (\bX_{ji} \con \theta_t) \equiv f \in \R$. Then, we find the quantity in condition \hyperlink{(GC7)}{(GC7)} is
		\begin{align*}
			 \frac{1}{\NC} \sumji \ind \big\{ | \randomf_t (\bX_{ji} \con \theta_t) - q_t | < |d_\NC| \big\}
			 =
			  \frac{1}{\NC} \sumji \ind \big\{ | f - q | < |d_\NC| \big\}
			  \leq
			M \ind \big\{ | f - q| < |d_\NC| \big\}
			= O(|d_\NC|) \ .
		\end{align*}

	\subsubsection{Proof of \protect\hyperlink{(GC8)}{(GC8)}}						\label{sec:GC8}
	
	We only show the first result of \hyperlink{(GC8)}{(GC8)}; the second result can be proven in similar manner. Let $[q_L , q_U]$ be a neighborhood of $q_t^*$ in condition (iv)-(b) of Assumption \ref{assp:4-1}. For sufficiently large $\NC$, we obtain $[q_t^* + e_\NC, q_t^* + e_\NC'] \subset \mathcal{E}_q^\circ$. Therefore, we obtain
	\begin{align*}
		 \frac{1}{\NC} \sumji \ind \big\{ e_\NC \leq \randomf_t (\bX_{ji} \con \theta_t^*) - q_t^* \leq e_\NC' \big\}
			 & = \frac{\NIT}{\NC} \Big\{ G_t\big(  e_\NC' + q_t^*  \con \NC,  \theta_t^* \big)  - G_t \big(  e_\NC + q_t^* \con \NC,  \theta_t^* \big) + O(\NC^{-1}) \Big\} 
			 \\
			 & \leq \frac{\NIT}{\NC}  \mu_t(\NC, \theta_t^*)  | \, e_\NC' - e_\NC \, |   + O(\NC^{-1}) 
			 = O \big( \NC^{-r} \big) \ .
	\end{align*}
	The first equality is from $ \NIT^{-1} \sumji  \ind \big\{ \randomf_t (\bX_{ji} \con \theta_t^*) - q_t^* < v \big\} =  \NIT^{-1} \sumji  \ind \big\{  \randomf_t (\bX_{ji} \con \theta_t^*) - q_t^* \leq v  \big\} + O(\NC^{-1})= G_t(v+q_t^* \con \NC, \theta_t^*) + O(\NC^{-1})$. The second inequality is from condition (iv)-(a) of Assumption \ref{assp:4-1}. The last asymptotic representation is from $\NIT/\NC = O(1)$ and $\mu_t(\NC, \theta_t^*) = O(1)$. 
	
	For sufficiently large $\NC$, we obtain $[q_t^* + e_\NC, q_t^* + e_\NC'] \subset [q_L , q_U]$ where $[q_L, q_U]$ is the neighborhood of $q_t^*$ satisfying condition (iv)-(b) of Assumption \ref{assp:4-1}. Thus, using condition (iv)-(b) of Assumption \ref{assp:4-1}, we get the following result for some $\kappa'>0$ when $\NC$ is sufficiently large. 
	\begin{align*}
		 \frac{1}{\NC} \sumji \ind \big\{ e_\NC \leq \randomf_t (\bX_{ji} \con \theta_t^*) - q_t^* \leq e_\NC' \big\}
			 & = \frac{\NIT}{\NC} \Big\{ G_t\big(  e_\NC' + q_t^*  \con \NC, \theta_t^* \big)  - G_t\big(  e_\NC + q_t^* \con \NC,  \theta_t^* \big) + O(\NC^{-1}) \Big\} 
			 \\
			 & \geq \frac{\NIT}{\NC}  \kappa_\NC | \, e_\NC' - e_\NC \, |  + O(\NC^{-1}) 
			 \geq  \kappa' \NC^{-r} \ .
	\end{align*}

\subsection{Proof of Lemma \ref{lmm:401}}					\label{sec:proofoflmm:401}

	First, we prove the result about $\widehat{\eta}_\NT$. We define a set $\mathcal{S}(\eta_\NT^*, r) = \{ \eta_\NT \cond \| \eta_\NT - \eta_\NT^* \| \geq r\}$ $(r \geq 0)$, which is compact. Therefore, there exists $\eta_\NT(r) \in \mathcal{S}(\eta_\NT^*,r)$ such that
		\begin{align*}
		&
		m_\NC(r) 
		= \inf_{\eta_\NT \in \mathcal{S}(\eta_\NT^*,r) } \Phi(\eta_\NT)
		= \min_{\eta_\NT \in \mathcal{S}(\eta_\NT^*,r) } \Phi(\eta_\NT)
		 = \Phi(\eta_\NT(r))
		 \\
		 &
			\Phi(\eta_\NT)
			= \frac{1}{\NC^2} \bigg[ 
			\bigg\| \underbrace{\sumji \nabla_\theta L \big( \NT_{ji} , f_\NT (\bX_{ji} \con \theta_\NT) \big) }_{\Phi_1(\theta_\NT)} \bigg\|^2	
			+
			\bigg[ \underbrace{\sumji  \Big\{ \NT_{ji} -  \SI \big( \randomf_\NT  (\bX_{ji} \con \theta_\NT) - q_\NT \big)  \Big\}}_{\Phi_2(\theta_\NT,q_\NT)} \bigg]^2  \bigg]	 \ .
		\end{align*}
		Note that $ m_\NC(0) = 0$ where the minimum is achieved at $\eta_\NT^*$ because $\mathcal{S}(\eta_\NT^* , 0) = \mathcal{E}$. 
		
		The derivative of $\Phi(\eta_\NT)$ with respect to $\eta$ is
		{\fontsize{10}{12} \selectfont
		\begin{align*}
			\frac{\partial }{\partial \eta} \Phi(\eta_\NT)
			=
			\frac{2}{\NC^2}  \begin{bmatrix}
				\begin{matrix}
				\big[ \sumji \nabla_\theta^2 L \big( \NT_{ji} , f_\NT (\bX_{ji} \con \theta_\NT) \big)
			 \big] \Phi_1(\theta_\NT)
			 \quad \quad \quad
			 \hfill
			 \\[-0.15cm]
			  \quad \quad \quad
			 -
				\big[ \sumji \SI ' \big( \randomf_\NT (\bX_{ji} \con \theta_\NT) - q_\NT \big) \big\{ \nabla_\theta f_\NT (\bX_{ji} \con \theta_\NT)  \big\} \big] \Phi_2(\theta_\NT,q_\NT)
				\end{matrix}
			 \\
			 \\[-0.5cm]
				{  \big[ \sumji \SI ' \big( \randomf_\NT (\bX_{ji} \con \theta_\NT) - q_\NT \big) \big] \Phi_2(\theta_\NT,q_\NT)  }
			 \end{bmatrix} \ .
		\end{align*}}%
	Here $\nabla_\theta^2 L$ is the Hessian of $L$ with respect to $\theta$ and $\SI'$ is the derivative of $\SI$:
	\begin{align}									\label{eq-SIdiff}
		\SI' ( t ) 
		= 
		\begin{cases}
			{ \frac{1-2c}{2h} \exp \big\{ - \frac{1-2c}{2ch} (t-h) \big\}  }
			& \text{if } h \leq t \\[-0.25cm]
			{  \frac{1-2c}{2h} } & \text{if } -h \leq t \leq h \\[-0.25cm]
			{  \frac{1-2c}{2h} \exp \big\{ \frac{1-2c}{2ch} (t+h) \big\} }
			& \text{if } t \leq -h \\
		\end{cases} \ .
	\end{align}
	Note that  $\Phi(\eta_\NT)$	is continuously differentiable for all $\eta_\NT$ in $\mathcal{E}^\circ$, the interior or $\mathcal{E}$. Combining the compactness of $\mathcal{S}(\eta_\NT^*,r_0)$ for any $r_0 \geq 0$ and the continuity of $\Phi(\eta_\NT)$, we obtain the continuity of $m_\NC(r)$: $\displaystyle{
		\lim_{r \rightarrow r_0} m_\NC(r) = \min_{\eta_\NT \in \mathcal{S}(\eta_\NT^*,r_0)} \Phi(\eta_\NT) 
		=  m_\NC(r_0)}$. Taking $r_0=0$ gives $\lim_{r \rightarrow 0} m_\NC(r) = 0$; i.e. we can take $\mathcal{S}(\eta_\NT^*,r)$ so that $m_\NC(r)$ is arbitrarily close to zero. 
	
	Next we study stationary points of $\Phi(\eta_\NT)$. Since $\SI' ( t ) $ is positive for all $t$, $ \partial \Phi(\eta_\NT) / \partial \eta = 0$ implies that $\Phi_2(\theta_\NT,q_\NT)$ is zero from the identity of the second component. Moreover, $\Phi_1(\theta_\NT)$ is also zero because $ \sumji \nabla_\theta^2 L \big( \NT_{ji} , f_\NT(\bX_{ji} \con \theta_\NT) \big) $ is invertible from \hyperlink{(GC3)}{(GC3)}. This implies that the unique stationary point is $\eta_\NT^*$ because $\eta_\NT^*$ is the unique solution to $\Phi_1(\theta_\NT)=0$ and $\Phi_2(\theta_\NT,q_\NT)=0$.

		We fix a sufficiently small constant $\epsilon>0$ so that $\{ \eta_\NT \cond \| \eta_\NT - \eta_\NT^*\| < \epsilon\} \subset \mathcal{E}$. There is no local minimum of $\Phi(\eta_\NT)$ in $\mathcal{S}^\circ(\eta_\NT^*,\epsilon)$, the interior of $\mathcal{S}(\eta_\NT^*,\epsilon)$, because every points in $\mathcal{S}^\circ(\eta_\NT^*,\epsilon)$ is a non-stationary point. This implies that the local  minimum of $\Phi(\eta_\NT)$ must be in the boundary of $\mathcal{S}(\eta_\NT^*,\epsilon)$, which is equivalent to $\{ \eta_\NT \cond \| \eta_\NT - \eta_\NT^*\| = \epsilon\} \cup  \partial \mathcal{E}$ where $\partial \mathcal{E}$ is the boundary of $\mathcal{E}$. Note that $\min_{\eta_\NT \in \partial \mathcal{E}} \Phi(\eta_\NT) > 0$ due to the uniqueness of $\eta_\NT^*$. From the continuity of $m_\NC(r)$, we have $m_\NC(r) \rightarrow 0$ as $r \rightarrow 0$. Therefore, we may take sufficiently small $\epsilon$ to have the minimum in the set $\{ \eta_\NT \cond \| \eta_\NT - \eta_\NT^*\| = \epsilon\}$; i.e.
		\begin{align*}
			m_\NC(\epsilon) = \min_{\eta_\NT \in \mathcal{S}(\eta_\NT^*,\epsilon) } \Phi(\eta_\NT)
		 = \min\Big\{ \min_{\eta_\NT : \| \eta_\NT - \eta_\NT^*\| = \epsilon } \Phi(\eta_\NT)  ,
		 \min_{\eta_\NT \in \partial \mathcal{E}}  \Phi(\eta_\NT)   \Big\}
		 =  \min_{\eta_\NT : \| \eta_\NT - \eta_\NT^*\| = \epsilon } \Phi(\eta_\NT)   \ .
		\end{align*}
		By \hyperlink{(GC6)}{(GC6)}, we find $m_\NC(\epsilon) \geq  B(\NC, \epsilon)^2 
		$ where $B(\NC,\epsilon) = \Theta \big( \NC^{-r} ( \log \NC )^{-v_3} \big)$. 
		Therefore, the event $\{ \| \eta_\NT - \eta_\NT^* \| > \epsilon \}$ is contained in the event $ \big\{ \| \Psi_\NT (\widehat{\eta}_\NT \con \NC) \| \geq B(\NC, \epsilon)
		 \big\} $ and, as a result, 
		\begin{align}									\label{proof-thm41-002}
			P \Big\{  \| \widehat{\eta}_\NT - \eta_\NT^* \| > \epsilon \, \Big| \, \extendF_\NC , \mathcal{\iv}_\NC\Big\}
			\leq
			P \Big\{ \| \Psi_\NT (\widehat{\eta}_\NT \con \NC) \| \geq B(\NC, \epsilon) 
			 \, \Big| \, \extendF_\NC  , \mathcal{\iv}_\NC\Big\} \ .
		\end{align}
	Thus, it suffices to show the right-hand side of \eqref{proof-thm41-002} converges to zero as $\NC \rightarrow \infty$. Suppose the following condition holds for any fixed constant $\xi>0$.
		\begin{align}							\label{proof-thm41-003}
		\lim_{\NC \rightarrow \infty}	P \bigg\{ \sup_{\eta_\NT \in \mathcal{E}^\circ } \big\{ \NC^r (\log\NC)^{v_3} \big\} \big  \| \Psi_\NT (\eta_\NT \con \NC) -  \widehat{\Psi}_\NT  (\eta_\NT \con \NC) \big \| > \xi +o(1) \, \bigg| \, \extendF_\NC , \mathcal{\iv}_\NC\bigg\} = 0 \ .
		\end{align}	
	Then, we have the following result as $\NC \rightarrow \infty$.
		\begin{align*}
			&
			P \Big\{ \big\| \Psi_\NT (\widehat{\eta}_\NT \con \NC) \big\| \geq B(\NC, \epsilon) 
			 \, \Big| \, \extendF_\NC , \mathcal{\iv}_\NC\Big\}
			 \\
			&
			 \leq
			P \Big\{ \big\{ \NC^r (\log\NC)^{v_3} \big\} \big\| \Psi_\NT (\widehat{\eta}_\NT \con \NC) - \widehat{\Psi}_\NT (\widehat{\eta}_\NT \con \NC) \big \| \geq B_L + o(1)  \, \Big| \, \extendF_\NC , \mathcal{\iv}_\NC\Big\} \\
			&
			\leq
			 P \bigg\{ \sup_{\eta_\NT \in \mathcal{E} } \big\{ \NC^r (\log\NC)^{v_3} \big\}  \big\| \Psi_\NT (\eta_\NT \con \NC) - \widehat{\Psi}_\NT (\eta_\NT \con \NC) \big \|> B_L + o(1) \, \bigg| \, \extendF_\NC , \mathcal{\iv}_\NC\bigg\} \\
			& =
				P \bigg\{ \sup_{\eta_\NT \in \mathcal{E}^\circ } \big\{ \NC^r (\log\NC)^{v_3} \big\} \big\| \Psi_\NT (\eta_\NT \con \NC) - \widehat{\Psi}_\NT (\eta_\NT \con \NC) \big \| > B_L + o(1) \, \bigg| \, \extendF_\NC , \mathcal{\iv}_\NC\bigg\} \rightarrow 0
		\end{align*}
		where $B_L>0$ is the limit inferior of $\big\{ \NC^r (\log\NC)^{v_3} \big\} B(\NC, \epsilon) $. The equality in the first line is from the definition of $\widehat{\eta}_\NT$, i.e. $\widehat{\Psi}_\NT (\widehat{\eta}_\NT \con \NC)=0$. The inequality in the second line is trivial. The equality in the third line is from the continuity of $\big\| \Psi_\NT (\eta_\NT \con \NC) - \widehat{\Psi}_\NT (\eta_\NT \con \NC) \big \| $ and the compactness of $\mathcal{E}$. The convergence to zero in the third line is from \eqref{proof-thm41-003}. Therefore, in conjunction with \eqref{proof-thm41-002}, we have the results of the theorem, i.e. $\displaystyle{
		P \Big\{  \| \widehat{\eta}_\NT - \eta_\NT^* \| > \epsilon \, \Big| \, \extendF_\NC , \mathcal{\iv}_\NC\Big\} \rightarrow 0}$ as $\NC \rightarrow \infty$. 
		
		To conclude the proof, we show that \eqref{proof-thm41-003} holds. Let $\delta$ be a fixed constant satisfying $1/2 < \delta<1-2r<1$. Since $\mathcal{E}$ is compact, we can partition $\mathcal{E}$ into $ \lceil \NC^\delta \rceil$ subsets such that $\displaystyle{
			\mathcal{E} = \bigcup_{k=1}^{ \lceil \NC^\delta \rceil} \mathcal{E}_k}$ where $\mathcal{E}_k$ satisfies that $\|\eta_k - \eta_k'\| \leq \| \overline{\eta}_\NC\|$ for all $\eta_k, \eta_k' \in \mathcal{E}_k$. For any $\eta_k \in \mathcal{E}_k$ $(k=1,\ldots, \lceil \NC^\delta \rceil)$, we have
		\begin{align}										\label{proof-thm41-004}
			& \sup_{\eta_\NT \in \mathcal{E}^\circ } \big\{ \NC^r (\log\NC)^{v_3} \big\} \big\| \Psi_\NT (\eta_\NT \con \NC) - \widehat{\Psi}_\NT (\eta_\NT \con \NC) \big \| 
			\\
			& \leq 
			\max_{k}  \big\{ \NC^r (\log\NC)^{v_3} \big\} \big\| \Psi_\NT (\eta_k \con \NC) - \widehat{\Psi}_\NT (\eta_k \con \NC) \big \| 
			\nonumber
			\\
			& \hspace*{1cm}
			+ 
			\max_{k} \sup_{\eta_\NT \in \mathcal{E}_k^\circ }  \big\{ \NC^r (\log\NC)^{v_3} \big\}  \big\| \{ \Psi_\NT(\eta_\NT \con \NC) - \widehat{\Psi}_\NT(\eta_\NT \con \NC)   \}  - \{ \Psi_\NT(\eta_k \con \NC) - \widehat{\Psi}_\NT(\eta_k \con \NC) \} \big\| \ .
			\nonumber
		\end{align}
		
		We study the first term of the right-hand side of \eqref{proof-thm41-004}. 
		From \hyperlink{(GC4)}{(GC4)}, the variance of $\widehat{\Psi}_\NT(\eta_k \con \NC) - \Psi_\NT(\eta_k \con \NC)$ given a population $ \extendF_\NC $ is represented by 
		\begin{align*}
		&
			\VAR \big\{ \Psi_\NT(\eta_k \con \NC) -  \widehat{\Psi}_\NT(\eta_k \con \NC) \cond \extendF_\NC , \mathcal{\iv}_\NC\big\}
			=
			 \frac{ (\NC -\tNC) }{\tNC  \NC ( \NC -1 ) } \sum_j \big\{ \psi_{\NT,j}(\eta_k) -   \Psi_\NT(\eta_k \con \NC) \big\}^{\otimes 2}
			= O\big( \NC^{-1} (\log \NC)^{2v_1} \big) \ .
		\end{align*}
	Therefore, we have the following result by the Chebyshev's inequality.
	\begin{align*}
	&
		P \Big\{ 
			\max_{k} \,  \big\{ \NC^r (\log\NC)^{v_3} \big\}  \big\| \Psi_\NT(\eta_k \con \NC) - \widehat{\Psi}_\NT(\eta_k \con \NC)  \big\| > \xi + o(1) \, \Big| \, \extendF_\NC , \mathcal{\iv}
		\Big\}
		\nonumber
		\\
		& \leq
		\frac{  \NC^{2r} (\log\NC)^{2 v_3} }{ \{\xi + o(1)\} ^2} \sum_{k=1}^{ \lceil \NC^\delta \rceil}\VAR \big\{ \Psi_\NT(\eta_k \con \NC) -  \widehat{\Psi}_\NT(\eta_k \con \NC)  \cond \mathcal{F}_\NC , \mathcal{\iv}_\NC\big\}
		= O \big( \NC^{\delta + 2r -1} (\log \NC)^{2v_1+2v_3} \big) 
		= o(1) \  .
	\end{align*}
	The last asymptotic representation is from $\delta + 2r - 1 < 0$. This concludes that the first term of the right-hand side of \eqref{proof-thm41-004} is $o_P(1)$.
	
	Next we study the second term of the right hand side of \eqref{proof-thm41-004}. From \hyperlink{(GC1)}{(GC1)}, we get $1-\iv_j \cdot J/m  \in [1-c^{-1} ,1]$ for all $j$. Thus, we have
	\begin{align*}
	\max_{k} \sup_{\eta_\NT \in \mathcal{E}_k^\circ } & \big\{ \NC^r (\log\NC)^{v_3} \big\} \big \| \{ \Psi_\NT(\eta_\NT \con \NC)  - \widehat{\Psi}_\NT(\eta_\NT \con \NC)   \}  - \{ \Psi_\NT(\eta_k \con \NC) - \widehat{\Psi}_\NT(\eta_k \con \NC) \} \big \|
	\nonumber	 
	 \\
	& = \max_{k} \sup_{\eta_\NT \in \mathcal{E}_k^\circ }
	\big\{ \NC^r (\log\NC)^{v_3} \big\}
	\Bigg\|
		\frac{1}{\NC} \sum_j \bigg( 1- \frac{\iv_j \NC}{\tNC} \bigg) \Big\{
			\psi_{\NT,j}(\eta_\NT) - \psi_{\NT,j}(\eta_k)
		\Big\}
	\Bigg\|
	\nonumber
			\\
	& \leq \max( c^{-1} -1 , 1 ) \cdot  \sup_{\eta_\NT \in \mathcal{E}^\circ } \frac{ \NC^r (\log\NC)^{v_3} }{\NC}
	\sum_j \big \|
			\psi_{\NT, j}(\eta_\NT) - \psi_{\NT, j}(\eta_k) \big \| 
		\leq 
			O \big( \NC^{r -1/2} (\log \NC)^{v_2+v_3} \big) \ .
 	\end{align*}
 	The first equality is from the definition of $\widehat{\Psi}_\NT(\cdot \con \NC)$ and $\Psi_\NT(\cdot \con \NC)$. The second equality is trivial. The inequality in the third line is from the boundedness of $1-\iv_j \cdot J/m $ and the relationship between $\mathcal{E}^\circ$ and $\mathcal{E}_k^\circ$. The last inequality is from \hyperlink{(GC5)}{(GC5)} because we have $\| \eta_\NT - \eta_k\| \leq \| \overline{\eta}_\NC \| =O(\NC^{-1/2})$ for any $\eta \in \mathcal{E}_k^\circ$ from the construction of $\mathcal{E}_k$.  This concludes that the second term of the right-hand side of \eqref{proof-thm41-004} is $o(1)$. Therefore, we obtain \eqref{proof-thm41-003} is $o_P(1)$.
		\begin{align*}
			&  \sup_{\eta_\NT \in \mathcal{E}^\circ } \big\{ \NC^r (\log\NC)^{v_3} \big\}  \big \| \Psi_\NT(\eta_\NT \con \NC) -  \widehat{\Psi}_\NT(\eta_\NT \con \NC) \big \| 
			 \\
			&
			\leq 
			\max_{k}  \big\{ \NC^r (\log\NC)^{v_3} \big\}  \big \|  \Psi_\NT(\eta_k \con \NC) - \widehat{\Psi}_\NT(\eta_k \con \NC) \big \|
			\\
			& \hspace*{1cm}
			+ 
			\max_{k} \sup_{\eta_\NT \in \mathcal{E}_k^\circ } \big\{ \NC^r (\log\NC)^{v_3} \big\}  \big \| \{ \Psi_\NT(\eta_\NT \con \NC) - \widehat{\Psi}_\NT(\eta_\NT \con \NC)   \}  - \{ \Psi_\NT(\eta_k \con \NC) - \widehat{\Psi}_\NT(\eta_k \con \NC) \} \big \|
			= o_P(1) \ .
		\end{align*}
		This concludes the proof about consistency of $\widehat{\eta}_\NT$. consistency of $\widehat{\eta}_\AT$ can be shown in a similar manner. 
		
		Next we prove consistency of $\widehat{\eta}_\CO$. First, $\randomf_\CO(\bX_{ji} \con \widehat{\theta}_\CO)$ is consistent as follows:
		\begin{align*}
			\Big\|
			\randomf_\CO(\bX_{ji} \con \widehat{\theta}_\CO)
			-
			\randomf_\CO(\bX_{ji} \con \theta_\CO^*)
			\Big\|
			&
			\leq 
			O (\log \NC) \cdot 
			\big\|
				\widehat{\theta}_\CO-\theta_\CO^*
			\big\|
			\quad (\because \ \randomf_t(\bX_{ji} \con \theta_t) = f_t(\bX_{ji} \con \theta_t) + \epsilon_{tji})
			\\
			&
			\leq 
			O (\log \NC) \cdot 
			\sum_{t \in \{\NT, \AT\} }
			\Big\{
			\big\|
				\widehat{w}_t-w_t^*
			\big\|
			+
			\big\|
				\widehat{w}_t-w_t^*
			\big\|
			\Big\}
			\\[-0.2cm]
			& =
			O_P \big( \NC^{-1/2} \log \NC \big) +
			O_P \big( \NC^{-r} \log \NC \big) \ , \ r \in (0,1/4) \ .
			\end{align*}
			The inequality in the firstline holds from \hyperlink{(GC4)}{(GC4)}. The inequality in the second line holds from $\theta_\CO = (w_\NT,w_\AT,\theta_\NT,\theta_\AT)$ and the triangle inequality. The last line holds from the convergence rate of $\widehat{w}_\NT$, $\widehat{w}_\AT$, $\widehat{\theta}_\NT$, and $\widehat{\theta}_\AT$.

		Let $\widetilde{q}_\CO$ be the solution to $\sum_{ji} \SI \big( \randomf_\CO(\bX_{ji} \con \theta_\CO^*) - \widetilde{q}_\CO \big)
		=
		\widehat{\NIT}_\CO = \NIT - \widehat{\NIT}_\NT - \widehat{\NIT}_\AT$. 		Let $B_\NC$ be the quantity that satisfy $B_\NC = O(\NC^{-r} (\log \NC) )$ and $
			\big\|
			\randomf_\CO(\bX_{ji} \con \widehat{\theta}_\CO)
			-
			\randomf_\CO(\bX_{ji} \con \theta_\CO^*)
			\big\| 
			\leq B_\NC$. Due to the increasing property of $\SI$, we find $
		\sumji \SI \big( \randomf_\CO(\bX_{ji} \con \theta_\CO^*) - B_\NC - q \big)
		\leq
		\sumji \SI \big( \randomf_\CO(\bX_{ji} \con \theta_\CO^*) + B_\NC - q \big)$. Since all three functions are decreasing in $q$, we find the solutions to the following three equations have the following relationships.
		\begin{align*}
		\left.
		\begin{matrix}
			\widetilde{q}_\CO \pm B_\NC \text{ solves }
		\sum_{ji} \SI \big( \randomf_\CO(\bX_{ji} \con \theta_\CO^*) \pm B_\NC - q \big)
		=
		\widehat{\NIT}_\CO
		\\
		\widehat{q}_\CO \text{ solves }
		\sum_{ji} \SI \big(\randomf_\CO(\bX_{ji} \con \widehat{\theta}_\CO) -  q \big)
		=
		\widehat{\NIT}_\CO
		\end{matrix}
		\right\}
		\quad \Rightarrow \quad		
		\widetilde{q}_\CO - B_\NC \leq \widehat{q}_\CO \leq \widetilde{q}_\CO + B_\NC \ .
		\end{align*}
		This implies $\big| \widetilde{q}_\CO - \widehat{q}_\CO \big| = O_P( \NC^{-r} \log \NC)$, $r \in (0,1/4)$.
		
		Let $H(q)$ be $H(q) = \sum_{ji} \big\{ \SI \big( \randomf_\CO(\bX_{ji} \con \theta_\CO^*)  - q \big) - 
			\SI \big( \randomf_\CO(\bX_{ji} \con \theta_\CO^*)- q_\CO^* \big)  \big\}$. Since $q_\CO^*$ solves the equation $\sum_{ji} \SI \big( \randomf_\CO(\bX_{ji} \con \theta_\CO^*) - q \big)
		=
		\NIT_\CO$, we find $	H(\widetilde{q}_\CO) =
			\widehat{\NIT}_\CO - \NIT_\CO$. Since $\SI$ is strictly decreasing in its argument, $q_\CO^*$ is the unique solution to $H(q_\CO^*)=0$. From Assumption \hyperlink{(GC8)}{(GC8)}, we have $\Theta(|q_2-q_1| \NC)$ units having $q_1 \leq \randomf_\CO(\bX_{ji} \con \theta_\CO^*) \leq q_2$ where there exists an interval $[q_L,q_U]$ satisfying $(q_1, q_2) \subset [q_L,q_U]$ and $q_\CO^* \in [q_L,q_U]$. 
		
		Suppose $\widetilde{q}_\CO$ does not converge to $q_\CO^*$ in probability, i.e. . $| \widetilde{q}_\CO - q_\CO^*| \geq C$ for some $C>0$ with positive probability. Without loss of generality, let $\widetilde{q}_\CO \leq q_\CO^*-C$. This implies $
		\SI \big(  \randomf_\CO(\bX_{ji} \con \theta_\CO^*) - \widetilde{q}_\CO \big)
		-
		\SI \big( \randomf_\CO(\bX_{ji} \con \theta_\CO^*) -  q_\CO^*\big) >0$. 	By taking small enough $C$, we can have $(q_\CO^*-C , q_\CO^*+C) \in [q_L, q_U]$. From the previous result, there are $\Theta(\NC)$ units satisfying $\randomf_\CO(\bX_{ji} \con \theta_\CO^* )  \in  (q_\CO^*-C/2,q_\CO^*+C/2)$. These individuals satisfy
	\begin{align*}
	\SI \big( \randomf_\CO(\bX_{ji} \con \theta_\CO^* ) - \widetilde{q}_\CO \big) 
	&
	\geq
	\SI \big( \randomf_\CO(\bX_{ji} \con \theta_\CO^* ) + C - q_\CO^* \big) 
	\geq
	\SI \big( C/2 \big) 
	\geq
	1-c \quad \text{ if } C/2 \geq h  \ .
	\end{align*}
	The last inequality holds from the form of $\SI$ with parameters $c= \Theta\big( (\log \NC)^{-1} \big)$ and $h = \Theta(\NC^{-1})$. Thus, the last inequality holds as $\NC \rightarrow \infty$. Thus, if $C \geq 2h$, we have
	{\fontsize{10}{12} \selectfont
	\begin{align*}
		H(\widetilde{q}_\CO)
		&
		\geq
		\sum_{ji}
		\ind \Big\{  \randomf_\CO(\bX_{ji} \con \theta_\CO^* ) \in( q_\CO^*-\frac{C}{2}, q_\CO^*+\frac{C}{2}) \Big\}
		\Big\{ \SI \big( \randomf_\CO(\bX_{ji} \con \theta_\CO^*)  - \widetilde{q}_\CO \big) - 
			\SI \big( \randomf_\CO(\bX_{ji} \con \theta_\CO^*)- q_\CO^* \big)  \Big\} 
		\\
		&
		\geq
		(1-c)
		\sum_{ji}
		\ind \Big\{  \randomf_\CO(\bX_{ji} \con \theta_\CO^*) \in( q_\CO^*-\frac{C}{2}, q_\CO^*+\frac{C}{2}) \Big\}	=
		\Theta(\NC) \ .
	\end{align*}}%
			Dividing by $\NIT$, we have $	H(\widetilde{q}_\CO)/\NIT = \Theta(1)$ with positive probability, but $H(\widetilde{q}_\CO)/\NIT= (\widehat{\NIT}_\CO -\NIT_\CO)/\NIT = O_P(\NC^{-1/2})$, which is a contradiction. This implies the assumption that $\widetilde{q}_\CO$ does not converge to $q_\CO^*$ in probability is wrong; i.e. $\widehat{q}_\CO \stackrel{P}{\rightarrow} q_\CO^*$.
	
	From Tailor expansion of $H(q)$ at $q_\CO^*$, we find
		\begin{align*}
			\Big| 
		\widehat{\NIT}_\CO -\NIT_\CO
		\Big|
		&
		=
		\bigg|
		\sum_{ji} \Big\{ \SI \big( \randomf_\CO(\bX_{ji} \con \theta_\CO^*)  - \widetilde{q}_\CO \big) - 
			\SI \big( \randomf_\CO(\bX_{ji} \con \theta_\CO^*)- q_\CO^* \big)  \Big\} 
		\bigg|
		\\
		&
		=
		\bigg|
		\big( \widetilde{q}_\CO  - q_\CO^* \big)
		\bigg[
			\sum_{ji} \SI '  \big( \randomf_\CO(\bX_{ji} \con \theta_\CO^*) - q_M \big)
		\bigg]
		\bigg|
		\quad \Leftarrow \ q_M \text{: intermediate point of } \widetilde{q}_\CO, q_\CO^*
		\\
		&
		\geq
		\big| \widetilde{q}_\CO  - q_\CO^* \big|
		\bigg[
			\sum_{ji} \frac{1-2c}{2h}  \ind \Big\{ \randomf_\CO(\bX_{ji} \con \theta_\CO^*) - q_M \in [-h,h]
			\Big\}
		\bigg]
		\\
		&
		=
		\big| \widetilde{q}_\CO  - q_\CO^* \big|
		 \frac{1-2c}{2h}
		\bigg[
			\sum_{ji}  \ind \Big\{ \randomf_\CO(\bX_{ji} \con \theta_\CO^*) - q_\CO^* \in [-h + q_M - q_\CO^*,h+ q_M - q_\CO^*]
			\Big\}
		\bigg] \ .
		\end{align*}
		The inequality is from the form of $\SI'$. Since the length of the interval $[-h + q_\CO^* - q_M, h + q_\CO^* - q_M]$ converges to $0$, we can use the assumption in the main paper to bound the quantity in the bracket: $\displaystyle{
		\frac{1}{\NIT}
		\sum_{ji}  \ind \Big\{ \randomf_\CO(\bX_{ji} \con \theta_\CO^*) - q_\CO^* \in [-h + q_M - q_\CO^*,h+ q_M - q_\CO^*]
			\Big\}
		=
		\Theta (h) 
		=
		\Theta (\NC^{-1})}$. Since $(1-2c)/2h = \Theta(\NC)$, we have
		{\fontsize{10}{12} \selectfont
			\begin{align*}
				\big| \widetilde{q}_\CO  - q_\CO^* \big|
				\leq
				\bigg|
				\frac{\widehat{\NIT}_\CO -\NIT_\CO}{\NIT}
				\bigg|
				\bigg[
				\underbrace{ \frac{1-2c}{2h}
		\frac{\sum_{ji}   \ind \{ \randomf_\CO(\bX_{ji} \con \theta_\CO^*) - q_\CO^* \in [-h + q_M - q_\CO^*,h+ q_M - q_\CO^*]
			\}}{\NIT}
		}_{\Theta(1)}
		\bigg]^{-1}
		= O_P(\NC^{-1/2}) \ .
			\end{align*}}%
	  Combining the above results, we find $
	 \big|	\widehat{q}_\CO - q_\CO^* \big|
	 \leq
	 \big| \widehat{q}_\CO - \widetilde{q}_\CO \big| 
	 +
	 \big| \widetilde{q}_\CO - q_\CO^* \big|
	 =
	 	O_P \big( \NC^{-r} \log \NC \big) + O_P(\NC^{-1/2})$ for any  $r \in (0,1/4)$. Therefore, we have $\big\| \widehat{\eta}_\CO - \eta_\CO^* \big\|
	 	\leq
	 	\big\| \widehat{\theta}_\CO - \theta_\CO^* \big\|
	 	+
	 	\big\| \widehat{q}_\CO - q_\CO^* \big\|
	 	=
	 	O_P \big( \NC^{-r} \log \NC \big)$.

\subsection{Proof of Lemma \ref{lemma:WLLN}}									\label{sec:C6}

Let $\bsS(\bw) = (\mathcal{S}_1(\bw),\ldots, \mathcal{S}_\NC(\bw))\T$ and $\IV = (\iv_1 , \ldots, \iv_\NC)\T$. From Lemma \ref{lem:301},  the expectation and the variance of the above quantity are given by
\begin{align*}
	& \EXP \bigg\{ \frac{\NC}{\NIT \tNC} \bsS(\bw)\T \IV \, \bigg| \, \F_\NC , \mathcal{\iv}_\NC\bigg\}
		= 
		\frac{\NC}{\NIT \tNC}  \bsS(\bw) \T \frac{\tNC}{\NC} \bm{1} 
		=
		\frac{1}{\NIT}  \bsS(\bw) \T  \bm{1} 
		=
		\frac{1}{\NIT} \sumji S_{ji} \ind(\bW_{ji} = \bw) \ ,
		\\
		& \VAR \bigg\{ \frac{\NC}{\NIT \tNC} \bsS(\bw)\T \IV \, \bigg| \, \F_\NC , \mathcal{\iv}_\NC\bigg\}
		= 
		\frac{\NC^2}{\NIT^2 \tNC^2}  \bsS(\bw) \T \frac{\tNC(\NC - \tNC)}{\NC(\NC-1)} \Pi_\NC \bsS(\bw) 
		=
		\frac{\NC(\NC-\tNC)}{\NIT^2(\NC-1)\tNC} \sum_j \big( \mathcal{S}_j(\bw) - \overline{\mathcal{S}}(\bw) \big)^2
\end{align*}
where $\overline{\mathcal{S}}(\bw) = \sum_j \mathcal{S}_j(\bw) / \NC$. Since $\mathcal{S}_j$ is upper-bounded by a constant $M$ and $\NIT$ is greater than $\NC$, we get 
\begin{align*}
	\VAR \bigg\{ \frac{\NC}{\NIT \tNC} \bsS(\bw)\T \IV \, \bigg| \, \F_\NC , \mathcal{\iv}_\NC\bigg\}
		\leq
		\frac{\NC(\NC-\tNC)}{\NC^2(\NC-1)\tNC} \NC M^2 = \frac{(\NC - \tNC)M^2}{(\NC-1) \tNC}
		= o(1) \ .
\end{align*}
Therefore, we obtain the following result as for any $\epsilon>0$ by the Chebyshev's inequality.
	\begin{align*}
		\lim_{\NC \rightarrow \infty}	P \bigg\{ \bigg|  
		\frac{\NC}{\NIT \tNC} \sumji \ind(\iv_j = 1) \mathcal{S}_j(\bw) 
		-
		\frac{1}{\NIT} \sumji \mathcal{S}_j(\bw) \bigg| > \epsilon \, \Bigg| \, \F_\NC , \mathcal{\iv}_\NC \bigg\} = 0  \ .
\end{align*}
We have the second result by replacing $\tNC$ and $\IV$ with $\NC-\tNC$ and $\bm{1} - \IV$, respectively.

\bibliographystyle{chicago}

\bibliography{IVnet.bib}

\end{document}